\newif\ifcomments     
\newif\ifanonymous    
\newif\ifsource       
\newif\ifextended     
\newif\ifsubmission   
\newif\ifpublic       

\commentsfalse        
\anonymousfalse
\extendedtrue
\sourcetrue

\submissionfalse      
\publictrue

\ifsubmission
\publicfalse
\commentsfalse
\extendedfalse
\anonymoustrue
\documentclass[acmlarge,review,anonymous]{acmart}
\else
\ifanonymous
\documentclass[acmlarge,anonymous]{acmart}
\else
\documentclass[acmsmall]{acmart}
\fi
\fi

\ifpublic
\submissionfalse
\commentsfalse
\anonymousfalse
\fi

\ifsource
\fi

\usepackage{amsmath}
\usepackage{amssymb}
\usepackage{supertabular}
\usepackage{stmaryrd}
\usepackage{color}
\usepackage{multirow}
\usepackage{calc}
\usepackage[T1]{fontenc}
\usepackage{graphicx}
\usepackage{xspace}
\usepackage{dashrule}  
\usepackage{stackrel}
\usepackage{enumerate}
\usepackage{framed}
\usepackage{bussproofs}
\usepackage{mdframed}  
\usepackage{hyperref}
\usepackage{comment}
\usepackage{ifthen}
\usepackage{pdftexcmds}
\usepackage{mathpartir}
\usepackage{ottalt}
\usepackage{fancyvrb}
\usepackage{listings}
\usepackage{url}

\newcommand\cd[1]{\lstinline{#1}}

\newif\ifadmissible\admissiblefalse
\newcommand\suppress[1]{\ifadmissible{[#1]}\else{}\fi}

\newcommand{\ottdrule}[4][]{{\displaystyle\frac{\begin{array}{l}#2\end{array}}{#3}\quad\ottdrulename{#4}}}
\newcommand{\ottusedrule}[1]{\[#1\]}
\newcommand{\ottpremise}[1]{ #1 \\}
\newenvironment{ottdefnblock}[3][]{ \framebox{\mbox{#2}} \quad #3 \\[0pt]}{}

\newcommand{\ottnt}[1]{\mathit{#1}}
\newcommand{\ottmv}[1]{\mathit{#1}}
\newcommand{\ottkw}[1]{\mathbf{#1}}
\newcommand{\ottsym}[1]{#1}
\newcommand{\ottcom}[1]{\text{#1}}
\newcommand{\ottdrulename}[1]{\textsc{#1}}

\newcommand{\ottafterlastrule}{\\}

\newcommand{\ottdruleNomBot}[1]{\ottdrule[#1]{%
}{
 \ottkw{Nom}  \leq  \ottnt{R} }{%
{\ottdrulename{NomBot}}{}%
}}

\newcommand{\ottdruleRepTop}[1]{\ottdrule[#1]{%
}{
 \ottnt{R}  \leq  \ottkw{Rep} }{%
{\ottdrulename{RepTop}}{}%
}}

\newcommand{\ottdruleRefl}[1]{\ottdrule[#1]{%
}{
 \ottnt{R}  \leq  \ottnt{R} }{%
{\ottdrulename{Refl}}{}%
}}

\newcommand{\ottdruleTrans}[1]{\ottdrule[#1]{%
\ottpremise{ \ottnt{R_{{\mathrm{1}}}}  \leq  \ottnt{R_{{\mathrm{2}}}} }%
\ottpremise{ \ottnt{R_{{\mathrm{2}}}}  \leq  \ottnt{R_{{\mathrm{3}}}} }%
}{
 \ottnt{R_{{\mathrm{1}}}}  \leq  \ottnt{R_{{\mathrm{3}}}} }{%
{\ottdrulename{Trans}}{}%
}}

\newcommand{\ottdefnSubRole}[1]{\begin{ottdefnblock}[#1]{$ \ottnt{R_{{\mathrm{1}}}}  \leq  \ottnt{R_{{\mathrm{2}}}} $}{\ottcom{Subroling judgement}}
\ottusedrule{\ottdruleNomBot{}}
\ottusedrule{\ottdruleRepTop{}}
\ottusedrule{\ottdruleRefl{}}
\ottusedrule{\ottdruleTrans{}}
\end{ottdefnblock}}

\newcommand{\ottdefnsJSubRole}{
\ottdefnSubRole{}}

\newcommand{\ottdruleRolePathXXAbsConst}[1]{\ottdrule[#1]{%
\ottpremise{\ottmv{F}  \ottsym{:}   \ottnt{A} \  \ottsym{@} \  \overline{R}  \, \in \,  \Sigma_0 }%
}{
 \mathsf{Roles}\; ( \ottmv{F} )  =  \overline{R} }{%
{\ottdrulename{RolePath\_AbsConst}}{}%
}}

\newcommand{\ottdruleRolePathXXConst}[1]{\ottdrule[#1]{%
\ottpremise{\ottmv{F}  \ottsym{:}   \ottnt{A} \  \ottsym{@} \  \overline{R} \ \mathsf{where}\  \ottnt{p}  \sim_{ \ottnt{R_{{\mathrm{1}}}} }  \ottnt{a}  \, \in \,  \Sigma_0 }%
}{
 \mathsf{Roles}\; ( \ottmv{F} )  =  \overline{R} }{%
{\ottdrulename{RolePath\_Const}}{}%
}}

\newcommand{\ottdruleRolePathXXTApp}[1]{\ottdrule[#1]{%
\ottpremise{ \mathsf{Roles}\; ( \ottnt{a} )  =  \ottnt{R_{{\mathrm{1}}}}  \ottsym{,}  \overline{R} }%
}{
 \mathsf{Roles}\; (  \ottnt{a} \  \ottnt{b} ^{ \ottnt{R_{{\mathrm{1}}}} }  )  =  \overline{R} }{%
{\ottdrulename{RolePath\_TApp}}{}%
}}

\newcommand{\ottdruleRolePathXXApp}[1]{\ottdrule[#1]{%
\ottpremise{ \mathsf{Roles}\; ( \ottnt{a} )  =  \ottkw{Nom}  \ottsym{,}  \overline{R} }%
}{
 \mathsf{Roles}\; (  \ottnt{a} \  \ottnt{b} ^{ \ottsym{+} }  )  =  \overline{R} }{%
{\ottdrulename{RolePath\_App}}{}%
}}

\newcommand{\ottdruleRolePathXXIApp}[1]{\ottdrule[#1]{%
\ottpremise{ \mathsf{Roles}\; ( \ottnt{a} )  =  \overline{R} }%
}{
 \mathsf{Roles}\; (  \ottnt{a} \  \Box ^{ \ottsym{-} }  )  =  \overline{R} }{%
{\ottdrulename{RolePath\_IApp}}{}%
}}

\newcommand{\ottdruleRolePathXXCApp}[1]{\ottdrule[#1]{%
\ottpremise{ \mathsf{Roles}\; ( \ottnt{a} )  =  \overline{R} }%
}{
 \mathsf{Roles}\; (  \ottnt{a} \; \bullet  )  =  \overline{R} }{%
{\ottdrulename{RolePath\_CApp}}{}%
}}

\newcommand{\ottdefnRolePath}[1]{\begin{ottdefnblock}[#1]{$ \mathsf{Roles}\; ( \ottnt{a} )  =  \overline{R} $}{\ottcom{Calculate remaining roles for path (partial function)}}
\ottusedrule{\ottdruleRolePathXXAbsConst{}}
\ottusedrule{\ottdruleRolePathXXConst{}}
\ottusedrule{\ottdruleRolePathXXTApp{}}
\ottusedrule{\ottdruleRolePathXXApp{}}
\ottusedrule{\ottdruleRolePathXXIApp{}}
\ottusedrule{\ottdruleRolePathXXCApp{}}
\end{ottdefnblock}}

\newcommand{\ottdruleAppsPathXXAbsConst}[1]{\ottdrule[#1]{%
\ottpremise{\ottmv{F}  \ottsym{:}   \ottnt{A} \  \ottsym{@} \  \overline{R}  \, \in \,  \Sigma_0 }%
}{
 \ottmv{F}  \leftrightarrow_{ \ottnt{R} }  \ottmv{F} \;    }{%
{\ottdrulename{AppsPath\_AbsConst}}{}%
}}

\newcommand{\ottdruleAppsPathXXConst}[1]{\ottdrule[#1]{%
\ottpremise{\ottmv{F}  \ottsym{:}   \ottnt{A} \  \ottsym{@} \  \overline{R} \ \mathsf{where}\  \ottnt{p}  \sim_{ \ottnt{R_{{\mathrm{1}}}} }  \ottnt{a}  \, \in \,  \Sigma_0 }%
\ottpremise{ \neg  \ottsym{(}   \ottnt{R_{{\mathrm{1}}}}  \leq  \ottnt{R}   \ottsym{)} }%
}{
 \ottmv{F}  \leftrightarrow_{ \ottnt{R} }  \ottmv{F} \;    }{%
{\ottdrulename{AppsPath\_Const}}{}%
}}

\newcommand{\ottdruleAppsPathXXApp}[1]{\ottdrule[#1]{%
\ottpremise{ \ottnt{a}  \leftrightarrow_{ \ottnt{R} }  \ottmv{F} \; \overline{\upsilon} }%
}{
 \ottsym{(}   \ottnt{a} \  \ottnt{b'} ^{ \ottnt{R_{{\mathrm{1}}}} }   \ottsym{)}  \leftrightarrow_{ \ottnt{R} }  \ottmv{F} \; \ottsym{(}   \overline{\upsilon} \;  \ottnt{R_{{\mathrm{1}}}}    \ottsym{)} }{%
{\ottdrulename{AppsPath\_App}}{}%
}}

\newcommand{\ottdruleAppsPathXXIApp}[1]{\ottdrule[#1]{%
\ottpremise{ \ottnt{a}  \leftrightarrow_{ \ottnt{R} }  \ottmv{F} \; \overline{\upsilon} }%
}{
 \ottsym{(}   \ottnt{a} \  \ottnt{b} ^{ \ottsym{-} }   \ottsym{)}  \leftrightarrow_{ \ottnt{R} }  \ottmv{F} \; \ottsym{(}   \overline{\upsilon} \;  \ottsym{-}    \ottsym{)} }{%
{\ottdrulename{AppsPath\_IApp}}{}%
}}

\newcommand{\ottdruleAppsPathXXCApp}[1]{\ottdrule[#1]{%
\ottpremise{ \ottnt{a}  \leftrightarrow_{ \ottnt{R} }  \ottmv{F} \; \overline{\upsilon} }%
}{
 \ottsym{(}   \ottnt{a} \; \bullet   \ottsym{)}  \leftrightarrow_{ \ottnt{R} }  \ottmv{F} \; \ottsym{(}   \overline{\upsilon} \;  \bullet    \ottsym{)} }{%
{\ottdrulename{AppsPath\_CApp}}{}%
}}

\newcommand{\ottdefnAppsPath}[1]{\begin{ottdefnblock}[#1]{$ \ottnt{a}  \leftrightarrow_{ \ottnt{R} }  \ottmv{F} \; \overline{\upsilon} $}{\ottcom{Calculate the application pattern of a, a value at role R (partial function)}}
\ottusedrule{\ottdruleAppsPathXXAbsConst{}}
\ottusedrule{\ottdruleAppsPathXXConst{}}
\ottusedrule{\ottdruleAppsPathXXApp{}}
\ottusedrule{\ottdruleAppsPathXXIApp{}}
\ottusedrule{\ottdruleAppsPathXXCApp{}}
\end{ottdefnblock}}

\newcommand{\ottdruleARXXnil}[1]{\ottdrule[#1]{%
}{
    \leftrightarrow   \cdot }{%
{\ottdrulename{AR\_nil}}{}%
}}

\newcommand{\ottdruleARXXconsTApp}[1]{\ottdrule[#1]{%
\ottpremise{\overline{\upsilon}  \leftrightarrow  \overline{R}}%
}{
  \ottnt{R_{{\mathrm{1}}}}  \; \overline{\upsilon}   \leftrightarrow  \ottnt{R_{{\mathrm{1}}}}  \ottsym{,}  \overline{R}}{%
{\ottdrulename{AR\_consTApp}}{}%
}}

\newcommand{\ottdruleARXXconsApp}[1]{\ottdrule[#1]{%
\ottpremise{\overline{\upsilon}  \leftrightarrow  \overline{R}}%
}{
  \ottsym{+}  \; \overline{\upsilon}   \leftrightarrow  \ottkw{Nom}  \ottsym{,}  \overline{R}}{%
{\ottdrulename{AR\_consApp}}{}%
}}

\newcommand{\ottdruleARXXconsIApp}[1]{\ottdrule[#1]{%
\ottpremise{\overline{\upsilon}  \leftrightarrow  \overline{R}}%
}{
  \ottsym{-}  \; \overline{\upsilon}   \leftrightarrow  \overline{R}}{%
{\ottdrulename{AR\_consIApp}}{}%
}}

\newcommand{\ottdruleARXXconsCApp}[1]{\ottdrule[#1]{%
\ottpremise{\overline{\upsilon}  \leftrightarrow  \overline{R}}%
}{
  \bullet  \; \overline{\upsilon}   \leftrightarrow  \overline{R}}{%
{\ottdrulename{AR\_consCApp}}{}%
}}

\newcommand{\ottdefnAppRoles}[1]{\begin{ottdefnblock}[#1]{$\overline{\upsilon}  \leftrightarrow  \overline{R}$}{}
\ottusedrule{\ottdruleARXXnil{}}
\ottusedrule{\ottdruleARXXconsTApp{}}
\ottusedrule{\ottdruleARXXconsApp{}}
\ottusedrule{\ottdruleARXXconsIApp{}}
\ottusedrule{\ottdruleARXXconsCApp{}}
\end{ottdefnblock}}

\newcommand{\ottdruleSatXXConst}[1]{\ottdrule[#1]{%
\ottpremise{\ottmv{F}  \ottsym{:}   \ottnt{A} \  \ottsym{@} \  \overline{R}  \, \in \,  \Sigma_0 }%
\ottpremise{\overline{\upsilon}  \leftrightarrow  \overline{R}}%
}{
\ottkw{Sat} \, \ottmv{F} \, \overline{\upsilon}}{%
{\ottdrulename{Sat\_Const}}{}%
}}

\newcommand{\ottdruleSatXXAxiom}[1]{\ottdrule[#1]{%
\ottpremise{\ottmv{F}  \ottsym{:}   \ottnt{A_{{\mathrm{1}}}} \  \ottsym{@} \  \overline{R} \ \mathsf{where}\  \ottnt{p}  \sim_{ \ottnt{R_{{\mathrm{1}}}} }  \ottnt{a_{{\mathrm{0}}}}  \, \in \,  \Sigma_0 }%
\ottpremise{ \neg  \ottsym{(}   \ottnt{R_{{\mathrm{1}}}}  \leq  \ottkw{Nom}   \ottsym{)} }%
\ottpremise{\overline{\upsilon}  \leftrightarrow  \overline{R}}%
}{
\ottkw{Sat} \, \ottmv{F} \, \overline{\upsilon}}{%
{\ottdrulename{Sat\_Axiom}}{}%
}}

\newcommand{\ottdefnSatApp}[1]{\begin{ottdefnblock}[#1]{$\ottkw{Sat} \, \ottmv{F} \, \overline{\upsilon}$}{}
\ottusedrule{\ottdruleSatXXConst{}}
\ottusedrule{\ottdruleSatXXAxiom{}}
\end{ottdefnblock}}

\newcommand{\ottdrulePatCtxXXConst}[1]{\ottdrule[#1]{%
}{
 \mathsf{PatCtx}\; ( \ottmv{F} , \ottmv{F} : \ottnt{A} ) =  \varnothing  ;  \ottnt{A}  ;  \varnothing  ;  \ottkw{emptylist} }{%
{\ottdrulename{PatCtx\_Const}}{}%
}}

\newcommand{\ottdrulePatCtxXXPiRel}[1]{\ottdrule[#1]{%
\ottpremise{ \mathsf{PatCtx}\; ( \ottnt{p} , \ottmv{F} : \ottnt{B} ) =  \Gamma  ;   \mathrm{\Pi}^ \ottsym{+} \ottmv{x} \!:\! \ottnt{A'} \to \ottnt{A}   ;  \Omega  ;  V }%
}{
 \mathsf{PatCtx}\; (  \ottnt{p} \  \ottmv{x} ^{ \ottnt{R} }  , \ottmv{F} : \ottnt{B} ) =   \Gamma ,  \ottmv{x} \!:\! \ottnt{A'}   ;  \ottnt{A}  ;   \Omega , \ottmv{x} \!:\! \ottnt{R}   ;  V }{%
{\ottdrulename{PatCtx\_PiRel}}{}%
}}

\newcommand{\ottdrulePatCtxXXPiIrr}[1]{\ottdrule[#1]{%
\ottpremise{ \mathsf{PatCtx}\; ( \ottnt{p} , \ottmv{F} : \ottnt{B} ) =  \Gamma  ;   \mathrm{\Pi}^ \ottsym{-} \ottmv{x} \!:\! \ottnt{A'} \to \ottnt{A}   ;  \Omega  ;  V }%
}{
 \mathsf{PatCtx}\; (  \ottnt{p} \  \Box ^{ \ottsym{-} }  , \ottmv{F} : \ottnt{B} ) =   \Gamma ,  \ottmv{x} \!:\! \ottnt{A'}   ;  \ottnt{A}  ;  \Omega  ;  V  \ottsym{,}  \ottmv{x} }{%
{\ottdrulename{PatCtx\_PiIrr}}{}%
}}

\newcommand{\ottdrulePatCtxXXCPi}[1]{\ottdrule[#1]{%
\ottpremise{ \mathsf{PatCtx}\; ( \ottnt{p} , \ottmv{F} : \ottnt{B} ) =  \Gamma  ;   \forall \ottmv{c} \!:\! \phi . \ottnt{A}   ;  \Omega  ;  V }%
}{
 \mathsf{PatCtx}\; (  \ottnt{p} \; \bullet  , \ottmv{F} : \ottnt{B} ) =   \Gamma ,  \ottmv{c} \!:\! \phi   ;  \ottnt{A}  ;  \Omega  ;  V }{%
{\ottdrulename{PatCtx\_CPi}}{}%
}}

\newcommand{\ottdefnPatternContexts}[1]{\begin{ottdefnblock}[#1]{$ \mathsf{PatCtx}\; ( \ottnt{p} , \ottmv{F} : \ottnt{A} ) =  \Gamma  ;  \ottnt{B}  ;  \Omega  ;  V $}{\ottcom{Contexts generated by a pattern (variables bound by the pattern)}}
\ottusedrule{\ottdrulePatCtxXXConst{}}
\ottusedrule{\ottdrulePatCtxXXPiRel{}}
\ottusedrule{\ottdrulePatCtxXXPiIrr{}}
\ottusedrule{\ottdrulePatCtxXXCPi{}}
\end{ottdefnblock}}

\newcommand{\ottdruleRenameXXBase}[1]{\ottdrule[#1]{%
}{
 \mathsf{rename} \  \ottmv{F}  \rightarrow  \ottnt{a}  \ \mathsf{to} \  \ottmv{F}  \rightarrow  \ottnt{a}  \ \mathsf{excluding} \  \Delta  \ \mathsf{and} \   \varnothing  }{%
{\ottdrulename{Rename\_Base}}{}%
}}

\newcommand{\ottdruleRenameXXAppRel}[1]{\ottdrule[#1]{%
\ottpremise{ \mathsf{rename} \  \ottnt{p_{{\mathrm{1}}}}  \rightarrow  \ottnt{a_{{\mathrm{1}}}}  \ \mathsf{to} \  \ottnt{p_{{\mathrm{2}}}}  \rightarrow  \ottnt{a_{{\mathrm{2}}}}  \ \mathsf{excluding} \  \Delta  \ \mathsf{and} \  \Delta' }%
\ottpremise{\ottmv{y} \, \not\in \, \ottsym{(}   \Delta ,  \Delta'   \ottsym{)}}%
}{
 \mathsf{rename} \  \ottsym{(}   \ottnt{p_{{\mathrm{1}}}} \  \ottmv{x} ^{ \ottnt{R} }   \ottsym{)}  \rightarrow  \ottnt{a_{{\mathrm{1}}}}  \ \mathsf{to} \  \ottsym{(}   \ottnt{p_{{\mathrm{2}}}} \  \ottmv{y} ^{ \ottnt{R} }   \ottsym{)}  \rightarrow  \ottsym{(}   \ottnt{a_{{\mathrm{2}}}}  \lbrace  \ottmv{y}   \ottsym{/}   \ottmv{x}   \rbrace   \ottsym{)}  \ \mathsf{excluding} \  \Delta  \ \mathsf{and} \  \ottsym{(}  \Delta'  \ottsym{,}  \ottmv{y}  \ottsym{)} }{%
{\ottdrulename{Rename\_AppRel}}{}%
}}

\newcommand{\ottdruleRenameXXAppIrrel}[1]{\ottdrule[#1]{%
\ottpremise{ \mathsf{rename} \  \ottnt{p_{{\mathrm{1}}}}  \rightarrow  \ottnt{a_{{\mathrm{1}}}}  \ \mathsf{to} \  \ottnt{p_{{\mathrm{2}}}}  \rightarrow  \ottnt{a_{{\mathrm{2}}}}  \ \mathsf{excluding} \  \Delta  \ \mathsf{and} \  \Delta' }%
}{
 \mathsf{rename} \  \ottsym{(}   \ottnt{p_{{\mathrm{1}}}} \  \Box ^{ \ottsym{-} }   \ottsym{)}  \rightarrow  \ottnt{a_{{\mathrm{1}}}}  \ \mathsf{to} \  \ottsym{(}   \ottnt{p_{{\mathrm{2}}}} \  \Box ^{ \ottsym{-} }   \ottsym{)}  \rightarrow  \ottnt{a_{{\mathrm{2}}}}  \ \mathsf{excluding} \  \Delta  \ \mathsf{and} \  \Delta' }{%
{\ottdrulename{Rename\_AppIrrel}}{}%
}}

\newcommand{\ottdruleRenameXXCApp}[1]{\ottdrule[#1]{%
\ottpremise{ \mathsf{rename} \  \ottnt{p_{{\mathrm{1}}}}  \rightarrow  \ottnt{a_{{\mathrm{1}}}}  \ \mathsf{to} \  \ottnt{p_{{\mathrm{2}}}}  \rightarrow  \ottnt{a_{{\mathrm{2}}}}  \ \mathsf{excluding} \  \Delta  \ \mathsf{and} \  \Delta' }%
}{
 \mathsf{rename} \  \ottsym{(}   \ottnt{p_{{\mathrm{1}}}} \; \bullet   \ottsym{)}  \rightarrow  \ottnt{a_{{\mathrm{1}}}}  \ \mathsf{to} \  \ottsym{(}   \ottnt{p_{{\mathrm{2}}}} \; \bullet   \ottsym{)}  \rightarrow  \ottnt{a_{{\mathrm{2}}}}  \ \mathsf{excluding} \  \Delta  \ \mathsf{and} \  \Delta' }{%
{\ottdrulename{Rename\_CApp}}{}%
}}

\newcommand{\ottdefnRename}[1]{\begin{ottdefnblock}[#1]{$ \mathsf{rename} \  \ottnt{p}  \rightarrow  \ottnt{a}  \ \mathsf{to} \  \ottnt{p'}  \rightarrow  \ottnt{a'}  \ \mathsf{excluding} \  \Delta  \ \mathsf{and} \  \Delta' $}{\ottcom{rename with fresh variables}}
\ottusedrule{\ottdruleRenameXXBase{}}
\ottusedrule{\ottdruleRenameXXAppRel{}}
\ottusedrule{\ottdruleRenameXXAppIrrel{}}
\ottusedrule{\ottdruleRenameXXCApp{}}
\end{ottdefnblock}}

\newcommand{\ottdefnsJRename}{
\ottdefnRename{}}

\newcommand{\ottdruleMatchSubstXXConst}[1]{\ottdrule[#1]{%
}{
 \mathsf{MatchSubst}\; (  \ottmv{F}  ,  \ottmv{F}  ,  \ottnt{b}  ) =  \ottnt{b} }{%
{\ottdrulename{MatchSubst\_Const}}{}%
}}

\newcommand{\ottdruleMatchSubstXXAppRelR}[1]{\ottdrule[#1]{%
\ottpremise{ \mathsf{MatchSubst}\; (  \ottnt{a_{{\mathrm{1}}}}  ,  \ottnt{p_{{\mathrm{1}}}}  ,  \ottnt{b_{{\mathrm{1}}}}  ) =  \ottnt{b_{{\mathrm{2}}}} }%
}{
 \mathsf{MatchSubst}\; (  \ottsym{(}   \ottnt{a_{{\mathrm{1}}}} \  \ottnt{a} ^{ \ottnt{R} }   \ottsym{)}  ,  \ottsym{(}   \ottnt{p_{{\mathrm{1}}}} \  \ottmv{x} ^{ \ottnt{R} }   \ottsym{)}  ,  \ottnt{b_{{\mathrm{1}}}}  ) =  \ottsym{(}   \ottnt{b_{{\mathrm{2}}}}  \lbrace  \ottnt{a}   \ottsym{/}   \ottmv{x}   \rbrace   \ottsym{)} }{%
{\ottdrulename{MatchSubst\_AppRelR}}{}%
}}

\newcommand{\ottdruleMatchSubstXXAppIrrel}[1]{\ottdrule[#1]{%
\ottpremise{ \mathsf{MatchSubst}\; (  \ottnt{a_{{\mathrm{1}}}}  ,  \ottnt{p_{{\mathrm{1}}}}  ,  \ottnt{b_{{\mathrm{1}}}}  ) =  \ottnt{b_{{\mathrm{2}}}} }%
}{
 \mathsf{MatchSubst}\; (  \ottsym{(}   \ottnt{a_{{\mathrm{1}}}} \  \ottnt{a} ^{ \ottsym{-} }   \ottsym{)}  ,  \ottsym{(}   \ottnt{p_{{\mathrm{1}}}} \  \Box ^{ \ottsym{-} }   \ottsym{)}  ,  \ottnt{b_{{\mathrm{1}}}}  ) =  \ottnt{b_{{\mathrm{2}}}} }{%
{\ottdrulename{MatchSubst\_AppIrrel}}{}%
}}

\newcommand{\ottdruleMatchSubstXXCApp}[1]{\ottdrule[#1]{%
\ottpremise{ \mathsf{MatchSubst}\; (  \ottnt{a_{{\mathrm{1}}}}  ,  \ottnt{a_{{\mathrm{2}}}}  ,  \ottnt{b_{{\mathrm{1}}}}  ) =  \ottnt{b_{{\mathrm{2}}}} }%
}{
 \mathsf{MatchSubst}\; (  \ottsym{(}   \ottnt{a_{{\mathrm{1}}}} \; \bullet   \ottsym{)}  ,  \ottsym{(}   \ottnt{a_{{\mathrm{2}}}} \; \bullet   \ottsym{)}  ,  \ottnt{b_{{\mathrm{1}}}}  ) =  \ottnt{b_{{\mathrm{2}}}} }{%
{\ottdrulename{MatchSubst\_CApp}}{}%
}}

\newcommand{\ottdefnMatchSubst}[1]{\begin{ottdefnblock}[#1]{$ \mathsf{MatchSubst}\; (  \ottnt{a_{{\mathrm{1}}}}  ,  \ottnt{p}  ,  \ottnt{b_{{\mathrm{1}}}}  ) =  \ottnt{b_{{\mathrm{2}}}} $}{\ottcom{match and substitute}}
\ottusedrule{\ottdruleMatchSubstXXConst{}}
\ottusedrule{\ottdruleMatchSubstXXAppRelR{}}
\ottusedrule{\ottdruleMatchSubstXXAppIrrel{}}
\ottusedrule{\ottdruleMatchSubstXXCApp{}}
\end{ottdefnblock}}

\newcommand{\ottdefnsJMatchSubst}{
\ottdefnMatchSubst{}}

\newcommand{\ottdruletmXXpatternXXagreeXXConst}[1]{\ottdrule[#1]{%
}{
 \ottmv{F}   \leftrightarrow   \ottmv{F} }{%
{\ottdrulename{tm\_pattern\_agree\_Const}}{}%
}}

\newcommand{\ottdruletmXXpatternXXagreeXXAppRelR}[1]{\ottdrule[#1]{%
\ottpremise{ \ottnt{a_{{\mathrm{1}}}}   \leftrightarrow   \ottnt{p_{{\mathrm{1}}}} }%
}{
 \ottsym{(}   \ottnt{a_{{\mathrm{1}}}} \  \ottnt{a_{{\mathrm{2}}}} ^{ \ottnt{R} }   \ottsym{)}   \leftrightarrow   \ottsym{(}   \ottnt{p_{{\mathrm{1}}}} \  \ottmv{x} ^{ \ottnt{R} }   \ottsym{)} }{%
{\ottdrulename{tm\_pattern\_agree\_AppRelR}}{}%
}}

\newcommand{\ottdruletmXXpatternXXagreeXXAppIrrel}[1]{\ottdrule[#1]{%
\ottpremise{ \ottnt{a_{{\mathrm{1}}}}   \leftrightarrow   \ottnt{p_{{\mathrm{1}}}} }%
}{
 \ottsym{(}   \ottnt{a_{{\mathrm{1}}}} \  \ottnt{a} ^{ \ottsym{-} }   \ottsym{)}   \leftrightarrow   \ottsym{(}   \ottnt{p_{{\mathrm{1}}}} \  \Box ^{ \ottsym{-} }   \ottsym{)} }{%
{\ottdrulename{tm\_pattern\_agree\_AppIrrel}}{}%
}}

\newcommand{\ottdruletmXXpatternXXagreeXXCApp}[1]{\ottdrule[#1]{%
\ottpremise{ \ottnt{a_{{\mathrm{1}}}}   \leftrightarrow   \ottnt{p_{{\mathrm{1}}}} }%
}{
 \ottsym{(}   \ottnt{a_{{\mathrm{1}}}} \; \bullet   \ottsym{)}   \leftrightarrow   \ottsym{(}   \ottnt{p_{{\mathrm{1}}}} \; \bullet   \ottsym{)} }{%
{\ottdrulename{tm\_pattern\_agree\_CApp}}{}%
}}

\newcommand{\ottdefntmXXpatternXXagree}[1]{\begin{ottdefnblock}[#1]{$ \ottnt{a}   \leftrightarrow   \ottnt{p} $}{\ottcom{term and pattern agree}}
\ottusedrule{\ottdruletmXXpatternXXagreeXXConst{}}
\ottusedrule{\ottdruletmXXpatternXXagreeXXAppRelR{}}
\ottusedrule{\ottdruletmXXpatternXXagreeXXAppIrrel{}}
\ottusedrule{\ottdruletmXXpatternXXagreeXXCApp{}}
\end{ottdefnblock}}

\newcommand{\ottdefnsJTmPatternAgree}{
\ottdefntmXXpatternXXagree{}}

\newcommand{\ottdrulesubtmXXpatternXXagreeXXBase}[1]{\ottdrule[#1]{%
\ottpremise{ \ottnt{a}   \leftrightarrow   \ottnt{p} }%
}{
 \ottnt{a}  \sqsupseteq  \ottnt{p} }{%
{\ottdrulename{subtm\_pattern\_agree\_Base}}{}%
}}

\newcommand{\ottdrulesubtmXXpatternXXagreeXXApp}[1]{\ottdrule[#1]{%
\ottpremise{ \ottnt{a}  \sqsupseteq  \ottnt{p} }%
}{
  \ottnt{a} \  \ottnt{a_{{\mathrm{2}}}} ^{ \nu }   \sqsupseteq  \ottnt{p} }{%
{\ottdrulename{subtm\_pattern\_agree\_App}}{}%
}}

\newcommand{\ottdrulesubtmXXpatternXXagreeXXCAppp}[1]{\ottdrule[#1]{%
\ottpremise{ \ottnt{a}  \sqsupseteq  \ottnt{p} }%
}{
  \ottnt{a} \; \bullet   \sqsupseteq  \ottnt{p} }{%
{\ottdrulename{subtm\_pattern\_agree\_CAppp}}{}%
}}

\newcommand{\ottdefnsubtmXXpatternXXagree}[1]{\begin{ottdefnblock}[#1]{$ \ottnt{a}  \sqsupseteq  \ottnt{p} $}{\ottcom{sub-term agrees with pattern}}
\ottusedrule{\ottdrulesubtmXXpatternXXagreeXXBase{}}
\ottusedrule{\ottdrulesubtmXXpatternXXagreeXXApp{}}
\ottusedrule{\ottdrulesubtmXXpatternXXagreeXXCAppp{}}
\end{ottdefnblock}}

\newcommand{\ottdefnsJSubTmPatternAgree}{
\ottdefnsubtmXXpatternXXagree{}}

\newcommand{\ottdruleValuePathXXAbsConst}[1]{\ottdrule[#1]{%
\ottpremise{\ottmv{F}  \ottsym{:}   \ottnt{A} \  \ottsym{@} \  \overline{R}  \, \in \,  \Sigma_0 }%
}{
 \mathsf{Head}( \ottmv{F} )  \mathrel{+\!\!\!\!\!\!\rightarrow}   \ottmv{F} }{%
{\ottdrulename{ValuePath\_AbsConst}}{}%
}}

\newcommand{\ottdruleValuePathXXConst}[1]{\ottdrule[#1]{%
\ottpremise{\ottmv{F}  \ottsym{:}   \ottnt{A} \  \ottsym{@} \  \overline{R} \ \mathsf{where}\  \ottnt{p}  \sim_{ \ottnt{R_{{\mathrm{1}}}} }  \ottnt{a}  \, \in \,  \Sigma_0 }%
}{
 \mathsf{Head}( \ottmv{F} )  \mathrel{+\!\!\!\!\!\!\rightarrow}   \ottmv{F} }{%
{\ottdrulename{ValuePath\_Const}}{}%
}}

\newcommand{\ottdruleValuePathXXApp}[1]{\ottdrule[#1]{%
\ottpremise{ \mathsf{Head}( \ottnt{w} )  \mathrel{+\!\!\!\!\!\!\rightarrow}   \ottmv{F} }%
}{
 \mathsf{Head}(  \ottnt{w} \  \ottnt{b'} ^{ \nu }  )  \mathrel{+\!\!\!\!\!\!\rightarrow}   \ottmv{F} }{%
{\ottdrulename{ValuePath\_App}}{}%
}}

\newcommand{\ottdruleValuePathXXCApp}[1]{\ottdrule[#1]{%
\ottpremise{ \mathsf{Head}( \ottnt{w} )  \mathrel{+\!\!\!\!\!\!\rightarrow}   \ottmv{F} }%
}{
 \mathsf{Head}(  \ottnt{w} \; \bullet  )  \mathrel{+\!\!\!\!\!\!\rightarrow}   \ottmv{F} }{%
{\ottdrulename{ValuePath\_CApp}}{}%
}}

\newcommand{\ottdefnValuePath}[1]{\begin{ottdefnblock}[#1]{$ \mathsf{Head}( \ottnt{a} )  \mathrel{+\!\!\!\!\!\!\rightarrow}   \ottmv{F} $}{\ottcom{Application path headed by valid constructor}}
\ottusedrule{\ottdruleValuePathXXAbsConst{}}
\ottusedrule{\ottdruleValuePathXXConst{}}
\ottusedrule{\ottdruleValuePathXXApp{}}
\ottusedrule{\ottdruleValuePathXXCApp{}}
\end{ottdefnblock}}

\newcommand{\ottdefnsJValuePath}{
\ottdefnValuePath{}}

\newcommand{\ottdruleCasePathXXAbsConst}[1]{\ottdrule[#1]{%
\ottpremise{ \mathsf{Head}( \ottnt{a} )  \mathrel{+\!\!\!\!\!\!\rightarrow}   \ottmv{F} }%
\ottpremise{\ottmv{F}  \ottsym{:}   \ottnt{A} \  \ottsym{@} \  \overline{R}  \, \in \,  \Sigma_0 }%
}{
 \mathsf{CasePath}_{ \ottnt{R} }\;  \ottnt{a}  =  \ottmv{F} }{%
{\ottdrulename{CasePath\_AbsConst}}{}%
}}

\newcommand{\ottdruleCasePathXXConst}[1]{\ottdrule[#1]{%
\ottpremise{ \mathsf{Head}( \ottnt{a} )  \mathrel{+\!\!\!\!\!\!\rightarrow}   \ottmv{F} }%
\ottpremise{\ottmv{F}  \ottsym{:}   \ottnt{A} \  \ottsym{@} \  \overline{R} \ \mathsf{where}\  \ottnt{p}  \sim_{ \ottnt{R_{{\mathrm{1}}}} }  \ottnt{b}  \, \in \,  \Sigma_0 }%
\ottpremise{ \neg  \ottsym{(}   \ottnt{R_{{\mathrm{1}}}}  \leq  \ottnt{R}   \ottsym{)} }%
}{
 \mathsf{CasePath}_{ \ottnt{R} }\;  \ottnt{a}  =  \ottmv{F} }{%
{\ottdrulename{CasePath\_Const}}{}%
}}

\newcommand{\ottdruleCasePathXXUnMatch}[1]{\ottdrule[#1]{%
\ottpremise{ \mathsf{Head}( \ottnt{a} )  \mathrel{+\!\!\!\!\!\!\rightarrow}   \ottmv{F} }%
\ottpremise{\ottmv{F}  \ottsym{:}   \ottnt{A} \  \ottsym{@} \  \overline{R} \ \mathsf{where}\  \ottnt{p}  \sim_{ \ottnt{R_{{\mathrm{1}}}} }  \ottnt{b}  \, \in \,  \Sigma_0 }%
\ottpremise{ \neg  \ottsym{(}   \ottnt{a}  \sqsupseteq  \ottnt{p}   \ottsym{)} }%
}{
 \mathsf{CasePath}_{ \ottnt{R} }\;  \ottnt{a}  =  \ottmv{F} }{%
{\ottdrulename{CasePath\_UnMatch}}{}%
}}

\newcommand{\ottdefnCasePath}[1]{\begin{ottdefnblock}[#1]{$ \mathsf{CasePath}_{ \ottnt{R} }\;  \ottnt{a}  =  \ottmv{F} $}{\ottcom{Path that is a value}}
\ottusedrule{\ottdruleCasePathXXAbsConst{}}
\ottusedrule{\ottdruleCasePathXXConst{}}
\ottusedrule{\ottdruleCasePathXXUnMatch{}}
\end{ottdefnblock}}

\newcommand{\ottdefnsJCasePath}{
\ottdefnCasePath{}}

\newcommand{\ottdruleApplyArgsXXConst}[1]{\ottdrule[#1]{%
}{
 \mathsf{ApplyArgs}( \ottmv{F} ,  \ottnt{b} ) =  \ottnt{b} }{%
{\ottdrulename{ApplyArgs\_Const}}{}%
}}

\newcommand{\ottdruleApplyArgsXXAppRole}[1]{\ottdrule[#1]{%
\ottpremise{ \mathsf{ApplyArgs}( \ottnt{a} ,  \ottnt{b} ) =  \ottnt{b'} }%
}{
 \mathsf{ApplyArgs}( \ottsym{(}   \ottnt{a} \  \ottnt{a'} ^{ \ottnt{R} }   \ottsym{)} ,  \ottnt{b} ) =  \ottsym{(}   \ottnt{b'} \  \ottnt{a'} ^{ \ottsym{+} }   \ottsym{)} }{%
{\ottdrulename{ApplyArgs\_AppRole}}{}%
}}

\newcommand{\ottdruleApplyArgsXXAppRho}[1]{\ottdrule[#1]{%
\ottpremise{ \mathsf{ApplyArgs}( \ottnt{a} ,  \ottnt{b} ) =  \ottnt{b'} }%
}{
 \mathsf{ApplyArgs}( \ottsym{(}   \ottnt{a} \  \ottnt{a'} ^{ \rho }   \ottsym{)} ,  \ottnt{b} ) =  \ottsym{(}   \ottnt{b'} \  \ottnt{a'} ^{ \rho }   \ottsym{)} }{%
{\ottdrulename{ApplyArgs\_AppRho}}{}%
}}

\newcommand{\ottdruleApplyArgsXXCApp}[1]{\ottdrule[#1]{%
\ottpremise{ \mathsf{ApplyArgs}( \ottnt{a} ,  \ottnt{b} ) =  \ottnt{b'} }%
}{
 \mathsf{ApplyArgs}(  \ottnt{a} \; \bullet  ,  \ottnt{b} ) =   \ottnt{b'} \; \bullet  }{%
{\ottdrulename{ApplyArgs\_CApp}}{}%
}}

\newcommand{\ottdefnApplyArgs}[1]{\begin{ottdefnblock}[#1]{$ \mathsf{ApplyArgs}( \ottnt{a} ,  \ottnt{b} ) =  \ottnt{b'} $}{\ottcom{Apply arguments of path a to b producing b'}}
\ottusedrule{\ottdruleApplyArgsXXConst{}}
\ottusedrule{\ottdruleApplyArgsXXAppRole{}}
\ottusedrule{\ottdruleApplyArgsXXAppRho{}}
\ottusedrule{\ottdruleApplyArgsXXCApp{}}
\end{ottdefnblock}}

\newcommand{\ottdefnsJApplyArgs}{
\ottdefnApplyArgs{}}

\newcommand{\ottdruleValueXXStar}[1]{\ottdrule[#1]{%
}{
 \mathsf{Value}_{ \ottnt{R} }\   \star  }{%
{\ottdrulename{Value\_Star}}{}%
}}

\newcommand{\ottdruleValueXXPi}[1]{\ottdrule[#1]{%
}{
 \mathsf{Value}_{ \ottnt{R} }\   \mathrm{\Pi}^ \rho \ottmv{x} \!:\! \ottnt{A} \to \ottnt{B}  }{%
{\ottdrulename{Value\_Pi}}{}%
}}

\newcommand{\ottdruleValueXXCPi}[1]{\ottdrule[#1]{%
}{
 \mathsf{Value}_{ \ottnt{R} }\   \forall \ottmv{c} \!:\! \phi . \ottnt{B}  }{%
{\ottdrulename{Value\_CPi}}{}%
}}

\newcommand{\ottdruleValueXXUAbsRel}[1]{\ottdrule[#1]{%
}{
 \mathsf{Value}_{ \ottnt{R} }\   \mathrm{\lambda}^{ \ottsym{+} } \ottmv{x} . \ottnt{a}  }{%
{\ottdrulename{Value\_UAbsRel}}{}%
}}

\newcommand{\ottdruleValueXXUAbsIrrel}[1]{\ottdrule[#1]{%
\ottpremise{ \mathsf{Value}_{ \ottnt{R} }\  \ottnt{a} }%
}{
 \mathsf{Value}_{ \ottnt{R} }\   \mathrm{\lambda}^{ \ottsym{-} } \ottmv{x} . \ottnt{a}  }{%
{\ottdrulename{Value\_UAbsIrrel}}{}%
}}

\newcommand{\ottdruleValueXXUCAbs}[1]{\ottdrule[#1]{%
}{
 \mathsf{Value}_{ \ottnt{R} }\   \mathrm{\Lambda} \ottmv{c} . \ottnt{a}  }{%
{\ottdrulename{Value\_UCAbs}}{}%
}}

\newcommand{\ottdruleValueXXPath}[1]{\ottdrule[#1]{%
\ottpremise{ \mathsf{CasePath}_{ \ottnt{R} }\;  \ottnt{a}  =  \ottmv{F} }%
}{
 \mathsf{Value}_{ \ottnt{R} }\  \ottnt{a} }{%
{\ottdrulename{Value\_Path}}{}%
}}

\newcommand{\ottdefnValue}[1]{\begin{ottdefnblock}[#1]{$ \mathsf{Value}_{ \ottnt{R} }\  \ottnt{A} $}{\ottcom{values}}
\ottusedrule{\ottdruleValueXXStar{}}
\ottusedrule{\ottdruleValueXXPi{}}
\ottusedrule{\ottdruleValueXXCPi{}}
\ottusedrule{\ottdruleValueXXUAbsRel{}}
\ottusedrule{\ottdruleValueXXUAbsIrrel{}}
\ottusedrule{\ottdruleValueXXUCAbs{}}
\ottusedrule{\ottdruleValueXXPath{}}
\end{ottdefnblock}}

\newcommand{\ottdefnsJValue}{
\ottdefnValue{}}

\newcommand{\ottdruleroleXXaXXBullet}[1]{\ottdrule[#1]{%
\ottpremise{ \emph{uniq}( \Omega ) }%
}{
 \Omega  \vDash  \Box  :  \ottnt{R} }{%
{\ottdrulename{role\_a\_Bullet}}{}%
}}

\newcommand{\ottdruleroleXXaXXStar}[1]{\ottdrule[#1]{%
\ottpremise{ \emph{uniq}( \Omega ) }%
}{
 \Omega  \vDash   \star   :  \ottnt{R} }{%
{\ottdrulename{role\_a\_Star}}{}%
}}

\newcommand{\ottdruleroleXXaXXVar}[1]{\ottdrule[#1]{%
\ottpremise{ \emph{uniq}( \Omega ) }%
\ottpremise{\ottmv{x}  \ottsym{:}  \ottnt{R} \, \in \, \Omega}%
\ottpremise{ \ottnt{R}  \leq  \ottnt{R_{{\mathrm{1}}}} }%
}{
 \Omega  \vDash  \ottmv{x}  :  \ottnt{R_{{\mathrm{1}}}} }{%
{\ottdrulename{role\_a\_Var}}{}%
}}

\newcommand{\ottdruleroleXXaXXAbs}[1]{\ottdrule[#1]{%
\ottpremise{  \Omega , \ottmv{x} \!:\! \ottkw{Nom}   \vDash  \ottnt{a}  :  \ottnt{R} }%
}{
 \Omega  \vDash  \ottsym{(}   \mathrm{\lambda}^{ \rho } \ottmv{x} . \ottnt{a}   \ottsym{)}  :  \ottnt{R} }{%
{\ottdrulename{role\_a\_Abs}}{}%
}}

\newcommand{\ottdruleroleXXaXXApp}[1]{\ottdrule[#1]{%
\ottpremise{ \Omega  \vDash  \ottnt{a}  :  \ottnt{R} }%
\ottpremise{ \Omega  \vDash  \ottnt{b}  :  \ottkw{Nom} }%
}{
 \Omega  \vDash  \ottsym{(}   \ottnt{a} \  \ottnt{b} ^{ \rho }   \ottsym{)}  :  \ottnt{R} }{%
{\ottdrulename{role\_a\_App}}{}%
}}

\newcommand{\ottdruleroleXXaXXTApp}[1]{\ottdrule[#1]{%
\ottpremise{ \Omega  \vDash  \ottnt{a}  :  \ottnt{R} }%
\ottpremise{ \Omega  \vDash  \ottnt{b}  :   \ottnt{R_{{\mathrm{1}}}} \wedge \ottnt{R}  }%
}{
 \Omega  \vDash   \ottnt{a} \  \ottnt{b} ^{ \ottnt{R_{{\mathrm{1}}}} }   :  \ottnt{R} }{%
{\ottdrulename{role\_a\_TApp}}{}%
}}

\newcommand{\ottdruleroleXXaXXPi}[1]{\ottdrule[#1]{%
\ottpremise{ \Omega  \vDash  \ottnt{A}  :  \ottnt{R} }%
\ottpremise{  \Omega , \ottmv{x} \!:\! \ottkw{Nom}   \vDash  \ottnt{B}  :  \ottnt{R} }%
}{
 \Omega  \vDash  \ottsym{(}   \mathrm{\Pi}^ \rho \ottmv{x} \!:\! \ottnt{A} \to \ottnt{B}   \ottsym{)}  :  \ottnt{R} }{%
{\ottdrulename{role\_a\_Pi}}{}%
}}

\newcommand{\ottdruleroleXXaXXCPi}[1]{\ottdrule[#1]{%
\ottpremise{ \Omega  \vDash  \ottnt{a}  :  \ottnt{R_{{\mathrm{1}}}} }%
\ottpremise{ \Omega  \vDash  \ottnt{b}  :  \ottnt{R_{{\mathrm{1}}}} }%
\ottpremise{ \Omega  \vDash  \ottnt{A}  :  \ottkw{Rep} }%
\ottpremise{ \Omega  \vDash  \ottnt{B}  :  \ottnt{R} }%
}{
 \Omega  \vDash  \ottsym{(}   \forall \ottmv{c} \!:\!  \ottnt{a}   \sim _{ \ottnt{R_{{\mathrm{1}}}} }  \ottnt{b}  :  \ottnt{A}  . \ottnt{B}   \ottsym{)}  :  \ottnt{R} }{%
{\ottdrulename{role\_a\_CPi}}{}%
}}

\newcommand{\ottdruleroleXXaXXCAbs}[1]{\ottdrule[#1]{%
\ottpremise{ \Omega  \vDash  \ottnt{b}  :  \ottnt{R} }%
}{
 \Omega  \vDash  \ottsym{(}   \mathrm{\Lambda} \ottmv{c} . \ottnt{b}   \ottsym{)}  :  \ottnt{R} }{%
{\ottdrulename{role\_a\_CAbs}}{}%
}}

\newcommand{\ottdruleroleXXaXXCApp}[1]{\ottdrule[#1]{%
\ottpremise{ \Omega  \vDash  \ottnt{a}  :  \ottnt{R} }%
}{
 \Omega  \vDash  \ottsym{(}   \ottnt{a} \; \bullet   \ottsym{)}  :  \ottnt{R} }{%
{\ottdrulename{role\_a\_CApp}}{}%
}}

\newcommand{\ottdruleroleXXaXXConst}[1]{\ottdrule[#1]{%
\ottpremise{ \emph{uniq}( \Omega ) }%
\ottpremise{\ottmv{F}  \ottsym{:}   \ottnt{A} \  \ottsym{@} \  \overline{R}  \, \in \,  \Sigma_0 }%
}{
 \Omega  \vDash  \ottmv{F}  :  \ottnt{R} }{%
{\ottdrulename{role\_a\_Const}}{}%
}}

\newcommand{\ottdruleroleXXaXXFam}[1]{\ottdrule[#1]{%
\ottpremise{ \emph{uniq}( \Omega ) }%
\ottpremise{\ottmv{F}  \ottsym{:}   \ottnt{A} \  \ottsym{@} \  \overline{R} \ \mathsf{where}\  \ottnt{p}  \sim_{ \ottnt{R} }  \ottnt{a}  \, \in \,  \Sigma_0 }%
}{
 \Omega  \vDash  \ottmv{F}  :  \ottnt{R_{{\mathrm{1}}}} }{%
{\ottdrulename{role\_a\_Fam}}{}%
}}

\newcommand{\ottdruleroleXXaXXPattern}[1]{\ottdrule[#1]{%
\ottpremise{ \Omega  \vDash  \ottnt{a}  :  \ottkw{Nom} }%
\ottpremise{ \Omega  \vDash  \ottnt{b_{{\mathrm{1}}}}  :  \ottnt{R_{{\mathrm{1}}}} }%
\ottpremise{ \Omega  \vDash  \ottnt{b_{{\mathrm{2}}}}  :  \ottnt{R_{{\mathrm{1}}}} }%
}{
 \Omega  \vDash   \mathsf{case} \hspace{3pt}  \ottnt{a}  \hspace{3pt} \mathsf{of} \hspace{3pt}  \ottmv{F} \  \overline{\upsilon}  \rightarrow  \ottnt{b_{{\mathrm{1}}}}  \| \_ \rightarrow  \ottnt{b_{{\mathrm{2}}}}   :  \ottnt{R_{{\mathrm{1}}}} }{%
{\ottdrulename{role\_a\_Pattern}}{}%
}}

\newcommand{\ottdefnroleing}[1]{\begin{ottdefnblock}[#1]{$ \Omega  \vDash  \ottnt{a}  :  \ottnt{R} $}{\ottcom{Roleing judgment}}
\ottusedrule{\ottdruleroleXXaXXBullet{}}
\ottusedrule{\ottdruleroleXXaXXStar{}}
\ottusedrule{\ottdruleroleXXaXXVar{}}
\ottusedrule{\ottdruleroleXXaXXAbs{}}
\ottusedrule{\ottdruleroleXXaXXApp{}}
\ottusedrule{\ottdruleroleXXaXXTApp{}}
\ottusedrule{\ottdruleroleXXaXXPi{}}
\ottusedrule{\ottdruleroleXXaXXCPi{}}
\ottusedrule{\ottdruleroleXXaXXCAbs{}}
\ottusedrule{\ottdruleroleXXaXXCApp{}}
\ottusedrule{\ottdruleroleXXaXXConst{}}
\ottusedrule{\ottdruleroleXXaXXFam{}}
\ottusedrule{\ottdruleroleXXaXXPattern{}}
\end{ottdefnblock}}

\newcommand{\ottdefnsJroleing}{
\ottdefnroleing{}}

\newcommand{\ottdruleRhoXXRel}[1]{\ottdrule[#1]{%
\ottpremise{  }%
}{
 ( \ottsym{+}  = +) \vee ( \ottmv{x} \not\in\mathsf{fv}\;  \ottnt{A} ) }{%
{\ottdrulename{Rho\_Rel}}{}%
}}

\newcommand{\ottdruleRhoXXIrrRel}[1]{\ottdrule[#1]{%
\ottpremise{\ottmv{x} \, \not\in \, \mathsf{fv}\! \, \ottnt{A}}%
}{
 ( \ottsym{-}  = +) \vee ( \ottmv{x} \not\in\mathsf{fv}\;  \ottnt{A} ) }{%
{\ottdrulename{Rho\_IrrRel}}{}%
}}

\newcommand{\ottdefnRhoCheck}[1]{\begin{ottdefnblock}[#1]{$ ( \rho  = +) \vee ( \ottmv{x} \not\in\mathsf{fv}\;  \ottnt{A} ) $}{\ottcom{Irrelevant argument check}}
\ottusedrule{\ottdruleRhoXXRel{}}
\ottusedrule{\ottdruleRhoXXIrrRel{}}
\end{ottdefnblock}}

\newcommand{\ottdefnsJChk}{
\ottdefnRhoCheck{}}

\newcommand{\ottdruleParXXRefl}[1]{\ottdrule[#1]{%
\ottpremise{ \Omega  \vDash  \ottnt{a}  :  \ottnt{R} }%
}{
 \Omega  \vDash  \ottnt{a}   \Rightarrow _{ \ottnt{R} }  \ottnt{a} }{%
{\ottdrulename{Par\_Refl}}{}%
}}

\newcommand{\ottdruleParXXBeta}[1]{\ottdrule[#1]{%
\ottpremise{ \Omega  \vDash  \ottnt{a}   \Rightarrow _{ \ottnt{R} }  \ottsym{(}   \mathrm{\lambda}^{ \rho } \ottmv{x} . \ottnt{a'}   \ottsym{)} }%
\ottpremise{ \Omega  \vDash  \ottnt{b}   \Rightarrow _{ \ottkw{Nom} }  \ottnt{b'} }%
}{
 \Omega  \vDash   \ottnt{a} \  \ottnt{b} ^{ \rho }    \Rightarrow _{ \ottnt{R} }  \ottnt{a'}  \lbrace  \ottnt{b'}  \ottsym{/}  \ottmv{x}  \rbrace }{%
{\ottdrulename{Par\_Beta}}{}%
}}

\newcommand{\ottdruleParXXApp}[1]{\ottdrule[#1]{%
\ottpremise{ \Omega  \vDash  \ottnt{a}   \Rightarrow _{ \ottnt{R} }  \ottnt{a'} }%
\ottpremise{ \Omega  \vDash  \ottnt{b}   \Rightarrow _{ \ottsym{(}  \ottsym{app\_role}  \nu \, \ottnt{R}  \ottsym{)} }  \ottnt{b'} }%
}{
 \Omega  \vDash   \ottnt{a} \  \ottnt{b} ^{ \nu }    \Rightarrow _{ \ottnt{R} }   \ottnt{a'} \  \ottnt{b'} ^{ \nu }  }{%
{\ottdrulename{Par\_App}}{}%
}}

\newcommand{\ottdruleParXXCBeta}[1]{\ottdrule[#1]{%
\ottpremise{ \Omega  \vDash  \ottnt{a}   \Rightarrow _{ \ottnt{R} }  \ottsym{(}   \mathrm{\Lambda} \ottmv{c} . \ottnt{a'}   \ottsym{)} }%
}{
 \Omega  \vDash   \ottnt{a} \; \bullet    \Rightarrow _{ \ottnt{R} }  \ottnt{a'}  \lbrace  \bullet  \ottsym{/}  \ottmv{c}  \rbrace }{%
{\ottdrulename{Par\_CBeta}}{}%
}}

\newcommand{\ottdruleParXXCApp}[1]{\ottdrule[#1]{%
\ottpremise{ \Omega  \vDash  \ottnt{a}   \Rightarrow _{ \ottnt{R} }  \ottnt{a'} }%
}{
 \Omega  \vDash   \ottnt{a} \; \bullet    \Rightarrow _{ \ottnt{R} }   \ottnt{a'} \; \bullet  }{%
{\ottdrulename{Par\_CApp}}{}%
}}

\newcommand{\ottdruleParXXAbs}[1]{\ottdrule[#1]{%
\ottpremise{  \Omega , \ottmv{x} \!:\! \ottkw{Nom}   \vDash  \ottnt{a}   \Rightarrow _{ \ottnt{R} }  \ottnt{a'} }%
}{
 \Omega  \vDash   \mathrm{\lambda}^{ \rho } \ottmv{x} . \ottnt{a}    \Rightarrow _{ \ottnt{R} }   \mathrm{\lambda}^{ \rho } \ottmv{x} . \ottnt{a'}  }{%
{\ottdrulename{Par\_Abs}}{}%
}}

\newcommand{\ottdruleParXXPi}[1]{\ottdrule[#1]{%
\ottpremise{ \Omega  \vDash  \ottnt{A}   \Rightarrow _{ \ottnt{R} }  \ottnt{A'} }%
\ottpremise{  \Omega , \ottmv{x} \!:\! \ottkw{Nom}   \vDash  \ottnt{B}   \Rightarrow _{ \ottnt{R} }  \ottnt{B'} }%
}{
 \Omega  \vDash   \mathrm{\Pi}^ \rho \ottmv{x} \!:\! \ottnt{A} \to \ottnt{B}    \Rightarrow _{ \ottnt{R} }   \mathrm{\Pi}^ \rho \ottmv{x} \!:\! \ottnt{A'} \to \ottnt{B'}  }{%
{\ottdrulename{Par\_Pi}}{}%
}}

\newcommand{\ottdruleParXXCAbs}[1]{\ottdrule[#1]{%
\ottpremise{ \Omega  \vDash  \ottnt{a}   \Rightarrow _{ \ottnt{R} }  \ottnt{a'} }%
}{
 \Omega  \vDash   \mathrm{\Lambda} \ottmv{c} . \ottnt{a}    \Rightarrow _{ \ottnt{R} }   \mathrm{\Lambda} \ottmv{c} . \ottnt{a'}  }{%
{\ottdrulename{Par\_CAbs}}{}%
}}

\newcommand{\ottdruleParXXCPi}[1]{\ottdrule[#1]{%
\ottpremise{ \Omega  \vDash  \ottnt{A}   \Rightarrow _{ \ottkw{Rep} }  \ottnt{A'} }%
\ottpremise{ \Omega  \vDash  \ottnt{a}   \Rightarrow _{ \ottnt{R_{{\mathrm{1}}}} }  \ottnt{a'} }%
\ottpremise{ \Omega  \vDash  \ottnt{b}   \Rightarrow _{ \ottnt{R_{{\mathrm{1}}}} }  \ottnt{b'} }%
\ottpremise{ \Omega  \vDash  \ottnt{B}   \Rightarrow _{ \ottnt{R} }  \ottnt{B'} }%
}{
 \Omega  \vDash   \forall \ottmv{c} \!:\!  \ottnt{a}   \sim _{ \ottnt{R_{{\mathrm{1}}}} }  \ottnt{b}  :  \ottnt{A}  . \ottnt{B}    \Rightarrow _{ \ottnt{R} }   \forall \ottmv{c} \!:\!  \ottnt{a'}   \sim _{ \ottnt{R_{{\mathrm{1}}}} }  \ottnt{b'}  :  \ottnt{A'}  . \ottnt{B'}  }{%
{\ottdrulename{Par\_CPi}}{}%
}}

\newcommand{\ottdruleParXXAxiomBase}[1]{\ottdrule[#1]{%
\ottpremise{\ottmv{F}  \ottsym{:}   \ottnt{A} \  \ottsym{@} \  \overline{R} \ \mathsf{where}\  \ottmv{F}  \sim_{ \ottnt{R_{{\mathrm{1}}}} }  \ottnt{b}  \, \in \,  \Sigma_0 }%
\ottpremise{ \ottnt{R_{{\mathrm{1}}}}  \leq  \ottnt{R} }%
\ottpremise{ \emph{uniq}( \Omega ) }%
}{
 \Omega  \vDash  \ottmv{F}   \Rightarrow _{ \ottnt{R} }  \ottnt{b} }{%
{\ottdrulename{Par\_AxiomBase}}{}%
}}

\newcommand{\ottdruleParXXAxiomApp}[1]{\ottdrule[#1]{%
\ottpremise{\ottmv{F}  \ottsym{:}   \ottnt{A} \  \ottsym{@} \  \overline{R} \ \mathsf{where}\  \ottnt{p}  \sim_{ \ottnt{R_{{\mathrm{1}}}} }  \ottnt{b}  \, \in \,  \Sigma_0 }%
\ottpremise{  \ottnt{a}  \sqsubseteq  \ottnt{p}   \wedge   \neg  \ottsym{(}   \ottnt{a}   \leftrightarrow   \ottnt{p}   \ottsym{)}  }%
\ottpremise{ \Omega  \vDash  \ottnt{a}   \Rightarrow _{ \ottnt{R} }  \ottnt{a'} }%
\ottpremise{ \Omega  \vDash  \ottnt{a_{{\mathrm{1}}}}   \Rightarrow _{ \ottsym{(}  \ottsym{app\_role}  \nu \, \ottnt{R}  \ottsym{)} }  \ottnt{a'_{{\mathrm{1}}}} }%
\ottpremise{ \mathsf{rename} \  \ottnt{p}  \rightarrow  \ottnt{b}  \ \mathsf{to} \  \ottnt{p'}  \rightarrow  \ottnt{b'}  \ \mathsf{excluding} \  \ottsym{(}    \widetilde { \Omega }  ,  \mathsf{fv}\! \, \ottnt{p}   \ottsym{)}  \ \mathsf{and} \  \Delta' }%
\ottpremise{ \mathsf{MatchSubst}\; (  \ottsym{(}   \ottnt{a'} \  \ottnt{a'_{{\mathrm{1}}}} ^{ \nu }   \ottsym{)}  ,  \ottnt{p'}  ,  \ottnt{b'}  ) =  \ottnt{a_{{\mathrm{2}}}} }%
\ottpremise{ \ottnt{R_{{\mathrm{1}}}}  \leq  \ottnt{R} }%
}{
 \Omega  \vDash   \ottnt{a} \  \ottnt{a_{{\mathrm{1}}}} ^{ \nu }    \Rightarrow _{ \ottnt{R} }  \ottnt{a_{{\mathrm{2}}}} }{%
{\ottdrulename{Par\_AxiomApp}}{}%
}}

\newcommand{\ottdruleParXXAxiomCApp}[1]{\ottdrule[#1]{%
\ottpremise{\ottmv{F}  \ottsym{:}   \ottnt{A} \  \ottsym{@} \  \overline{R} \ \mathsf{where}\  \ottnt{p}  \sim_{ \ottnt{R_{{\mathrm{1}}}} }  \ottnt{b}  \, \in \,  \Sigma_0 }%
\ottpremise{  \ottnt{a}  \sqsubseteq  \ottnt{p}   \wedge   \neg  \ottsym{(}   \ottnt{a}   \leftrightarrow   \ottnt{p}   \ottsym{)}  }%
\ottpremise{ \Omega  \vDash  \ottnt{a}   \Rightarrow _{ \ottnt{R} }  \ottnt{a'} }%
\ottpremise{ \mathsf{rename} \  \ottnt{p}  \rightarrow  \ottnt{b}  \ \mathsf{to} \  \ottnt{p'}  \rightarrow  \ottnt{b'}  \ \mathsf{excluding} \  \ottsym{(}    \widetilde { \Omega }  ,  \mathsf{fv}\! \, \ottnt{p}   \ottsym{)}  \ \mathsf{and} \  \Delta' }%
\ottpremise{ \mathsf{MatchSubst}\; (  \ottsym{(}   \ottnt{a'} \; \bullet   \ottsym{)}  ,  \ottnt{p'}  ,  \ottnt{b'}  ) =  \ottnt{a_{{\mathrm{2}}}} }%
\ottpremise{ \ottnt{R_{{\mathrm{1}}}}  \leq  \ottnt{R} }%
}{
 \Omega  \vDash   \ottnt{a} \; \bullet    \Rightarrow _{ \ottnt{R} }  \ottnt{a_{{\mathrm{2}}}} }{%
{\ottdrulename{Par\_AxiomCApp}}{}%
}}

\newcommand{\ottdruleParXXPattern}[1]{\ottdrule[#1]{%
\ottpremise{ \Omega  \vDash  \ottnt{a}   \Rightarrow _{ \ottkw{Nom} }  \ottnt{a'} }%
\ottpremise{ \Omega  \vDash  \ottnt{b_{{\mathrm{1}}}}   \Rightarrow _{ \ottnt{R_{{\mathrm{0}}}} }  \ottnt{b'_{{\mathrm{1}}}} }%
\ottpremise{ \Omega  \vDash  \ottnt{b_{{\mathrm{2}}}}   \Rightarrow _{ \ottnt{R_{{\mathrm{0}}}} }  \ottnt{b'_{{\mathrm{2}}}} }%
}{
 \Omega  \vDash  \ottsym{(}   \mathsf{case} \hspace{3pt}  \ottnt{a}  \hspace{3pt} \mathsf{of} \hspace{3pt}  \ottmv{F} \  \overline{\upsilon}  \rightarrow  \ottnt{b_{{\mathrm{1}}}}  \| \_ \rightarrow  \ottnt{b_{{\mathrm{2}}}}   \ottsym{)}   \Rightarrow _{ \ottnt{R_{{\mathrm{0}}}} }  \ottsym{(}   \mathsf{case} \hspace{3pt}  \ottnt{a'}  \hspace{3pt} \mathsf{of} \hspace{3pt}  \ottmv{F} \  \overline{\upsilon}  \rightarrow  \ottnt{b'_{{\mathrm{1}}}}  \| \_ \rightarrow  \ottnt{b'_{{\mathrm{2}}}}   \ottsym{)} }{%
{\ottdrulename{Par\_Pattern}}{}%
}}

\newcommand{\ottdruleParXXPatternTrue}[1]{\ottdrule[#1]{%
\ottpremise{ \Omega  \vDash  \ottnt{a}   \Rightarrow _{ \ottkw{Nom} }  \ottnt{a'} }%
\ottpremise{ \Omega  \vDash  \ottnt{b_{{\mathrm{1}}}}   \Rightarrow _{ \ottnt{R_{{\mathrm{0}}}} }  \ottnt{b'_{{\mathrm{1}}}} }%
\ottpremise{ \Omega  \vDash  \ottnt{b_{{\mathrm{2}}}}   \Rightarrow _{ \ottnt{R_{{\mathrm{0}}}} }  \ottnt{b'_{{\mathrm{2}}}} }%
\ottpremise{ \ottnt{a'}  \leftrightarrow_{ \ottkw{Nom} }  \ottmv{F} \; \overline{\upsilon} }%
\ottpremise{ \mathsf{ApplyArgs}( \ottnt{a'} ,  \ottnt{b'_{{\mathrm{1}}}} ) =  \ottnt{b} }%
\ottpremise{ \suppress{ \ottkw{Sat} \, \ottmv{F} \, \overline{\upsilon}' } }%
}{
 \Omega  \vDash  \ottsym{(}   \mathsf{case} \hspace{3pt}  \ottnt{a}  \hspace{3pt} \mathsf{of} \hspace{3pt}  \ottmv{F} \  \overline{\upsilon}  \rightarrow  \ottnt{b_{{\mathrm{1}}}}  \| \_ \rightarrow  \ottnt{b_{{\mathrm{2}}}}   \ottsym{)}   \Rightarrow _{ \ottnt{R_{{\mathrm{0}}}} }   \ottnt{b} \; \bullet  }{%
{\ottdrulename{Par\_PatternTrue}}{}%
}}

\newcommand{\ottdruleParXXPatternFalse}[1]{\ottdrule[#1]{%
\ottpremise{ \Omega  \vDash  \ottnt{a}   \Rightarrow _{ \ottkw{Nom} }  \ottnt{a'} }%
\ottpremise{ \Omega  \vDash  \ottnt{b_{{\mathrm{1}}}}   \Rightarrow _{ \ottnt{R_{{\mathrm{0}}}} }  \ottnt{b'_{{\mathrm{1}}}} }%
\ottpremise{ \Omega  \vDash  \ottnt{b_{{\mathrm{2}}}}   \Rightarrow _{ \ottnt{R_{{\mathrm{0}}}} }  \ottnt{b'_{{\mathrm{2}}}} }%
\ottpremise{ \mathsf{Value}_{ \ottkw{Nom} }\  \ottnt{a'} }%
\ottpremise{ \neg  \ottsym{(}   \ottnt{a'}  \leftrightarrow_{ \ottkw{Nom} }  \ottmv{F} \; \overline{\upsilon}   \ottsym{)} }%
}{
 \Omega  \vDash  \ottsym{(}   \mathsf{case} \hspace{3pt}  \ottnt{a}  \hspace{3pt} \mathsf{of} \hspace{3pt}  \ottmv{F} \  \overline{\upsilon}  \rightarrow  \ottnt{b_{{\mathrm{1}}}}  \| \_ \rightarrow  \ottnt{b_{{\mathrm{2}}}}   \ottsym{)}   \Rightarrow _{ \ottnt{R_{{\mathrm{0}}}} }  \ottnt{b'_{{\mathrm{2}}}} }{%
{\ottdrulename{Par\_PatternFalse}}{}%
}}

\newcommand{\ottdefnPar}[1]{\begin{ottdefnblock}[#1]{$ \Omega  \vDash  \ottnt{a}   \Rightarrow _{ \ottnt{R} }  \ottnt{b} $}{\ottcom{parallel reduction}}
\ottusedrule{\ottdruleParXXRefl{}}
\ottusedrule{\ottdruleParXXBeta{}}
\ottusedrule{\ottdruleParXXApp{}}
\ottusedrule{\ottdruleParXXCBeta{}}
\ottusedrule{\ottdruleParXXCApp{}}
\ottusedrule{\ottdruleParXXAbs{}}
\ottusedrule{\ottdruleParXXPi{}}
\ottusedrule{\ottdruleParXXCAbs{}}
\ottusedrule{\ottdruleParXXCPi{}}
\ottusedrule{\ottdruleParXXAxiomBase{}}
\ottusedrule{\ottdruleParXXAxiomApp{}}
\ottusedrule{\ottdruleParXXAxiomCApp{}}
\ottusedrule{\ottdruleParXXPattern{}}
\ottusedrule{\ottdruleParXXPatternTrue{}}
\ottusedrule{\ottdruleParXXPatternFalse{}}
\end{ottdefnblock}}

\newcommand{\ottdruleBetaXXAppAbs}[1]{\ottdrule[#1]{%
\ottpremise{ \mathsf{Value}_{ \ottnt{R} }\  \ottsym{(}   \mathrm{\lambda}^{ \rho } \ottmv{x} . \ottnt{a}   \ottsym{)} }%
}{
 \vDash    \ottsym{(}   \mathrm{\lambda}^{ \rho } \ottmv{x} . \ottnt{a}   \ottsym{)} \  \ottnt{b} ^{ \rho }   \rightarrow^{\beta}_{ \ottnt{R} }  \ottnt{a}  \lbrace  \ottnt{b}  \ottsym{/}  \ottmv{x}  \rbrace }{%
{\ottdrulename{Beta\_AppAbs}}{}%
}}

\newcommand{\ottdruleBetaXXCAppCAbs}[1]{\ottdrule[#1]{%
}{
 \vDash    \ottsym{(}   \mathrm{\Lambda} \ottmv{c} . \ottnt{a'}   \ottsym{)} \; \bullet   \rightarrow^{\beta}_{ \ottnt{R} }  \ottnt{a'}  \lbrace  \bullet  \ottsym{/}  \ottmv{c}  \rbrace }{%
{\ottdrulename{Beta\_CAppCAbs}}{}%
}}

\newcommand{\ottdruleBetaXXAxiom}[1]{\ottdrule[#1]{%
\ottpremise{\ottmv{F}  \ottsym{:}   \ottnt{A} \  \ottsym{@} \  \overline{R} \ \mathsf{where}\  \ottnt{p}  \sim_{ \ottnt{R_{{\mathrm{1}}}} }  \ottnt{b}  \, \in \,  \Sigma_0 }%
\ottpremise{ \mathsf{rename} \  \ottnt{p}  \rightarrow  \ottnt{b}  \ \mathsf{to} \  \ottnt{p_{{\mathrm{1}}}}  \rightarrow  \ottnt{b_{{\mathrm{1}}}}  \ \mathsf{excluding} \  \ottsym{(}   \mathsf{fv}\! \, \ottnt{a} ,  \mathsf{fv}\! \, \ottnt{p}   \ottsym{)}  \ \mathsf{and} \  \Delta' }%
\ottpremise{ \mathsf{MatchSubst}\; (  \ottnt{a}  ,  \ottnt{p_{{\mathrm{1}}}}  ,  \ottnt{b_{{\mathrm{1}}}}  ) =  \ottnt{b'} }%
\ottpremise{ \ottnt{R_{{\mathrm{1}}}}  \leq  \ottnt{R} }%
}{
 \vDash   \ottnt{a}  \rightarrow^{\beta}_{ \ottnt{R} }  \ottnt{b'} }{%
{\ottdrulename{Beta\_Axiom}}{}%
}}

\newcommand{\ottdruleBetaXXPatternTrue}[1]{\ottdrule[#1]{%
\ottpremise{ \ottnt{a}  \leftrightarrow_{ \ottkw{Nom} }  \ottmv{F} \; \overline{\upsilon} }%
\ottpremise{ \mathsf{ApplyArgs}( \ottnt{a} ,  \ottnt{b_{{\mathrm{1}}}} ) =  \ottnt{b'_{{\mathrm{1}}}} }%
\ottpremise{ \suppress{ \ottkw{Sat} \, \ottmv{F} \, \overline{\upsilon}' } }%
}{
 \vDash   \ottsym{(}   \mathsf{case} \hspace{3pt}  \ottnt{a}  \hspace{3pt} \mathsf{of} \hspace{3pt}  \ottmv{F} \  \overline{\upsilon}  \rightarrow  \ottnt{b_{{\mathrm{1}}}}  \| \_ \rightarrow  \ottnt{b_{{\mathrm{2}}}}   \ottsym{)}  \rightarrow^{\beta}_{ \ottnt{R} }   \ottnt{b'_{{\mathrm{1}}}} \; \bullet  }{%
{\ottdrulename{Beta\_PatternTrue}}{}%
}}

\newcommand{\ottdruleBetaXXPatternFalse}[1]{\ottdrule[#1]{%
\ottpremise{ \mathsf{Value}_{ \ottkw{Nom} }\  \ottnt{a} }%
\ottpremise{ \neg  \ottsym{(}   \ottnt{a}  \leftrightarrow_{ \ottkw{Nom} }  \ottmv{F} \; \overline{\upsilon}   \ottsym{)} }%
}{
 \vDash   \ottsym{(}   \mathsf{case} \hspace{3pt}  \ottnt{a}  \hspace{3pt} \mathsf{of} \hspace{3pt}  \ottmv{F} \  \overline{\upsilon}  \rightarrow  \ottnt{b_{{\mathrm{1}}}}  \| \_ \rightarrow  \ottnt{b_{{\mathrm{2}}}}   \ottsym{)}  \rightarrow^{\beta}_{ \ottnt{R} }  \ottnt{b_{{\mathrm{2}}}} }{%
{\ottdrulename{Beta\_PatternFalse}}{}%
}}

\newcommand{\ottdefnBeta}[1]{\begin{ottdefnblock}[#1]{$ \vDash   \ottnt{a}  \rightarrow^{\beta}_{ \ottnt{R} }  \ottnt{b} $}{\ottcom{primitive reductions}}
\ottusedrule{\ottdruleBetaXXAppAbs{}}
\ottusedrule{\ottdruleBetaXXCAppCAbs{}}
\ottusedrule{\ottdruleBetaXXAxiom{}}
\ottusedrule{\ottdruleBetaXXPatternTrue{}}
\ottusedrule{\ottdruleBetaXXPatternFalse{}}
\end{ottdefnblock}}

\newcommand{\ottdruleEXXAbsTerm}[1]{\ottdrule[#1]{%
\ottpremise{ \vDash   \ottnt{a}   \leadsto _{ \ottnt{R} }  \ottnt{a'} }%
}{
 \vDash    \mathrm{\lambda}^{ \ottsym{-} } \ottmv{x} . \ottnt{a}    \leadsto _{ \ottnt{R} }   \mathrm{\lambda}^{ \ottsym{-} } \ottmv{x} . \ottnt{a'}  }{%
{\ottdrulename{E\_AbsTerm}}{}%
}}

\newcommand{\ottdruleEXXAppLeft}[1]{\ottdrule[#1]{%
\ottpremise{ \vDash   \ottnt{a}   \leadsto _{ \ottnt{R} }  \ottnt{a'} }%
}{
 \vDash    \ottnt{a} \  \ottnt{b} ^{ \nu }    \leadsto _{ \ottnt{R} }   \ottnt{a'} \  \ottnt{b} ^{ \nu }  }{%
{\ottdrulename{E\_AppLeft}}{}%
}}

\newcommand{\ottdruleEXXCAppLeft}[1]{\ottdrule[#1]{%
\ottpremise{ \vDash   \ottnt{a}   \leadsto _{ \ottnt{R} }  \ottnt{a'} }%
}{
 \vDash    \ottnt{a} \; \bullet    \leadsto _{ \ottnt{R} }   \ottnt{a'} \; \bullet  }{%
{\ottdrulename{E\_CAppLeft}}{}%
}}

\newcommand{\ottdruleEXXPattern}[1]{\ottdrule[#1]{%
\ottpremise{ \vDash   \ottnt{a}   \leadsto _{ \ottkw{Nom} }  \ottnt{a'} }%
}{
 \vDash   \ottsym{(}   \mathsf{case} \hspace{3pt}  \ottnt{a}  \hspace{3pt} \mathsf{of} \hspace{3pt}  \ottmv{F} \  \overline{\upsilon}  \rightarrow  \ottnt{b_{{\mathrm{1}}}}  \| \_ \rightarrow  \ottnt{b_{{\mathrm{2}}}}   \ottsym{)}   \leadsto _{ \ottnt{R} }  \ottsym{(}   \mathsf{case} \hspace{3pt}  \ottnt{a'}  \hspace{3pt} \mathsf{of} \hspace{3pt}  \ottmv{F} \  \overline{\upsilon}  \rightarrow  \ottnt{b_{{\mathrm{1}}}}  \| \_ \rightarrow  \ottnt{b_{{\mathrm{2}}}}   \ottsym{)} }{%
{\ottdrulename{E\_Pattern}}{}%
}}

\newcommand{\ottdruleEXXPrim}[1]{\ottdrule[#1]{%
\ottpremise{ \vDash   \ottnt{a}  \rightarrow^{\beta}_{ \ottnt{R} }  \ottnt{b} }%
}{
 \vDash   \ottnt{a}   \leadsto _{ \ottnt{R} }  \ottnt{b} }{%
{\ottdrulename{E\_Prim}}{}%
}}

\newcommand{\ottdefnreductionXXinXXone}[1]{\begin{ottdefnblock}[#1]{$ \vDash   \ottnt{a}   \leadsto _{ \ottnt{R} }  \ottnt{b} $}{\ottcom{Single-step head reduction}}
\ottusedrule{\ottdruleEXXAbsTerm{}}
\ottusedrule{\ottdruleEXXAppLeft{}}
\ottusedrule{\ottdruleEXXCAppLeft{}}
\ottusedrule{\ottdruleEXXPattern{}}
\ottusedrule{\ottdruleEXXPrim{}}
\end{ottdefnblock}}

\newcommand{\ottdruleBranchTypingXXBase}[1]{\ottdrule[#1]{%
\ottpremise{ \emph{uniq} \  \Gamma }%
\ottpremise{\ottnt{C_{{\mathrm{1}}}}  \lbrace  \bullet  \ottsym{/}  \ottmv{c}  \rbrace  \ottsym{=}  \ottnt{C_{{\mathrm{2}}}}}%
}{
 \Gamma  \vDash \mathsf{case} \hspace{2pt} ( \ottnt{a} \sim \ottnt{b} \; \overline{\mu} : \ottnt{A} ) \hspace{2pt} \mathsf{of} \hspace{2pt}  F\;{    } :  \ottnt{A}  \Rightarrow   \forall \ottmv{c} \!:\! \ottsym{(}   \ottnt{a}   \sim _{ \ottkw{Nom} }   \ottnt{b} \; \overline{\mu}   :  \ottnt{A}   \ottsym{)} . \ottnt{C_{{\mathrm{1}}}}   \ | \,  \ottnt{C_{{\mathrm{2}}}} }{%
{\ottdrulename{BranchTyping\_Base}}{}%
}}

\newcommand{\ottdruleBranchTypingXXPiRole}[1]{\ottdrule[#1]{%
\ottpremise{  \Gamma ,  \ottmv{x} \!:\! \ottnt{A}   \vDash \mathsf{case} \hspace{2pt} ( \ottnt{a} \sim \ottnt{b} \;  \overline{\mu} \;  \ottmv{x} ^{ \ottnt{R} }   : \ottnt{A_{{\mathrm{1}}}} ) \hspace{2pt} \mathsf{of} \hspace{2pt}  F\;{ \overline{\upsilon} } :  \ottnt{B}  \Rightarrow  \ottnt{C}  \ | \,  \ottnt{C'} }%
}{
 \Gamma  \vDash \mathsf{case} \hspace{2pt} ( \ottnt{a} \sim \ottnt{b} \; \overline{\mu} : \ottnt{A_{{\mathrm{1}}}} ) \hspace{2pt} \mathsf{of} \hspace{2pt}  F\;{ \ottsym{(}    \ottnt{R}  \; \overline{\upsilon}   \ottsym{)} } :   \mathrm{\Pi}^ \ottsym{+} \ottmv{x} \!:\! \ottnt{A} \to \ottnt{B}   \Rightarrow   \mathrm{\Pi}^ \ottsym{+} \ottmv{x} \!:\! \ottnt{A} \to \ottnt{C}   \ | \,  \ottnt{C'} }{%
{\ottdrulename{BranchTyping\_PiRole}}{}%
}}

\newcommand{\ottdruleBranchTypingXXPiRel}[1]{\ottdrule[#1]{%
\ottpremise{  \Gamma ,  \ottmv{x} \!:\! \ottnt{A}   \vDash \mathsf{case} \hspace{2pt} ( \ottnt{a} \sim \ottnt{b} \;  \overline{\mu} \;  \ottmv{x} ^{ \ottsym{+} }   : \ottnt{A_{{\mathrm{1}}}} ) \hspace{2pt} \mathsf{of} \hspace{2pt}  F\;{ \overline{\upsilon} } :  \ottnt{B}  \Rightarrow  \ottnt{C}  \ | \,  \ottnt{C'} }%
}{
 \Gamma  \vDash \mathsf{case} \hspace{2pt} ( \ottnt{a} \sim \ottnt{b} \; \overline{\mu} : \ottnt{A_{{\mathrm{1}}}} ) \hspace{2pt} \mathsf{of} \hspace{2pt}  F\;{ \ottsym{(}    \ottsym{+}  \; \overline{\upsilon}   \ottsym{)} } :   \mathrm{\Pi}^ \ottsym{+} \ottmv{x} \!:\! \ottnt{A} \to \ottnt{B}   \Rightarrow   \mathrm{\Pi}^ \ottsym{+} \ottmv{x} \!:\! \ottnt{A} \to \ottnt{C}   \ | \,  \ottnt{C'} }{%
{\ottdrulename{BranchTyping\_PiRel}}{}%
}}

\newcommand{\ottdruleBranchTypingXXPiIrrel}[1]{\ottdrule[#1]{%
\ottpremise{  \Gamma ,  \ottmv{x} \!:\! \ottnt{A}   \vDash \mathsf{case} \hspace{2pt} ( \ottnt{a} \sim \ottnt{b} \;  \overline{\mu} \;  \Box ^{ \ottsym{-} }   : \ottnt{A_{{\mathrm{1}}}} ) \hspace{2pt} \mathsf{of} \hspace{2pt}  F\;{ \overline{\upsilon} } :  \ottnt{B}  \Rightarrow  \ottnt{C}  \ | \,  \ottnt{C'} }%
}{
 \Gamma  \vDash \mathsf{case} \hspace{2pt} ( \ottnt{a} \sim \ottnt{b} \; \overline{\mu} : \ottnt{A_{{\mathrm{1}}}} ) \hspace{2pt} \mathsf{of} \hspace{2pt}  F\;{ \ottsym{(}    \ottsym{-}  \; \overline{\upsilon}   \ottsym{)} } :   \mathrm{\Pi}^ \ottsym{-} \ottmv{x} \!:\! \ottnt{A} \to \ottnt{B}   \Rightarrow   \mathrm{\Pi}^ \ottsym{-} \ottmv{x} \!:\! \ottnt{A} \to \ottnt{C}   \ | \,  \ottnt{C'} }{%
{\ottdrulename{BranchTyping\_PiIrrel}}{}%
}}

\newcommand{\ottdruleBranchTypingXXCPi}[1]{\ottdrule[#1]{%
\ottpremise{  \Gamma ,  \ottmv{c} \!:\! \phi   \vDash \mathsf{case} \hspace{2pt} ( \ottnt{a} \sim \ottnt{b} \;  \overline{\mu} \;  \bullet   : \ottnt{A} ) \hspace{2pt} \mathsf{of} \hspace{2pt}  F\;{ \overline{\upsilon} } :  \ottnt{B}  \Rightarrow  \ottnt{C}  \ | \,  \ottnt{C'} }%
}{
 \Gamma  \vDash \mathsf{case} \hspace{2pt} ( \ottnt{a} \sim \ottnt{b} \; \overline{\mu} : \ottnt{A} ) \hspace{2pt} \mathsf{of} \hspace{2pt}  F\;{ \ottsym{(}    \bullet  \; \overline{\upsilon}   \ottsym{)} } :   \forall \ottmv{c} \!:\! \phi . \ottnt{B}   \Rightarrow   \forall \ottmv{c} \!:\! \phi . \ottnt{C}   \ | \,  \ottnt{C'} }{%
{\ottdrulename{BranchTyping\_CPi}}{}%
}}

\newcommand{\ottdefnBranchTyping}[1]{\begin{ottdefnblock}[#1]{$ \Gamma  \vDash \mathsf{case} \hspace{2pt} ( \ottnt{a} \sim \ottnt{b} \; \overline{\mu} : \ottnt{A} ) \hspace{2pt} \mathsf{of} \hspace{2pt}  F\;{ \overline{\upsilon} } :  \ottnt{B}  \Rightarrow  \ottnt{C}  \ | \,  \ottnt{C'} $}{\ottcom{Branch Typing (aligning the types of case)}}
\ottusedrule{\ottdruleBranchTypingXXBase{}}
\ottusedrule{\ottdruleBranchTypingXXPiRole{}}
\ottusedrule{\ottdruleBranchTypingXXPiRel{}}
\ottusedrule{\ottdruleBranchTypingXXPiIrrel{}}
\ottusedrule{\ottdruleBranchTypingXXCPi{}}
\end{ottdefnblock}}

\newcommand{\ottdruleEXXWff}[1]{\ottdrule[#1]{%
\ottpremise{\Gamma  \vDash  \ottnt{a}  \ottsym{:}  \ottnt{A}}%
\ottpremise{\Gamma  \vDash  \ottnt{b}  \ottsym{:}  \ottnt{A}}%
\ottpremise{ \suppress{ \Gamma  \vDash  \ottnt{A}  \ottsym{:}   \star  } }%
}{
\Gamma  \vDash   \ottnt{a}   \sim _{ \ottnt{R} }  \ottnt{b}  :  \ottnt{A}  \, \ \mathsf{ok}}{%
{\ottdrulename{E\_Wff}}{}%
}}

\newcommand{\ottdefnPropWff}[1]{\begin{ottdefnblock}[#1]{$\Gamma  \vDash  \phi \, \ \mathsf{ok}$}{\ottcom{Prop wellformedness}}
\ottusedrule{\ottdruleEXXWff{}}
\end{ottdefnblock}}

\newcommand{\ottdruleEXXStar}[1]{\ottdrule[#1]{%
\ottpremise{\vDash  \Gamma}%
}{
\Gamma  \vDash   \star   \ottsym{:}   \star }{%
{\ottdrulename{E\_Star}}{}%
}}

\newcommand{\ottdruleEXXVar}[1]{\ottdrule[#1]{%
\ottpremise{\vDash  \Gamma}%
\ottpremise{\ottmv{x}  \ottsym{:}  \ottnt{A} \, \in \, \Gamma}%
}{
\Gamma  \vDash  \ottmv{x}  \ottsym{:}  \ottnt{A}}{%
{\ottdrulename{E\_Var}}{}%
}}

\newcommand{\ottdruleEXXPi}[1]{\ottdrule[#1]{%
\ottpremise{ \Gamma ,  \ottmv{x} \!:\! \ottnt{A}   \vDash  \ottnt{B}  \ottsym{:}   \star }%
\ottpremise{\Gamma  \vDash  \ottnt{A}  \ottsym{:}   \star }%
}{
\Gamma  \vDash   \mathrm{\Pi}^ \rho \ottmv{x} \!:\! \ottnt{A} \to \ottnt{B}   \ottsym{:}   \star }{%
{\ottdrulename{E\_Pi}}{}%
}}

\newcommand{\ottdruleEXXAbs}[1]{\ottdrule[#1]{%
\ottpremise{ \Gamma ,  \ottmv{x} \!:\! \ottnt{A}   \vDash  \ottnt{a}  \ottsym{:}  \ottnt{B}}%
\ottpremise{ \suppress{ \Gamma  \vDash  \ottnt{A}  \ottsym{:}   \star  } }%
\ottpremise{ ( \rho  = +) \vee ( \ottmv{x} \not\in\mathsf{fv}\;  \ottnt{a} ) }%
}{
\Gamma  \vDash   \mathrm{\lambda}^{ \rho } \ottmv{x} . \ottnt{a}   \ottsym{:}  \ottsym{(}   \mathrm{\Pi}^ \rho \ottmv{x} \!:\! \ottnt{A} \to \ottnt{B}   \ottsym{)}}{%
{\ottdrulename{E\_Abs}}{}%
}}

\newcommand{\ottdruleEXXApp}[1]{\ottdrule[#1]{%
\ottpremise{\Gamma  \vDash  \ottnt{b}  \ottsym{:}   \mathrm{\Pi}^ \ottsym{+} \ottmv{x} \!:\! \ottnt{A} \to \ottnt{B} }%
\ottpremise{\Gamma  \vDash  \ottnt{a}  \ottsym{:}  \ottnt{A}}%
}{
\Gamma  \vDash   \ottnt{b} \  \ottnt{a} ^{ \ottsym{+} }   \ottsym{:}  \ottnt{B}  \lbrace  \ottnt{a}  \ottsym{/}  \ottmv{x}  \rbrace}{%
{\ottdrulename{E\_App}}{}%
}}

\newcommand{\ottdruleEXXTApp}[1]{\ottdrule[#1]{%
\ottpremise{\Gamma  \vDash  \ottnt{b}  \ottsym{:}   \mathrm{\Pi}^ \ottsym{+} \ottmv{x} \!:\! \ottnt{A} \to \ottnt{B} }%
\ottpremise{\Gamma  \vDash  \ottnt{a}  \ottsym{:}  \ottnt{A}}%
\ottpremise{ \mathsf{Roles}\; ( \ottnt{b} )  =  \ottnt{R}  \ottsym{,}  \overline{R} }%
}{
\Gamma  \vDash   \ottnt{b} \  \ottnt{a} ^{ \ottnt{R} }   \ottsym{:}  \ottnt{B}  \lbrace  \ottnt{a}  \ottsym{/}  \ottmv{x}  \rbrace}{%
{\ottdrulename{E\_TApp}}{}%
}}

\newcommand{\ottdruleEXXIApp}[1]{\ottdrule[#1]{%
\ottpremise{\Gamma  \vDash  \ottnt{b}  \ottsym{:}   \mathrm{\Pi}^ \ottsym{-} \ottmv{x} \!:\! \ottnt{A} \to \ottnt{B} }%
\ottpremise{\Gamma  \vDash  \ottnt{a}  \ottsym{:}  \ottnt{A}}%
}{
\Gamma  \vDash   \ottnt{b} \  \Box ^{ \ottsym{-} }   \ottsym{:}  \ottnt{B}  \lbrace  \ottnt{a}  \ottsym{/}  \ottmv{x}  \rbrace}{%
{\ottdrulename{E\_IApp}}{}%
}}

\newcommand{\ottdruleEXXConv}[1]{\ottdrule[#1]{%
\ottpremise{\Gamma  \vDash  \ottnt{a}  \ottsym{:}  \ottnt{A}}%
\ottpremise{ \Gamma  ;   \widetilde { \Gamma }    \vDash   \ottnt{A}   \equiv _{ \ottkw{Rep} }  \ottnt{B}  :   \star  }%
\ottpremise{ \suppress{ \Gamma  \vDash  \ottnt{B}  \ottsym{:}   \star  } }%
}{
\Gamma  \vDash  \ottnt{a}  \ottsym{:}  \ottnt{B}}{%
{\ottdrulename{E\_Conv}}{}%
}}

\newcommand{\ottdruleEXXCPi}[1]{\ottdrule[#1]{%
\ottpremise{ \Gamma ,  \ottmv{c} \!:\! \phi   \vDash  \ottnt{B}  \ottsym{:}   \star }%
\ottpremise{ \suppress{ \Gamma  \vDash  \phi \, \ \mathsf{ok} } }%
}{
\Gamma  \vDash   \forall \ottmv{c} \!:\! \phi . \ottnt{B}   \ottsym{:}   \star }{%
{\ottdrulename{E\_CPi}}{}%
}}

\newcommand{\ottdruleEXXCAbs}[1]{\ottdrule[#1]{%
\ottpremise{ \Gamma ,  \ottmv{c} \!:\! \phi   \vDash  \ottnt{a}  \ottsym{:}  \ottnt{B}}%
\ottpremise{ \suppress{ \Gamma  \vDash  \phi \, \ \mathsf{ok} } }%
}{
\Gamma  \vDash   \mathrm{\Lambda} \ottmv{c} . \ottnt{a}   \ottsym{:}   \forall \ottmv{c} \!:\! \phi . \ottnt{B} }{%
{\ottdrulename{E\_CAbs}}{}%
}}

\newcommand{\ottdruleEXXCApp}[1]{\ottdrule[#1]{%
\ottpremise{\Gamma  \vDash  \ottnt{a_{{\mathrm{1}}}}  \ottsym{:}   \forall \ottmv{c} \!:\! \ottsym{(}   \ottnt{a}   \sim _{ \ottnt{R} }  \ottnt{b}  :  \ottnt{A}   \ottsym{)} . \ottnt{B_{{\mathrm{1}}}} }%
\ottpremise{ \Gamma  ;   \widetilde { \Gamma }    \vDash   \ottnt{a}   \equiv _{ \ottnt{R} }  \ottnt{b}  :  \ottnt{A} }%
}{
\Gamma  \vDash   \ottnt{a_{{\mathrm{1}}}} \; \bullet   \ottsym{:}  \ottnt{B_{{\mathrm{1}}}}  \lbrace  \bullet  \ottsym{/}  \ottmv{c}  \rbrace}{%
{\ottdrulename{E\_CApp}}{}%
}}

\newcommand{\ottdruleEXXConst}[1]{\ottdrule[#1]{%
\ottpremise{\vDash  \Gamma}%
\ottpremise{\ottmv{F}  \ottsym{:}   \ottnt{A} \  \ottsym{@} \  \overline{R}  \, \in \,  \Sigma_0 }%
\ottpremise{ \suppress{ \varnothing  \vDash  \ottnt{A}  \ottsym{:}   \star  } }%
}{
\Gamma  \vDash  \ottmv{F}  \ottsym{:}  \ottnt{A}}{%
{\ottdrulename{E\_Const}}{}%
}}

\newcommand{\ottdruleEXXFam}[1]{\ottdrule[#1]{%
\ottpremise{\vDash  \Gamma}%
\ottpremise{\ottmv{F}  \ottsym{:}   \ottnt{A} \  \ottsym{@} \  \overline{R} \ \mathsf{where}\  \ottnt{p}  \sim_{ \ottnt{R_{{\mathrm{1}}}} }  \ottnt{a}  \, \in \,  \Sigma_0 }%
\ottpremise{ \suppress{ \varnothing  \vDash  \ottnt{A}  \ottsym{:}   \star  } }%
}{
\Gamma  \vDash  \ottmv{F}  \ottsym{:}  \ottnt{A}}{%
{\ottdrulename{E\_Fam}}{}%
}}

\newcommand{\ottdruleEXXCase}[1]{\ottdrule[#1]{%
\ottpremise{\Gamma  \vDash  \ottnt{a}  \ottsym{:}  \ottnt{A}}%
\ottpremise{\Gamma  \vDash  \ottnt{b_{{\mathrm{1}}}}  \ottsym{:}  \ottnt{B}}%
\ottpremise{\Gamma  \vDash  \ottnt{b_{{\mathrm{2}}}}  \ottsym{:}  \ottnt{C}}%
\ottpremise{ \Gamma  \vDash \mathsf{case} \hspace{2pt} ( \ottnt{a} \sim \ottmv{F} \;    : \ottnt{A} ) \hspace{2pt} \mathsf{of} \hspace{2pt}  F\;{ \overline{\upsilon} } :  \ottnt{A_{{\mathrm{1}}}}  \Rightarrow  \ottnt{B}  \ | \,  \ottnt{C} }%
\ottpremise{\Gamma  \vDash  \ottmv{F}  \ottsym{:}  \ottnt{A_{{\mathrm{1}}}}}%
\ottpremise{\ottkw{Sat} \, \ottmv{F} \, \overline{\upsilon}}%
}{
\Gamma  \vDash   \mathsf{case} \hspace{3pt}  \ottnt{a}  \hspace{3pt} \mathsf{of} \hspace{3pt}  \ottmv{F} \  \overline{\upsilon}  \rightarrow  \ottnt{b_{{\mathrm{1}}}}  \| \_ \rightarrow  \ottnt{b_{{\mathrm{2}}}}   \ottsym{:}  \ottnt{C}}{%
{\ottdrulename{E\_Case}}{}%
}}

\newcommand{\ottdefnTyping}[1]{\begin{ottdefnblock}[#1]{$\Gamma  \vDash  \ottnt{a}  \ottsym{:}  \ottnt{A}$}{\ottcom{typing}}
\ottusedrule{\ottdruleEXXStar{}}
\ottusedrule{\ottdruleEXXVar{}}
\ottusedrule{\ottdruleEXXPi{}}
\ottusedrule{\ottdruleEXXAbs{}}
\ottusedrule{\ottdruleEXXApp{}}
\ottusedrule{\ottdruleEXXTApp{}}
\ottusedrule{\ottdruleEXXIApp{}}
\ottusedrule{\ottdruleEXXConv{}}
\ottusedrule{\ottdruleEXXCPi{}}
\ottusedrule{\ottdruleEXXCAbs{}}
\ottusedrule{\ottdruleEXXCApp{}}
\ottusedrule{\ottdruleEXXConst{}}
\ottusedrule{\ottdruleEXXFam{}}
\ottusedrule{\ottdruleEXXCase{}}
\end{ottdefnblock}}

\newcommand{\ottdruleEXXPropCong}[1]{\ottdrule[#1]{%
\ottpremise{ \Gamma  ;  \Delta   \vDash   \ottnt{A_{{\mathrm{1}}}}   \equiv _{ \ottnt{R} }  \ottnt{A_{{\mathrm{2}}}}  :  \ottnt{A} }%
\ottpremise{ \Gamma  ;  \Delta   \vDash   \ottnt{B_{{\mathrm{1}}}}   \equiv _{ \ottnt{R} }  \ottnt{B_{{\mathrm{2}}}}  :  \ottnt{A} }%
}{
\Gamma  \ottsym{;}  \Delta  \vDash   \ottnt{A_{{\mathrm{1}}}}   \sim _{ \ottnt{R} }  \ottnt{B_{{\mathrm{1}}}}  :  \ottnt{A}   \equiv   \ottnt{A_{{\mathrm{2}}}}   \sim _{ \ottnt{R} }  \ottnt{B_{{\mathrm{2}}}}  :  \ottnt{A} }{%
{\ottdrulename{E\_PropCong}}{}%
}}

\newcommand{\ottdruleEXXIsoConv}[1]{\ottdrule[#1]{%
\ottpremise{ \Gamma  ;  \Delta   \vDash   \ottnt{A}   \equiv _{ \ottnt{R_{{\mathrm{0}}}} }  \ottnt{B}  :   \star  }%
\ottpremise{\Gamma  \vDash   \ottnt{A_{{\mathrm{1}}}}   \sim _{ \ottnt{R} }  \ottnt{A_{{\mathrm{2}}}}  :  \ottnt{A}  \, \ \mathsf{ok}}%
\ottpremise{\Gamma  \vDash   \ottnt{A_{{\mathrm{1}}}}   \sim _{ \ottnt{R} }  \ottnt{A_{{\mathrm{2}}}}  :  \ottnt{B}  \, \ \mathsf{ok}}%
}{
\Gamma  \ottsym{;}  \Delta  \vDash   \ottnt{A_{{\mathrm{1}}}}   \sim _{ \ottnt{R} }  \ottnt{A_{{\mathrm{2}}}}  :  \ottnt{A}   \equiv   \ottnt{A_{{\mathrm{1}}}}   \sim _{ \ottnt{R} }  \ottnt{A_{{\mathrm{2}}}}  :  \ottnt{B} }{%
{\ottdrulename{E\_IsoConv}}{}%
}}

\newcommand{\ottdruleEXXCPiFst}[1]{\ottdrule[#1]{%
\ottpremise{ \Gamma  ;  \Delta   \vDash    \forall \ottmv{c} \!:\! \ottsym{(}   \ottnt{a_{{\mathrm{1}}}}   \sim _{ \ottnt{R_{{\mathrm{1}}}} }  \ottnt{a_{{\mathrm{2}}}}  :  \ottnt{A}   \ottsym{)} . \ottnt{B_{{\mathrm{1}}}}    \equiv _{ \ottnt{R'} }   \forall \ottmv{c} \!:\! \ottsym{(}   \ottnt{b_{{\mathrm{1}}}}   \sim _{ \ottnt{R_{{\mathrm{2}}}} }  \ottnt{b_{{\mathrm{2}}}}  :  \ottnt{B}   \ottsym{)} . \ottnt{B_{{\mathrm{2}}}}   :   \star  }%
}{
\Gamma  \ottsym{;}  \Delta  \vDash   \ottnt{a_{{\mathrm{1}}}}   \sim _{ \ottnt{R_{{\mathrm{1}}}} }  \ottnt{a_{{\mathrm{2}}}}  :  \ottnt{A}   \equiv   \ottnt{b_{{\mathrm{1}}}}   \sim _{ \ottnt{R_{{\mathrm{2}}}} }  \ottnt{b_{{\mathrm{2}}}}  :  \ottnt{B} }{%
{\ottdrulename{E\_CPiFst}}{}%
}}

\newcommand{\ottdefnIso}[1]{\begin{ottdefnblock}[#1]{$\Gamma  \ottsym{;}  \Delta  \vDash  \phi_{{\mathrm{1}}}  \equiv  \phi_{{\mathrm{2}}}$}{\ottcom{prop equality}}
\ottusedrule{\ottdruleEXXPropCong{}}
\ottusedrule{\ottdruleEXXIsoConv{}}
\ottusedrule{\ottdruleEXXCPiFst{}}
\end{ottdefnblock}}

\newcommand{\ottdruleEXXAssn}[1]{\ottdrule[#1]{%
\ottpremise{\vDash  \Gamma}%
\ottpremise{\ottmv{c}  \ottsym{:}  \ottsym{(}   \ottnt{a}   \sim _{ \ottnt{R} }  \ottnt{b}  :  \ottnt{A}   \ottsym{)} \, \in \, \Gamma}%
\ottpremise{\ottmv{c} \, \in \, \Delta}%
}{
 \Gamma  ;  \Delta   \vDash   \ottnt{a}   \equiv _{ \ottnt{R} }  \ottnt{b}  :  \ottnt{A} }{%
{\ottdrulename{E\_Assn}}{}%
}}

\newcommand{\ottdruleEXXRefl}[1]{\ottdrule[#1]{%
\ottpremise{\Gamma  \vDash  \ottnt{a}  \ottsym{:}  \ottnt{A}}%
}{
 \Gamma  ;  \Delta   \vDash   \ottnt{a}   \equiv _{ \ottnt{R} }  \ottnt{a}  :  \ottnt{A} }{%
{\ottdrulename{E\_Refl}}{}%
}}

\newcommand{\ottdruleEXXSym}[1]{\ottdrule[#1]{%
\ottpremise{ \Gamma  ;  \Delta   \vDash   \ottnt{b}   \equiv _{ \ottnt{R} }  \ottnt{a}  :  \ottnt{A} }%
}{
 \Gamma  ;  \Delta   \vDash   \ottnt{a}   \equiv _{ \ottnt{R} }  \ottnt{b}  :  \ottnt{A} }{%
{\ottdrulename{E\_Sym}}{}%
}}

\newcommand{\ottdruleEXXTrans}[1]{\ottdrule[#1]{%
\ottpremise{ \Gamma  ;  \Delta   \vDash   \ottnt{a}   \equiv _{ \ottnt{R} }  \ottnt{a_{{\mathrm{1}}}}  :  \ottnt{A} }%
\ottpremise{ \Gamma  ;  \Delta   \vDash   \ottnt{a_{{\mathrm{1}}}}   \equiv _{ \ottnt{R} }  \ottnt{b}  :  \ottnt{A} }%
}{
 \Gamma  ;  \Delta   \vDash   \ottnt{a}   \equiv _{ \ottnt{R} }  \ottnt{b}  :  \ottnt{A} }{%
{\ottdrulename{E\_Trans}}{}%
}}

\newcommand{\ottdruleEXXSub}[1]{\ottdrule[#1]{%
\ottpremise{ \Gamma  ;  \Delta   \vDash   \ottnt{a}   \equiv _{ \ottnt{R_{{\mathrm{1}}}} }  \ottnt{b}  :  \ottnt{A} }%
\ottpremise{ \ottnt{R_{{\mathrm{1}}}}  \leq  \ottnt{R_{{\mathrm{2}}}} }%
}{
 \Gamma  ;  \Delta   \vDash   \ottnt{a}   \equiv _{ \ottnt{R_{{\mathrm{2}}}} }  \ottnt{b}  :  \ottnt{A} }{%
{\ottdrulename{E\_Sub}}{}%
}}

\newcommand{\ottdruleEXXBeta}[1]{\ottdrule[#1]{%
\ottpremise{\Gamma  \vDash  \ottnt{a_{{\mathrm{1}}}}  \ottsym{:}  \ottnt{B}}%
\ottpremise{ \suppress{ \Gamma  \vDash  \ottnt{a_{{\mathrm{2}}}}  \ottsym{:}  \ottnt{B} } }%
\ottpremise{ \vDash   \ottnt{a_{{\mathrm{1}}}}  \rightarrow^{\beta}_{ \ottnt{R} }  \ottnt{a_{{\mathrm{2}}}} }%
}{
 \Gamma  ;  \Delta   \vDash   \ottnt{a_{{\mathrm{1}}}}   \equiv _{ \ottnt{R} }  \ottnt{a_{{\mathrm{2}}}}  :  \ottnt{B} }{%
{\ottdrulename{E\_Beta}}{}%
}}

\newcommand{\ottdruleEXXPiCong}[1]{\ottdrule[#1]{%
\ottpremise{ \Gamma  ;  \Delta   \vDash   \ottnt{A_{{\mathrm{1}}}}   \equiv _{ \ottnt{R'} }  \ottnt{A_{{\mathrm{2}}}}  :   \star  }%
\ottpremise{  \Gamma ,  \ottmv{x} \!:\! \ottnt{A_{{\mathrm{1}}}}   ;  \Delta   \vDash   \ottnt{B_{{\mathrm{1}}}}   \equiv _{ \ottnt{R'} }  \ottnt{B_{{\mathrm{2}}}}  :   \star  }%
\ottpremise{ \suppress{ \Gamma  \vDash  \ottnt{A_{{\mathrm{1}}}}  \ottsym{:}   \star  } }%
\ottpremise{ \suppress{ \Gamma  \vDash   \mathrm{\Pi}^ \rho \ottmv{x} \!:\! \ottnt{A_{{\mathrm{1}}}} \to \ottnt{B_{{\mathrm{1}}}}   \ottsym{:}   \star  } }%
\ottpremise{ \suppress{ \Gamma  \vDash   \mathrm{\Pi}^ \rho \ottmv{x} \!:\! \ottnt{A_{{\mathrm{2}}}} \to \ottnt{B_{{\mathrm{2}}}}   \ottsym{:}   \star  } }%
}{
 \Gamma  ;  \Delta   \vDash   \ottsym{(}   \mathrm{\Pi}^ \rho \ottmv{x} \!:\! \ottnt{A_{{\mathrm{1}}}} \to \ottnt{B_{{\mathrm{1}}}}   \ottsym{)}   \equiv _{ \ottnt{R'} }  \ottsym{(}   \mathrm{\Pi}^ \rho \ottmv{x} \!:\! \ottnt{A_{{\mathrm{2}}}} \to \ottnt{B_{{\mathrm{2}}}}   \ottsym{)}  :   \star  }{%
{\ottdrulename{E\_PiCong}}{}%
}}

\newcommand{\ottdruleEXXAbsCong}[1]{\ottdrule[#1]{%
\ottpremise{  \Gamma ,  \ottmv{x} \!:\! \ottnt{A_{{\mathrm{1}}}}   ;  \Delta   \vDash   \ottnt{b_{{\mathrm{1}}}}   \equiv _{ \ottnt{R'} }  \ottnt{b_{{\mathrm{2}}}}  :  \ottnt{B} }%
\ottpremise{ \suppress{ \Gamma  \vDash  \ottnt{A_{{\mathrm{1}}}}  \ottsym{:}   \star  } }%
\ottpremise{ ( \rho  = +) \vee ( \ottmv{x} \not\in\mathsf{fv}\;  \ottnt{b_{{\mathrm{1}}}} ) }%
\ottpremise{ ( \rho  = +) \vee ( \ottmv{x} \not\in\mathsf{fv}\;  \ottnt{b_{{\mathrm{2}}}} ) }%
}{
 \Gamma  ;  \Delta   \vDash   \ottsym{(}   \mathrm{\lambda}^{ \rho } \ottmv{x} . \ottnt{b_{{\mathrm{1}}}}   \ottsym{)}   \equiv _{ \ottnt{R'} }  \ottsym{(}   \mathrm{\lambda}^{ \rho } \ottmv{x} . \ottnt{b_{{\mathrm{2}}}}   \ottsym{)}  :  \ottsym{(}   \mathrm{\Pi}^ \rho \ottmv{x} \!:\! \ottnt{A_{{\mathrm{1}}}} \to \ottnt{B}   \ottsym{)} }{%
{\ottdrulename{E\_AbsCong}}{}%
}}

\newcommand{\ottdruleEXXAppCong}[1]{\ottdrule[#1]{%
\ottpremise{ \Gamma  ;  \Delta   \vDash   \ottnt{a_{{\mathrm{1}}}}   \equiv _{ \ottnt{R'} }  \ottnt{b_{{\mathrm{1}}}}  :  \ottsym{(}   \mathrm{\Pi}^ \ottsym{+} \ottmv{x} \!:\! \ottnt{A} \to \ottnt{B}   \ottsym{)} }%
\ottpremise{ \Gamma  ;  \Delta   \vDash   \ottnt{a_{{\mathrm{2}}}}   \equiv _{ \ottkw{Nom} }  \ottnt{b_{{\mathrm{2}}}}  :  \ottnt{A} }%
}{
 \Gamma  ;  \Delta   \vDash    \ottnt{a_{{\mathrm{1}}}} \  \ottnt{a_{{\mathrm{2}}}} ^{ \ottsym{+} }    \equiv _{ \ottnt{R'} }   \ottnt{b_{{\mathrm{1}}}} \  \ottnt{b_{{\mathrm{2}}}} ^{ \ottsym{+} }   :  \ottsym{(}  \ottnt{B}  \lbrace  \ottnt{a_{{\mathrm{2}}}}  \ottsym{/}  \ottmv{x}  \rbrace  \ottsym{)} }{%
{\ottdrulename{E\_AppCong}}{}%
}}

\newcommand{\ottdruleEXXTAppCong}[1]{\ottdrule[#1]{%
\ottpremise{ \Gamma  ;  \Delta   \vDash   \ottnt{a_{{\mathrm{1}}}}   \equiv _{ \ottnt{R'} }  \ottnt{b_{{\mathrm{1}}}}  :  \ottsym{(}   \mathrm{\Pi}^ \ottsym{+} \ottmv{x} \!:\! \ottnt{A} \to \ottnt{B}   \ottsym{)} }%
\ottpremise{ \Gamma  ;  \Delta   \vDash   \ottnt{a_{{\mathrm{2}}}}   \equiv _{  \ottnt{R} \wedge \ottnt{R'}  }  \ottnt{b_{{\mathrm{2}}}}  :  \ottnt{A} }%
\ottpremise{ \mathsf{Roles}\; ( \ottnt{a_{{\mathrm{1}}}} )  =  \ottnt{R}  \ottsym{,}  \overline{R} }%
\ottpremise{ \mathsf{Roles}\; ( \ottnt{b_{{\mathrm{1}}}} )  =  \ottnt{R}  \ottsym{,}  \overline{R} }%
\ottpremise{\Gamma  \vDash   \ottnt{b_{{\mathrm{1}}}} \  \ottnt{b_{{\mathrm{2}}}} ^{ \ottnt{R} }   \ottsym{:}  \ottnt{B}  \lbrace  \ottnt{a_{{\mathrm{2}}}}  \ottsym{/}  \ottmv{x}  \rbrace}%
}{
 \Gamma  ;  \Delta   \vDash    \ottnt{a_{{\mathrm{1}}}} \  \ottnt{a_{{\mathrm{2}}}} ^{ \ottnt{R} }    \equiv _{ \ottnt{R'} }   \ottnt{b_{{\mathrm{1}}}} \  \ottnt{b_{{\mathrm{2}}}} ^{ \ottnt{R} }   :  \ottsym{(}  \ottnt{B}  \lbrace  \ottnt{a_{{\mathrm{2}}}}  \ottsym{/}  \ottmv{x}  \rbrace  \ottsym{)} }{%
{\ottdrulename{E\_TAppCong}}{}%
}}

\newcommand{\ottdruleEXXIAppCong}[1]{\ottdrule[#1]{%
\ottpremise{ \Gamma  ;  \Delta   \vDash   \ottnt{a_{{\mathrm{1}}}}   \equiv _{ \ottnt{R} }  \ottnt{b_{{\mathrm{1}}}}  :  \ottsym{(}   \mathrm{\Pi}^ \ottsym{-} \ottmv{x} \!:\! \ottnt{A} \to \ottnt{B}   \ottsym{)} }%
\ottpremise{\Gamma  \vDash  \ottnt{a}  \ottsym{:}  \ottnt{A}}%
}{
 \Gamma  ;  \Delta   \vDash    \ottnt{a_{{\mathrm{1}}}} \  \Box ^{ \ottsym{-} }    \equiv _{ \ottnt{R} }   \ottnt{b_{{\mathrm{1}}}} \  \Box ^{ \ottsym{-} }   :  \ottsym{(}  \ottnt{B}  \lbrace  \ottnt{a}  \ottsym{/}  \ottmv{x}  \rbrace  \ottsym{)} }{%
{\ottdrulename{E\_IAppCong}}{}%
}}

\newcommand{\ottdruleEXXPiFst}[1]{\ottdrule[#1]{%
\ottpremise{ \Gamma  ;  \Delta   \vDash    \mathrm{\Pi}^ \rho \ottmv{x} \!:\! \ottnt{A_{{\mathrm{1}}}} \to \ottnt{B_{{\mathrm{1}}}}    \equiv _{ \ottnt{R'} }   \mathrm{\Pi}^ \rho \ottmv{x} \!:\! \ottnt{A_{{\mathrm{2}}}} \to \ottnt{B_{{\mathrm{2}}}}   :   \star  }%
}{
 \Gamma  ;  \Delta   \vDash   \ottnt{A_{{\mathrm{1}}}}   \equiv _{ \ottnt{R'} }  \ottnt{A_{{\mathrm{2}}}}  :   \star  }{%
{\ottdrulename{E\_PiFst}}{}%
}}

\newcommand{\ottdruleEXXPiSnd}[1]{\ottdrule[#1]{%
\ottpremise{ \Gamma  ;  \Delta   \vDash    \mathrm{\Pi}^ \rho \ottmv{x} \!:\! \ottnt{A_{{\mathrm{1}}}} \to \ottnt{B_{{\mathrm{1}}}}    \equiv _{ \ottnt{R} }   \mathrm{\Pi}^ \rho \ottmv{x} \!:\! \ottnt{A_{{\mathrm{2}}}} \to \ottnt{B_{{\mathrm{2}}}}   :   \star  }%
\ottpremise{ \Gamma  ;  \Delta   \vDash   \ottnt{a_{{\mathrm{1}}}}   \equiv _{ \ottkw{Nom} }  \ottnt{a_{{\mathrm{2}}}}  :  \ottnt{A_{{\mathrm{1}}}} }%
}{
 \Gamma  ;  \Delta   \vDash   \ottnt{B_{{\mathrm{1}}}}  \lbrace  \ottnt{a_{{\mathrm{1}}}}  \ottsym{/}  \ottmv{x}  \rbrace   \equiv _{ \ottnt{R} }  \ottnt{B_{{\mathrm{2}}}}  \lbrace  \ottnt{a_{{\mathrm{2}}}}  \ottsym{/}  \ottmv{x}  \rbrace  :   \star  }{%
{\ottdrulename{E\_PiSnd}}{}%
}}

\newcommand{\ottdruleEXXCPiCong}[1]{\ottdrule[#1]{%
\ottpremise{\Gamma  \ottsym{;}  \Delta  \vDash   \ottnt{a_{{\mathrm{1}}}}   \sim _{ \ottnt{R} }  \ottnt{b_{{\mathrm{1}}}}  :  \ottnt{A_{{\mathrm{1}}}}   \equiv   \ottnt{a_{{\mathrm{2}}}}   \sim _{ \ottnt{R} }  \ottnt{b_{{\mathrm{2}}}}  :  \ottnt{A_{{\mathrm{2}}}} }%
\ottpremise{  \Gamma ,  \ottmv{c} \!:\!  \ottnt{a_{{\mathrm{1}}}}   \sim _{ \ottnt{R} }  \ottnt{b_{{\mathrm{1}}}}  :  \ottnt{A_{{\mathrm{1}}}}    ;  \Delta   \vDash   \ottnt{A}   \equiv _{ \ottnt{R'} }  \ottnt{B}  :   \star  }%
\ottpremise{ \suppress{ \Gamma  \vDash   \ottnt{a_{{\mathrm{1}}}}   \sim _{ \ottnt{R} }  \ottnt{b_{{\mathrm{1}}}}  :  \ottnt{A_{{\mathrm{1}}}}  \, \ \mathsf{ok} } }%
\ottpremise{ \suppress{ \Gamma  \vDash   \forall \ottmv{c} \!:\!  \ottnt{a_{{\mathrm{1}}}}   \sim _{ \ottnt{R} }  \ottnt{b_{{\mathrm{1}}}}  :  \ottnt{A_{{\mathrm{1}}}}  . \ottnt{A}   \ottsym{:}   \star  } }%
\ottpremise{ \suppress{ \Gamma  \vDash   \forall \ottmv{c} \!:\!  \ottnt{a_{{\mathrm{2}}}}   \sim _{ \ottnt{R} }  \ottnt{b_{{\mathrm{2}}}}  :  \ottnt{A_{{\mathrm{2}}}}  . \ottnt{B}   \ottsym{:}   \star  } }%
}{
 \Gamma  ;  \Delta   \vDash    \forall \ottmv{c} \!:\!  \ottnt{a_{{\mathrm{1}}}}   \sim _{ \ottnt{R} }  \ottnt{b_{{\mathrm{1}}}}  :  \ottnt{A_{{\mathrm{1}}}}  . \ottnt{A}    \equiv _{ \ottnt{R'} }   \forall \ottmv{c} \!:\!  \ottnt{a_{{\mathrm{2}}}}   \sim _{ \ottnt{R} }  \ottnt{b_{{\mathrm{2}}}}  :  \ottnt{A_{{\mathrm{2}}}}  . \ottnt{B}   :   \star  }{%
{\ottdrulename{E\_CPiCong}}{}%
}}

\newcommand{\ottdruleEXXCAbsCong}[1]{\ottdrule[#1]{%
\ottpremise{  \Gamma ,  \ottmv{c} \!:\! \phi_{{\mathrm{1}}}   ;  \Delta   \vDash   \ottnt{a}   \equiv _{ \ottnt{R} }  \ottnt{b}  :  \ottnt{B} }%
\ottpremise{ \suppress{ \Gamma  \vDash  \phi_{{\mathrm{1}}} \, \ \mathsf{ok} } }%
}{
 \Gamma  ;  \Delta   \vDash   \ottsym{(}   \mathrm{\Lambda} \ottmv{c} . \ottnt{a}   \ottsym{)}   \equiv _{ \ottnt{R} }  \ottsym{(}   \mathrm{\Lambda} \ottmv{c} . \ottnt{b}   \ottsym{)}  :   \forall \ottmv{c} \!:\! \phi_{{\mathrm{1}}} . \ottnt{B}  }{%
{\ottdrulename{E\_CAbsCong}}{}%
}}

\newcommand{\ottdruleEXXCAppCong}[1]{\ottdrule[#1]{%
\ottpremise{ \Gamma  ;  \Delta   \vDash   \ottnt{a_{{\mathrm{1}}}}   \equiv _{ \ottnt{R'} }  \ottnt{b_{{\mathrm{1}}}}  :  \ottsym{(}   \forall \ottmv{c} \!:\! \ottsym{(}   \ottnt{a}   \sim _{ \ottnt{R} }  \ottnt{b}  :  \ottnt{A}   \ottsym{)} . \ottnt{B}   \ottsym{)} }%
\ottpremise{ \Gamma  ;   \widetilde { \Gamma }    \vDash   \ottnt{a}   \equiv _{ \ottnt{R} }  \ottnt{b}  :  \ottnt{A} }%
}{
 \Gamma  ;  \Delta   \vDash    \ottnt{a_{{\mathrm{1}}}} \; \bullet    \equiv _{ \ottnt{R'} }   \ottnt{b_{{\mathrm{1}}}} \; \bullet   :  \ottsym{(}  \ottnt{B}  \lbrace  \bullet  \ottsym{/}  \ottmv{c}  \rbrace  \ottsym{)} }{%
{\ottdrulename{E\_CAppCong}}{}%
}}

\newcommand{\ottdruleEXXCPiSnd}[1]{\ottdrule[#1]{%
\ottpremise{ \Gamma  ;  \Delta   \vDash    \forall \ottmv{c} \!:\! \ottsym{(}   \ottnt{a_{{\mathrm{1}}}}   \sim _{ \ottnt{R} }  \ottnt{a_{{\mathrm{2}}}}  :  \ottnt{A}   \ottsym{)} . \ottnt{B_{{\mathrm{1}}}}    \equiv _{ \ottnt{R_{{\mathrm{0}}}} }   \forall \ottmv{c} \!:\! \ottsym{(}   \ottnt{a'_{{\mathrm{1}}}}   \sim _{ \ottnt{R'} }  \ottnt{a'_{{\mathrm{2}}}}  :  \ottnt{A'}   \ottsym{)} . \ottnt{B_{{\mathrm{2}}}}   :   \star  }%
\ottpremise{ \Gamma  ;   \widetilde { \Gamma }    \vDash   \ottnt{a_{{\mathrm{1}}}}   \equiv _{ \ottnt{R} }  \ottnt{a_{{\mathrm{2}}}}  :  \ottnt{A} }%
\ottpremise{ \Gamma  ;   \widetilde { \Gamma }    \vDash   \ottnt{a'_{{\mathrm{1}}}}   \equiv _{ \ottnt{R'} }  \ottnt{a'_{{\mathrm{2}}}}  :  \ottnt{A'} }%
}{
 \Gamma  ;  \Delta   \vDash   \ottnt{B_{{\mathrm{1}}}}  \lbrace  \bullet  \ottsym{/}  \ottmv{c}  \rbrace   \equiv _{ \ottnt{R_{{\mathrm{0}}}} }  \ottnt{B_{{\mathrm{2}}}}  \lbrace  \bullet  \ottsym{/}  \ottmv{c}  \rbrace  :   \star  }{%
{\ottdrulename{E\_CPiSnd}}{}%
}}

\newcommand{\ottdruleEXXCast}[1]{\ottdrule[#1]{%
\ottpremise{ \Gamma  ;  \Delta   \vDash   \ottnt{a}   \equiv _{ \ottnt{R} }  \ottnt{b}  :  \ottnt{A} }%
\ottpremise{\Gamma  \ottsym{;}  \Delta  \vDash   \ottnt{a}   \sim _{ \ottnt{R} }  \ottnt{b}  :  \ottnt{A}   \equiv   \ottnt{a'}   \sim _{ \ottnt{R'} }  \ottnt{b'}  :  \ottnt{A'} }%
}{
 \Gamma  ;  \Delta   \vDash   \ottnt{a'}   \equiv _{ \ottnt{R'} }  \ottnt{b'}  :  \ottnt{A'} }{%
{\ottdrulename{E\_Cast}}{}%
}}

\newcommand{\ottdruleEXXEqConv}[1]{\ottdrule[#1]{%
\ottpremise{ \Gamma  ;  \Delta   \vDash   \ottnt{a}   \equiv _{ \ottnt{R} }  \ottnt{b}  :  \ottnt{A} }%
\ottpremise{ \Gamma  ;   \widetilde { \Gamma }    \vDash   \ottnt{A}   \equiv _{ \ottkw{Rep} }  \ottnt{B}  :   \star  }%
\ottpremise{ \suppress{ \Gamma  \vDash  \ottnt{B}  \ottsym{:}   \star  } }%
}{
 \Gamma  ;  \Delta   \vDash   \ottnt{a}   \equiv _{ \ottnt{R} }  \ottnt{b}  :  \ottnt{B} }{%
{\ottdrulename{E\_EqConv}}{}%
}}

\newcommand{\ottdruleEXXIsoSnd}[1]{\ottdrule[#1]{%
\ottpremise{\Gamma  \ottsym{;}  \Delta  \vDash   \ottnt{a}   \sim _{ \ottnt{R_{{\mathrm{1}}}} }  \ottnt{b}  :  \ottnt{A}   \equiv   \ottnt{a'}   \sim _{ \ottnt{R_{{\mathrm{1}}}} }  \ottnt{b'}  :  \ottnt{A'} }%
}{
 \Gamma  ;  \Delta   \vDash   \ottnt{A}   \equiv _{ \ottkw{Rep} }  \ottnt{A'}  :   \star  }{%
{\ottdrulename{E\_IsoSnd}}{}%
}}

\newcommand{\ottdruleEXXPatCong}[1]{\ottdrule[#1]{%
\ottpremise{ \Gamma  ;  \Delta   \vDash   \ottnt{a}   \equiv _{ \ottkw{Nom} }  \ottnt{a'}  :  \ottnt{A} }%
\ottpremise{ \Gamma  ;  \Delta   \vDash   \ottnt{b_{{\mathrm{1}}}}   \equiv _{ \ottnt{R_{{\mathrm{0}}}} }  \ottnt{b'_{{\mathrm{1}}}}  :  \ottnt{B} }%
\ottpremise{ \Gamma  ;  \Delta   \vDash   \ottnt{b_{{\mathrm{2}}}}   \equiv _{ \ottnt{R_{{\mathrm{0}}}} }  \ottnt{b'_{{\mathrm{2}}}}  :  \ottnt{C} }%
\ottpremise{ \Gamma  \vDash \mathsf{case} \hspace{2pt} ( \ottnt{a} \sim \ottmv{F} \;    : \ottnt{A} ) \hspace{2pt} \mathsf{of} \hspace{2pt}  F\;{ \overline{\upsilon} } :  \ottnt{A_{{\mathrm{1}}}}  \Rightarrow  \ottnt{B}  \ | \,  \ottnt{C} }%
\ottpremise{ \Gamma  \vDash \mathsf{case} \hspace{2pt} ( \ottnt{a'} \sim \ottmv{F} \;    : \ottnt{A} ) \hspace{2pt} \mathsf{of} \hspace{2pt}  F\;{ \overline{\upsilon} } :  \ottnt{A_{{\mathrm{1}}}}  \Rightarrow  \ottnt{B'}  \ | \,  \ottnt{C} }%
\ottpremise{ \Gamma  ;  \Delta   \vDash   \ottnt{B}   \equiv _{ \ottkw{Rep} }  \ottnt{B'}  :   \star  }%
\ottpremise{\ottkw{Sat} \, \ottmv{F} \, \overline{\upsilon}}%
\ottpremise{\Gamma  \vDash  \ottmv{F}  \ottsym{:}  \ottnt{A_{{\mathrm{1}}}}}%
}{
 \Gamma  ;  \Delta   \vDash    \mathsf{case} \hspace{3pt}  \ottnt{a}  \hspace{3pt} \mathsf{of} \hspace{3pt}  \ottmv{F} \  \overline{\upsilon}  \rightarrow  \ottnt{b_{{\mathrm{1}}}}  \| \_ \rightarrow  \ottnt{b_{{\mathrm{2}}}}    \equiv _{ \ottnt{R_{{\mathrm{0}}}} }   \mathsf{case} \hspace{3pt}  \ottnt{a'}  \hspace{3pt} \mathsf{of} \hspace{3pt}  \ottmv{F} \  \overline{\upsilon}  \rightarrow  \ottnt{b'_{{\mathrm{1}}}}  \| \_ \rightarrow  \ottnt{b'_{{\mathrm{2}}}}   :  \ottnt{C} }{%
{\ottdrulename{E\_PatCong}}{}%
}}

\newcommand{\ottdruleEXXLeftRel}[1]{\ottdrule[#1]{%
\ottpremise{ \mathsf{CasePath}_{ \ottnt{R'} }\;  \ottsym{(}   \ottnt{a} \  \ottnt{b} ^{ \ottnt{R_{{\mathrm{1}}}} }   \ottsym{)}  =  \ottmv{F} }%
\ottpremise{ \mathsf{CasePath}_{ \ottnt{R'} }\;  \ottsym{(}   \ottnt{a'} \  \ottnt{b'} ^{ \ottnt{R_{{\mathrm{1}}}} }   \ottsym{)}  =  \ottmv{F} }%
\ottpremise{\Gamma  \vDash  \ottnt{a}  \ottsym{:}   \mathrm{\Pi}^ \ottsym{+} \ottmv{x} \!:\! \ottnt{A} \to \ottnt{B} }%
\ottpremise{\Gamma  \vDash  \ottnt{b}  \ottsym{:}  \ottnt{A}}%
\ottpremise{\Gamma  \vDash  \ottnt{a'}  \ottsym{:}   \mathrm{\Pi}^ \ottsym{+} \ottmv{x} \!:\! \ottnt{A} \to \ottnt{B} }%
\ottpremise{\Gamma  \vDash  \ottnt{b'}  \ottsym{:}  \ottnt{A}}%
\ottpremise{ \Gamma  ;  \Delta   \vDash    \ottnt{a} \  \ottnt{b} ^{ \ottnt{R_{{\mathrm{1}}}} }    \equiv _{ \ottnt{R'} }   \ottnt{a'} \  \ottnt{b'} ^{ \ottnt{R_{{\mathrm{1}}}} }   :  \ottnt{B}  \lbrace  \ottnt{b}  \ottsym{/}  \ottmv{x}  \rbrace }%
\ottpremise{ \Gamma  ;   \widetilde { \Gamma }    \vDash   \ottnt{B}  \lbrace  \ottnt{b}  \ottsym{/}  \ottmv{x}  \rbrace   \equiv _{ \ottkw{Rep} }  \ottnt{B}  \lbrace  \ottnt{b'}  \ottsym{/}  \ottmv{x}  \rbrace  :   \star  }%
}{
 \Gamma  ;  \Delta   \vDash   \ottnt{a}   \equiv _{ \ottnt{R'} }  \ottnt{a'}  :   \mathrm{\Pi}^ \ottsym{+} \ottmv{x} \!:\! \ottnt{A} \to \ottnt{B}  }{%
{\ottdrulename{E\_LeftRel}}{}%
}}

\newcommand{\ottdruleEXXLeftIrrel}[1]{\ottdrule[#1]{%
\ottpremise{ \mathsf{CasePath}_{ \ottnt{R'} }\;  \ottsym{(}   \ottnt{a} \  \Box ^{ \ottsym{-} }   \ottsym{)}  =  \ottmv{F} }%
\ottpremise{ \mathsf{CasePath}_{ \ottnt{R'} }\;  \ottsym{(}   \ottnt{a'} \  \Box ^{ \ottsym{-} }   \ottsym{)}  =  \ottmv{F} }%
\ottpremise{\Gamma  \vDash  \ottnt{a}  \ottsym{:}   \mathrm{\Pi}^ \ottsym{-} \ottmv{x} \!:\! \ottnt{A} \to \ottnt{B} }%
\ottpremise{\Gamma  \vDash  \ottnt{b}  \ottsym{:}  \ottnt{A}}%
\ottpremise{\Gamma  \vDash  \ottnt{a'}  \ottsym{:}   \mathrm{\Pi}^ \ottsym{-} \ottmv{x} \!:\! \ottnt{A} \to \ottnt{B} }%
\ottpremise{\Gamma  \vDash  \ottnt{b'}  \ottsym{:}  \ottnt{A}}%
\ottpremise{ \Gamma  ;  \Delta   \vDash    \ottnt{a} \  \Box ^{ \ottsym{-} }    \equiv _{ \ottnt{R'} }   \ottnt{a'} \  \Box ^{ \ottsym{-} }   :  \ottnt{B}  \lbrace  \ottnt{b}  \ottsym{/}  \ottmv{x}  \rbrace }%
\ottpremise{ \Gamma  ;   \widetilde { \Gamma }    \vDash   \ottnt{B}  \lbrace  \ottnt{b}  \ottsym{/}  \ottmv{x}  \rbrace   \equiv _{ \ottkw{Rep} }  \ottnt{B}  \lbrace  \ottnt{b'}  \ottsym{/}  \ottmv{x}  \rbrace  :   \star  }%
}{
 \Gamma  ;  \Delta   \vDash   \ottnt{a}   \equiv _{ \ottnt{R'} }  \ottnt{a'}  :   \mathrm{\Pi}^ \ottsym{-} \ottmv{x} \!:\! \ottnt{A} \to \ottnt{B}  }{%
{\ottdrulename{E\_LeftIrrel}}{}%
}}

\newcommand{\ottdruleEXXRight}[1]{\ottdrule[#1]{%
\ottpremise{ \mathsf{CasePath}_{ \ottnt{R_{{\mathrm{2}}}} }\;  \ottsym{(}   \ottnt{a} \  \ottnt{b} ^{ \ottnt{R_{{\mathrm{1}}}} }   \ottsym{)}  =  \ottmv{F} }%
\ottpremise{ \mathsf{CasePath}_{ \ottnt{R_{{\mathrm{2}}}} }\;  \ottsym{(}   \ottnt{a'} \  \ottnt{b'} ^{ \ottnt{R_{{\mathrm{1}}}} }   \ottsym{)}  =  \ottmv{F} }%
\ottpremise{\Gamma  \vDash  \ottnt{a}  \ottsym{:}   \mathrm{\Pi}^ \ottsym{+} \ottmv{x} \!:\! \ottnt{A} \to \ottnt{B} }%
\ottpremise{\Gamma  \vDash  \ottnt{b}  \ottsym{:}  \ottnt{A}}%
\ottpremise{\Gamma  \vDash  \ottnt{a'}  \ottsym{:}   \mathrm{\Pi}^ \ottsym{+} \ottmv{x} \!:\! \ottnt{A} \to \ottnt{B} }%
\ottpremise{\Gamma  \vDash  \ottnt{b'}  \ottsym{:}  \ottnt{A}}%
\ottpremise{ \Gamma  ;  \Delta   \vDash    \ottnt{a} \  \ottnt{b} ^{ \ottnt{R_{{\mathrm{1}}}} }    \equiv _{ \ottnt{R_{{\mathrm{2}}}} }   \ottnt{a'} \  \ottnt{b'} ^{ \ottnt{R_{{\mathrm{1}}}} }   :  \ottnt{B}  \lbrace  \ottnt{b}  \ottsym{/}  \ottmv{x}  \rbrace }%
\ottpremise{ \Gamma  ;   \widetilde { \Gamma }    \vDash   \ottnt{B}  \lbrace  \ottnt{b}  \ottsym{/}  \ottmv{x}  \rbrace   \equiv _{ \ottkw{Rep} }  \ottnt{B}  \lbrace  \ottnt{b'}  \ottsym{/}  \ottmv{x}  \rbrace  :   \star  }%
}{
 \Gamma  ;  \Delta   \vDash   \ottnt{b}   \equiv _{  \ottnt{R_{{\mathrm{1}}}} \wedge \ottnt{R_{{\mathrm{2}}}}  }  \ottnt{b'}  :  \ottnt{A} }{%
{\ottdrulename{E\_Right}}{}%
}}

\newcommand{\ottdruleEXXCLeft}[1]{\ottdrule[#1]{%
\ottpremise{ \mathsf{CasePath}_{ \ottnt{R'} }\;  \ottsym{(}   \ottnt{a} \; \bullet   \ottsym{)}  =  \ottmv{F} }%
\ottpremise{ \mathsf{CasePath}_{ \ottnt{R'} }\;  \ottsym{(}   \ottnt{a'} \; \bullet   \ottsym{)}  =  \ottmv{F} }%
\ottpremise{\Gamma  \vDash  \ottnt{a}  \ottsym{:}   \forall \ottmv{c} \!:\! \ottsym{(}   \ottnt{a_{{\mathrm{1}}}}   \sim _{ \ottnt{R_{{\mathrm{1}}}} }  \ottnt{a_{{\mathrm{2}}}}  :  \ottnt{A}   \ottsym{)} . \ottnt{B} }%
\ottpremise{\Gamma  \vDash  \ottnt{a'}  \ottsym{:}   \forall \ottmv{c} \!:\! \ottsym{(}   \ottnt{a_{{\mathrm{1}}}}   \sim _{ \ottnt{R_{{\mathrm{1}}}} }  \ottnt{a_{{\mathrm{2}}}}  :  \ottnt{A}   \ottsym{)} . \ottnt{B} }%
\ottpremise{ \Gamma  ;   \widetilde { \Gamma }    \vDash   \ottnt{a_{{\mathrm{1}}}}   \equiv _{  \ottnt{R_{{\mathrm{1}}}} \wedge \ottnt{R'}  }  \ottnt{a_{{\mathrm{2}}}}  :  \ottnt{A} }%
\ottpremise{ \Gamma  ;  \Delta   \vDash    \ottnt{a} \; \bullet    \equiv _{ \ottnt{R'} }   \ottnt{a'} \; \bullet   :  \ottnt{B}  \lbrace  \bullet  \ottsym{/}  \ottmv{c}  \rbrace }%
}{
 \Gamma  ;  \Delta   \vDash   \ottnt{a}   \equiv _{ \ottnt{R'} }  \ottnt{a'}  :   \forall \ottmv{c} \!:\! \ottsym{(}   \ottnt{a_{{\mathrm{1}}}}   \sim _{ \ottnt{R_{{\mathrm{1}}}} }  \ottnt{a_{{\mathrm{2}}}}  :  \ottnt{A}   \ottsym{)} . \ottnt{B}  }{%
{\ottdrulename{E\_CLeft}}{}%
}}

\newcommand{\ottdefnDefEq}[1]{\begin{ottdefnblock}[#1]{$ \Gamma  ;  \Delta   \vDash   \ottnt{a}   \equiv _{ \ottnt{R} }  \ottnt{b}  :  \ottnt{A} $}{\ottcom{definitional equality}}
\ottusedrule{\ottdruleEXXAssn{}}
\ottusedrule{\ottdruleEXXRefl{}}
\ottusedrule{\ottdruleEXXSym{}}
\ottusedrule{\ottdruleEXXTrans{}}
\ottusedrule{\ottdruleEXXSub{}}
\ottusedrule{\ottdruleEXXBeta{}}
\ottusedrule{\ottdruleEXXPiCong{}}
\ottusedrule{\ottdruleEXXAbsCong{}}
\ottusedrule{\ottdruleEXXAppCong{}}
\ottusedrule{\ottdruleEXXTAppCong{}}
\ottusedrule{\ottdruleEXXIAppCong{}}
\ottusedrule{\ottdruleEXXPiFst{}}
\ottusedrule{\ottdruleEXXPiSnd{}}
\ottusedrule{\ottdruleEXXCPiCong{}}
\ottusedrule{\ottdruleEXXCAbsCong{}}
\ottusedrule{\ottdruleEXXCAppCong{}}
\ottusedrule{\ottdruleEXXCPiSnd{}}
\ottusedrule{\ottdruleEXXCast{}}
\ottusedrule{\ottdruleEXXEqConv{}}
\ottusedrule{\ottdruleEXXIsoSnd{}}
\ottusedrule{\ottdruleEXXPatCong{}}
\ottusedrule{\ottdruleEXXLeftRel{}}
\ottusedrule{\ottdruleEXXLeftIrrel{}}
\ottusedrule{\ottdruleEXXRight{}}
\ottusedrule{\ottdruleEXXCLeft{}}
\end{ottdefnblock}}

\newcommand{\ottdruleEXXEmpty}[1]{\ottdrule[#1]{%
}{
\vDash  \varnothing}{%
{\ottdrulename{E\_Empty}}{}%
}}

\newcommand{\ottdruleEXXConsTm}[1]{\ottdrule[#1]{%
\ottpremise{\vDash  \Gamma}%
\ottpremise{\Gamma  \vDash  \ottnt{A}  \ottsym{:}   \star }%
\ottpremise{\ottmv{x} \, \not\in \,  \widetilde { \Gamma } }%
}{
\vDash   \Gamma ,  \ottmv{x} \!:\! \ottnt{A} }{%
{\ottdrulename{E\_ConsTm}}{}%
}}

\newcommand{\ottdruleEXXConsCo}[1]{\ottdrule[#1]{%
\ottpremise{\vDash  \Gamma}%
\ottpremise{\Gamma  \vDash  \phi \, \ \mathsf{ok}}%
\ottpremise{\ottmv{c} \, \not\in \,  \widetilde { \Gamma } }%
}{
\vDash   \Gamma ,  \ottmv{c} \!:\! \phi }{%
{\ottdrulename{E\_ConsCo}}{}%
}}

\newcommand{\ottdefnCtx}[1]{\begin{ottdefnblock}[#1]{$\vDash  \Gamma$}{\ottcom{context wellformedness}}
\ottusedrule{\ottdruleEXXEmpty{}}
\ottusedrule{\ottdruleEXXConsTm{}}
\ottusedrule{\ottdruleEXXConsCo{}}
\end{ottdefnblock}}

\newcommand{\ottdruleSigXXEmpty}[1]{\ottdrule[#1]{%
}{
\vDash   \varnothing }{%
{\ottdrulename{Sig\_Empty}}{}%
}}

\newcommand{\ottdruleSigXXConsConst}[1]{\ottdrule[#1]{%
\ottpremise{\vDash  \Sigma}%
\ottpremise{\varnothing  \vDash  \ottnt{A}  \ottsym{:}   \star }%
\ottpremise{\ottmv{F} \, \not\in \, \mathsf{dom} \, \Sigma}%
}{
\vDash   \Sigma  \cup \{ \ottmv{F}  :   \ottnt{A} \  \ottsym{@} \  \overline{R}  \} }{%
{\ottdrulename{Sig\_ConsConst}}{}%
}}

\newcommand{\ottdruleSigXXConsAx}[1]{\ottdrule[#1]{%
\ottpremise{\vDash  \Sigma}%
\ottpremise{\ottmv{F} \, \not\in \, \mathsf{dom} \, \Sigma}%
\ottpremise{\varnothing  \vDash  \ottnt{A}  \ottsym{:}   \star }%
\ottpremise{ \mathsf{PatCtx}\; ( \ottnt{p} , \ottmv{F} : \ottnt{A} ) =  \Gamma  ;  \ottnt{B}  ;  \Omega  ;  V }%
\ottpremise{\Gamma  \vDash  \ottnt{a}  \ottsym{:}  \ottnt{B}}%
\ottpremise{V  \ottsym{\#}  \mathsf{fv}\! \, \ottnt{a}}%
\ottpremise{ \Omega  \vDash  \ottnt{a}  :  \ottnt{R} }%
}{
\vDash   \Sigma  \cup \{ \ottmv{F}  :   \ottnt{A} \  \ottsym{@} \   \mathsf{rng} \Omega  \ \mathsf{where}\  \ottnt{p}  \sim_{ \ottnt{R} }  \ottnt{a}  \} }{%
{\ottdrulename{Sig\_ConsAx}}{}%
}}

\newcommand{\ottdefnSig}[1]{\begin{ottdefnblock}[#1]{$\vDash  \Sigma$}{\ottcom{signature wellformedness}}
\ottusedrule{\ottdruleSigXXEmpty{}}
\ottusedrule{\ottdruleSigXXConsConst{}}
\ottusedrule{\ottdruleSigXXConsAx{}}
\end{ottdefnblock}}

\newcommand{\ottdruleARXXCast}[1]{\ottdrule[#1]{%
\ottpremise{ \Gamma   \vDash _{ \ottnt{R} }  \ottnt{a}  :  \ottnt{A} }%
\ottpremise{ \Gamma  \vDash  \ottnt{A}   \equiv _{ \ottkw{Rep} }  \ottnt{B}  :   \star  }%
}{
 \Gamma   \vDash _{ \ottnt{R} }   \ottkw{coerce} \  \ottnt{a}   :  \ottnt{B} }{%
{\ottdrulename{AR\_Cast}}{}%
}}

  \renewottcommands[ott]

\newcommand{\alt}{\ |\ }

\ifcomments
\newcommand{\scw}[1]{\textcolor{blue}{{SCW: #1}}}
\newcommand{\rae}[1]{\textcolor{magenta}{{RAE: #1}}}
\newcommand\av[1]{\textcolor{red}{{AV: #1}}}
\else
\newcommand{\scw}[1]{}
\newcommand{\rae}[1]{}
\newcommand\av[1]{}
\fi

\ifextended
\newcommand{\auxiliarymaterial}{the appendix}
\newcommand{\auxref}[1]{Appendix~\ref{#1}}
\else
\ifsubmission
\newcommand{\auxiliarymaterial}{the anonymized supplementary material}
\else
\newcommand{\auxiliarymaterial}{the extended version of this paper~\cite{dep-roles-extended}}
\fi
\newcommand{\auxref}[1]{\auxiliarymaterial}
\fi

\makeatletter\if@ACM@journal\makeatother
\acmJournal{PACMPL}
\acmVolume{1}
\acmNumber{1}
\acmArticle{1}
\acmYear{2019}
\acmMonth{1}
\startPage{1}
\else\makeatother
\acmConference[PL'19]{ACM SIGPLAN Conference on Programming Languages}{January 01--03, 2017}{New York, NY, USA}
\acmYear{2019}
\acmISBN{978-x-xxxx-xxxx-x/YY/MM}
\acmDOI{10.1145/nnnnnnn.nnnnnnn}
\startPage{1}
\fi

\setcopyright{acmlicensed}
\copyrightyear{2018}           

\begin{document}

\ifextended
\title{A Role for Dependent Types in Haskell (Extended version)}
\else
\title{A Role for Dependent Types in Haskell}
\fi

\author{Stephanie Weirich}
\affiliation{
  \department{Computer and Information Science}              
  \institution{University of Pennsylvania}            
  \streetaddress{3330 Walnut St}
  \city{Philadelphia}
  \state{PA}
  \postcode{19104}
  \country{USA}
}
\email{sweirich@cis.upenn.edu}

 \author{Pritam Choudhury}
 \affiliation{
   \position{}
   \department{Computer and Information Science}              
   \institution{University of Pennsylvania}            
   \country{USA}
 }
 \email{pritam@seas.upenn.edu}

 \author{Antoine Voizard}
 \affiliation{
   \position{}
   \department{Computer and Information Science}              
   \institution{University of Pennsylvania}            
   \country{USA}
 }
 \email{voizard@seas.upenn.edu}

 \author{Richard A.~Eisenberg}
 \affiliation{
   \position{Assistant Professor}
   \department{Computer Science}              
   \institution{Bryn Mawr College}            
   \streetaddress{101 N.~Merion Ave}
   \city{Bryn Mawr}
   \state{PA}
   \postcode{19010}
   \country{USA}
 }
 \email{rae@cs.brynmawr.edu}

\begin{abstract}
Modern Haskell supports \emph{zero-cost} coercions, a mechanism where types
that share the same run-time representation may be freely converted between.
To make sure such conversions are safe and desirable, this feature relies on a
mechanism of \emph{roles} to prohibit invalid coercions.
In this work, we show how to integrate roles with dependent type
systems and prove, using the Coq proof assistant, that the resulting system is
sound. We have designed this work as a foundation for the addition of
dependent types to the Glasgow Haskell Compiler, but we also expect that it
will be of use to designers of other dependently-typed languages who might
want to adopt Haskell's safe coercions feature.


\av{Being real nitpicky now, but shouldn't be either "integrate roles with
dependent types" or "integrate roles IN dependent(ly) type(d) systems"?}

\end{abstract}



\maketitle


\makeatletter
\mpr@andskip=1em plus 0.5fil minus 0.5em
\makeatother

\section{Combining safe coercions with dependent types}

\av{We still have this problem of the intro being a little dry/jumping right
into it. I think one of the missing link is that we don't mention that
we're about to introduce a mechanism, newtypes, that is the main (only?) way
to introduce types we want to (safely) coerce between. (we talked about
safely coercing (in the abstract + section title) but we jump right into newtypes,
not explaining how they fit in the picture).\\
Any suggestions for a space-efficient way to do that? I have one but I don't like it very much:
"In Haskell, the need for safe, zero-cost coercions arises from the presence of
newtypes --- type aliases that use the same representation than their original types,
yet are not equal to them."\\
Also, if possible it'd be nice to explain the spirit of newtypes (like in this suggestion)
before showing how
they're defined (the "with exactly one constructor and one argument" part is just
the technical way to define newtypes, not the spirit of what a newtype is)}

A \emph{newtype} in Haskell\footnote{In this paper, we use
  ``Haskell'' to refer to the language implemented by the Glasgow
  Haskell Compiler (GHC), version 8.6.}  is a user-defined algebraic datatype
with exactly one constructor; that constructor takes exactly one
argument. Here is an example:

\begin{lstlisting}
newtype HTML = MkHTML String
\end{lstlisting}

This declaration creates a generative abstraction; the \cd{HTML} type is
\emph{new} in the sense that it is not equal to any existing type.  We call
the argument type (\cd{String}) the \emph{representation type}.  Because a
newtype is isomorphic to its representation type, the Haskell \emph{compiler} uses
the same in-memory format for values of these types. Thus, creating a value of
a newtype (i.e., calling \cd{MkHTML}) is free at runtime, as is unpacking it
(i.e., using a pattern-match).

However, the Haskell \emph{type checker} treats the newtype and its representation
type as wholly distinct, meaning programmers cannot accidentally confuse
\cd{HTML} objects with \cd{String} objects. We thus call newtypes a
\emph{zero-cost} abstraction: a convenient compile-time distinction with no
cost to runtime efficiency. A newtype exported from a module without its
constructor is an abstract datatype; clients do not have access to its
representation.

Inside the defining module, a newtype is a translucent
abstraction: you can see through it with effort.
The \emph{safe coercions}~\cite{breitner2016} extension to the Glasgow Haskell
Compiler (GHC) reduces this effort through the availability of the
\cd{coerce} primitive.  For example, as \cd{HTML} and \cd{String} are
represented by the same bits in memory, so are the lists \cd{[HTML]} and
\cd{[String]}. Therefore, we can define a no-cost operation that converts the
former to the latter by coercing between representationally equal types.
\begin{lstlisting}
unpackList :: [HTML] -> [String]
unpackList = coerce
\end{lstlisting}

However, \cd{coerce} must be used with care. Not every structure is
appropriate for conversion. For example, converting a \cd{Map HTML Int} to a
\cd{Map String Int} would be disastrous if the ordering relation used for keys
differs between \cd{HTML} and \cd{String}.  Even worse, allowing
\cd{coerce} on types that use the \cd{type family}
feature~\cite{DBLP:conf/icfp/ChakravartyKJ05} leads to unsoundness.  Haskell
thus includes \emph{role annotations} for type constructors that indicate
whether it is appropriate to lift newtype-coercions through abstract
structures, such as \cd{Map}.

The key idea of \emph{safe coercions} is that there are two different notions
of type equality at play---the usual definition of type equality, called
\emph{nominal} equality, that distinguishes between the types \cd{HTML} and
\cd{String}, and \emph{representational} equality that identifies types
\av{"with the same representation,"}
suitable for coercion.  Some type constructor arguments are not congruent with
respect to representational equivalence, so role annotations prohibit the
derivation of these undesired \av{unsafe?}\scw{the map coercions are safe, but
still undesired}\av{Oh, I see, you're right!} equalities.

\subsection{Extending GHC with Dependent Types}

Recent work has laid out a design for Haskell extended with
\emph{dependent
  types}~\cite{weirich:dwk,gundry:phd,eisenberg:phd,weirich:systemd} and
there is ongoing work dedicated to implementing this
theory~\cite{xie-coercion-quantification}.\footnote{Also, see \url{https://github.com/ghc-proposals/ghc-proposals/pull/102}} Dependent
types are desirable for Haskell because they increase the ability to create
abstractions. Indexing types by terms allows datatypes to maintain
application-specific invariants, increasing program reliability and
expressiveness~\cite{swierstra:power-of-pi,weirich:icfp14,weirich:popl17}.

However, even though dependent type theories are fundamentally based on a rich
definition of type equality, it is not clear how best to incorporate roles and
safe coercions with these systems. In the context of GHC, this omission has
interfered with the incorporation of dependent types.  To make progress we
must reconcile the practical efficiency of safe, zero-cost coercions with the
power of dependent types. We need to know how a use of \cd{coerce} interacts with
type equality, and we must resolve how roles can be assigned in the presence
of type dependency.

\subsection{Contribution}

The contribution of this work is System~DR, a language that safely integrates
dependent types and roles. The starting point for our design is System~D, the
core language for Dependent Haskell from \citet{weirich:systemd}. Though this
language has full-spectrum dependent types, it is not a standard dependent
type theory: it admits logical inconsistency and the $ \star  :  \star $
axiom, along with support for equality assumptions and type erasure. Our
integration of roles is meant to be realizable in GHC and is thus based on 
the existing design of \citet{breitner2016}.

Integrating roles with Dependent Haskell's type system is not straightforward.
Unpacking the point above, our paper describes the following aspects of our
contribution:

\begin{itemize}
\item To model the two different notions of type equality, we index the
  type system's definition of equality by \emph{roles}, using the declared
  roles of abstract constructors to control the sorts of equivalences that may
  be derived. In Section~\ref{sec:key-idea} we describe how roles and newtype
  axioms interact with a minimal dependent type system. In particular, type
  equality is based on computation, so we also update the operational
  semantics to be role-sensitive. Newtypes evaluate to their representations
  only at the representational role; at the nominal role, they are
  values. In contrast, type family axioms step to their definitions at all
  roles.

\item
In Section~\ref{sec:full} we extend the basic system with support for
  the features of GHC. We start with a discussion of the interaction of roles
  with System~D's features of \emph{coercion abstraction}
  (Section~\ref{sec:coercion-abstraction}) and \emph{irrelevant arguments}
  (Section~\ref{sec:irrelevant-arguments}).

\item Supporting GHC's type families requires an operation for intensional
  type-analysis as type families branch on the head constructors of
  types. Therefore, in Section~\ref{sec:case} we add a \emph{case} expression
  to model this behavior. Because our language is dependently-typed, this
  expression supports both compile-time type analysis (as in type
  families) and run-time type analysis (i.e. typecase).

\item Our type equality includes \emph{nth} projections, a way to decompose
  equalities between type constants. We describe the rules that support these
  projections and how they interact with roles in
  Section~\ref{sec:decomposition}.

\item Our mechanized proof in Coq, available online,\footnote{\url{https://github.com/sweirich/corespec}}
 is presented in
  Section~\ref{sec:metatheory}. Proving safety is important because the
  combination of \cd{coerce} and type families, without roles, is known to
  violate type safety.\footnote{This safety violation was originally reported
    as \url{http://ghc.haskell.org/trac/ghc/ticket/1496}.} This work provides
  the first mechanically checked proof of the safety of this combination.

\item Our work solves a longstanding issue in GHC, known as the
  \emph{Constraint vs. Type problem}. In Section~\ref{sec:constraint-vs-type} we
  describe this problem and how defining \cd{Constraint} as a \cd{newkind}
  resolves this tension.

\item Our work sheds new light on the semantics of safe-coercions. Prior
  work~\cite{breitner2016}, includes a \emph{phantom} role, in addition to the
  nominal and representational roles. This role allows free conversion between
  the parameters of type constructors that do not use their arguments. In
  \ref{sec:phantom} we show that this role need not be made primitive, but
  instead can be encoded using irrelevant arguments.

\item We also observe that although our work integrates roles and dependent
  types at the level of GHC's core intermediate language, we lack a direct
  specification of source Haskell augmented with the \cd{coerce}
  primitive. The problem, which we describe in detail in
  Section~\ref{sec:no-popl11}, is that it is difficult to give an operational
  semantics of \cd{coerce}: reducing it away would violate type preservation,
  but it quite literally has no runtime behavior. Instead, in
  ~\ref{sec:src-language} we argue that our core language can provide a
  (type-sound) specification through elaboration.
\end{itemize}

Although our work is tailored to our goal of adding dependent types to
Haskell, an existing language with safe-coercions, we also view it as a
blueprint for adding safe-coercions to dependently-typed languages. Many
dependently-typed languages include features related to the ones discussed
here.  For example, some support semi-opaque definitions, such as Coq's
\cd{Opaque} and \cd{Transparent} commands. Such definitions often guide
type-class resolution~\cite{coq-classes,agda-classes,idris}, so precise
control over their unfolding is important. Cedille includes zero-cost
coercions~\cite{stump2018} and Idris has recently added experimental support
for
typecase\footnote{\url{https://gist.github.com/edwinb/25cd0449aab932bdf49456d426960fed}}.
Because the design considerations of these languages differ from that
of GHC, we compare our treatment of roles to modal dependent type theory in
Section~\ref{sec:modal}.

Our Coq 
is a
significant extension of prior work~\cite{weirich:systemd}.  Our work is fully
integrated---the same source is used to generate the \LaTeX inference rules,
Coq definitions, and variable binding infrastructure.  We use the tools
Ott~\cite{ott} and LNgen~\cite{aydemir:lngen} to represent the binding
structure using a \emph{locally nameless
  representation}~\cite{aydemir:popl-binders}.  Our code includes over 21k
nonblank, noncomment lines of Coq definitions and proofs.  This total includes
1.8k lines generated directly from our Ott definitions, and 7k lines generated
by LNgen.

In the next section, we review existing mechanisms for newtypes and safe
coercions in GHC in more detail and lay out the considerations that govern
their design. We present our new system starting in
Section~\ref{sec:key-idea}.

\section{Newtypes and safe coercions in Haskell}
\label{sec:example}


\subsection{Newtypes provide zero-cost abstractions}

We first flesh out the \cd{HTML} example of the introduction, by considering
this Haskell module:

\begin{lstlisting}[language=Haskell]
module Html( HTML, text, unHTML, ... ) where
newtype HTML = MkHTML String

unHTML :: HTML -> String
unHTML (MkHTML s) = s

text :: String -> HTML
text s = MkHTML (escapeSpecialCharacters s)

instance IsString HTML where
  fromString = text
\end{lstlisting}

As above, \cd{HTML} is a newtype; its representation type is \cd{String}.
This means that \cd{HTML}s and \cd{String}s are represented identically
at runtime and that the \cd{MkHTML} constructor compiles into the
identity function. However, the type system keeps \cd{HTML} and \cd{String}
separate: a function that expects an \cd{HTML} will \emph{not} accept
something of type \cd{String}.

Even in this small example, the Haskell programmer benefits from the newtype abstraction.
The module exports the type \cd{HTML} but not its data
constructor \cd{MkHTML}. Accordingly, outside the module, the only way to construct a
value of this type is to use the \cd{text} function, which enforces the
invariant of the data structure.  By exporting the type
\cd{HTML} without its data constructor, the module ensures that the type is
abstract---clients cannot make arbitrary strings into \cd{HTML}---and thereby
prevents, for instance, cross-site scripting attacks.

Naturally, the author of this module will want to reuse functions that
work with \cd{String}s to also work with values of type \cd{HTML}---even if
those functions actually work with, say, lists of \cd{String}s.
\av{Don't think that sentence works now (reuse/use/work with). Personally
liked the previous one better (heavy, but working)}\scw{pritam found the
original confusing. not sure how to rewrite.}
To support this reuse,
certain types, including functions \cd{(->)} and lists \cd{[]}, allow us to
\emph{lift} coercions between \cd{String} and \cd{HTML}. For example, suppose
we wish to break up chunks of \cd{HTML} into their constituent lines. We define

\begin{lstlisting}
linesH :: HTML -> [HTML]
linesH = coerce lines
\end{lstlisting}

Using Haskell's standard \cd{lines :: String -> [String]} function,
we have now, with minimal effort, lifted its action to work over \cd{HTML}.
Critically, the use of \cd{coerce} above is allowed only when
the \cd{MkHTML} constructor is in scope; in other words, \cd{linesH} can be written
in the \cd{Html} module but not outside it. In this way, the author of \cd{HTML}
has a chance to ensure that any invariants of the \cd{HTML} type are maintained
if an \cd{HTML} chunk is split into lines.

\subsection{Newtypes guide type-directed programming}
\rae{This description of type-directed programming seems less central to our
argument than other sections. If we need space, we can consider cutting.}

Newtypes allow more than just abstraction; they may also be used to guide
type-directed programming. For example, the sorting function in the base library
has the following type.
\begin{lstlisting}
sort :: Ord a => [a] -> [a]
\end{lstlisting}
The \cd{Ord} type class constraint means that sorting requires a comparison
function. When this function is called, the standard comparison function for
the element type will be used. In other words, the type of the list determines
how it is sorted.

Suppose our application sometimes must work with sorted lists of \cd{HTML}
chunks.  For efficiency reasons, we wish to partition our sorted lists into a
region where all chunks start with a tag (that is, the \cd{'<'} character) and
a region where no chunk starts with a tag. To that end, we define
a custom
\cd{Ord} instance that will sort all \cd{HTML} chunks that begin with
\cd{<} before all those that do not.

\begin{lstlisting}
instance Ord HTML where
  MkHTML left `compare` MkHTML right
    | tagged left == tagged right = left `compare` right
    | tagged left                 = LT
    | otherwise                   = GT
    where
      tagged ('<':_) = True
      tagged _       = False
\end{lstlisting}

Now, when we \cd{sort} a list of chunks, we can be confident that
the sorting algorithm will use our custom comparison operation. The validity
of this approach vitally depends on the generative nature of newtypes: if the
type-checker could confuse \cd{HTML} with \cd{String}, we could not be sure
whether type inference would select our custom ordering or the default
lexicographic ordering.

Newtypes can also be used to locally override the behavior of the sorting
operation. For example, the newtype \cd{Down a}, defined in the Haskell base
library, is isomorphic to its representation \cd{a}, but reverses the
comparison in its instance for \cd{Ord}.  Therefore, to sort a list in reverse
order, use \cd{coerce} to change the element type from \cd{a} to \cd{Down a},
thus modifying the comparison operation used by \cd{sort}.

\begin{lstlisting}
sortDown :: forall a. Ord a => [a] -> [a]
sortDown x = coerce (sort (coerce x :: [Down a]))
\end{lstlisting}

More generally, GHC's recent \cd{DerivingVia} extension~\cite{deriving-via},
based on \cd{coerce}, uses newtypes and their zero-cost coercions to extend
this idea.  This extension allows programmers to effectively write templates
for instances; individual types need not write their own class instances but
can select among the templates, each one embodied in a newtype.

\subsection{The problem with unfettered coerce}
\label{sec:discern}

We have shown that functions and lists support lifting coercions, but doing so
is not safe for all types. Consider this (contrived) example:

\begin{lstlisting}
type family Discern t where
  Discern String = Bool
  Discern HTML   = Char
data D a = MkD (Discern a)
\end{lstlisting}

The \cd{Discern} type family~\cite{DBLP:conf/icfp/ChakravartyKJ05,closed-type-families} behaves
like a function, where \cd{Discern String} is \cd{Bool} and \cd{Discern HTML} is
\cd{Char}. Thus, a \cd{D String} wraps a \cd{Bool}, while a \cd{D HTML} wraps
a \cd{Char}. Being able to use \cd{coerce} to go between \cd{D String} and \cd{D HTML}
would be disastrous: those two types have \emph{different} runtime representations
(in contrast to \cd{[String]} and \cd{[HTML]}). The goal of roles is to permit
safe liftings (like for lists) and rule out unsafe ones (like for \cd{D}).

Therefore, to control the use of \cd{coerce}, all datatype and newtype parameters
are assigned one of two roles: \emph{nominal} and \emph{representational}.\footnote{The implementation
in GHC supports a third role, \emph{phantom}, which behaves somewhat differently from
the other two. We ignore it for the bulk of this paper, returning to it
in Section~\ref{sec:phantom}.}
In a nominally-roled parameter, the \emph{name} of the type provided is
important \av{relevant?}\scw{too loaded of a word, don't want to actually
  connect to relevant/irrelevant arguments} to the definition, not just its representation.
The one parameter of \cd{D} is assigned a
nominal role because the definition of \cd{D} distinguishes between the names
\cd{String} and \cd{HTML}. We cannot safely coerce between \cd{D String} and
 \cd{D HTML}, because these two types have different representations. In contrast,
the type parameter of list has a representational role; coercing
between \cd{[String]} and \cd{[HTML]} is indeed safe.

Roles are assigned either by user annotation or by \emph{role
  inference}~\cite[Section~4.6]{breitner2016}. The safety of user-provided role
annotations is ensured by the compiler; the user would be unable to assign
a representational role to the parameter of \cd{D}.

\subsection{Representational equality}
\label{sec:coercible}

The full type of \cd{coerce} is \cd{Coercible a b => a -> b}. That is, we can
safely coerce between two types if those types are \cd{Coercible}. The pseudo-class
\cd{Coercible} (it has custom solving rules and is thus not a proper type class)
is an equivalence relation; we call it \emph{representational equality}. We
can thus coerce between any two representationally equal types. Representational
equality is coarser than Haskell's standard type equality (also called \emph{nominal
equality}): not only does it relate all pairs of types that are traditionally understood
to be equal, it also relates newtypes to their representation types.

Crucially, representational equality relates datatypes whose parameters have
the relationship indicated by the datatype's parameters' roles. Thus, because
the list type's parameter has a representational role, \cd{[ty1]} is
representationally equal to \cd{[ty2]} iff \cd{ty1} is representationally
equal to \cd{ty2}. And because \cd{D}'s parameter has a nominal role, \cd{D
  ty1} is representationally equal to \cd{D ty2} iff \cd{ty1} is nominally
equal to \cd{ty2}.

In addition to the \emph{lifting} rules sketched above, representational equality
also relates a newtype to its representation type, but with this caveat: this
relationship holds only when the newtype's constructor is in scope. This caveat
is added to allow the \cd{Html} module to enforce its abstraction barrier. If
a newtype were \emph{always} representationally equal to its representation,
then any client of \cd{Html} could use \cd{coerce} in place of the unavailable
constructor \cd{MkHTML}, defeating the goal of abstraction.

\subsection{Design considerations}
\label{sec:design-constraints}

The system for safe-coercions laid out in \citet{breitner2016} is subject to
design constraints that arise from the context of integration with the Haskell
language. In particular, safe coercions are considered an advanced feature and
should have minimal interaction with the rest of the language. In other words,
Haskell programmers should not need to think about roles if they never
use \cd{coerce}.

This separation between types and kinds was not present in the first design of
a role system for Haskell~\cite{weirich:newtypes}. Due to its complexity,
that first system was never integrated into GHC. Instead, by keeping roles
separate from types, \citet{breitner2016} simplified both the implementation of
\cd{coerce} (i.e., it was easier to extend the compiler) and the language
specification, as programmers who do not use \cd{coerce} need not understand
roles.

Keeping types and roles separate imposes two major constraints on the design
of System~DR.
\begin{itemize}
\item First, the type checking judgment should not also check roles.  In the
  system that we present in the next section, the type checking judgment
  $\Gamma  \vDash  \ottnt{a}  \ottsym{:}  \ottnt{A}$ does not depend on the role-checking judgment
  $ \Omega  \vDash  \ottnt{a}  :  \ottnt{R} $. The only interaction between these two systems is confined
  to checking the role annotations on top-level axioms. (In contrast, in the
  first version of the system, type and role checking occurred together in a
  single judgment.)

\item Second, the syntax of types and kinds should not include roles. In the
  first version of the system, the kinds of type constructors included role
  information for parameters. However, this means that all users of
  higher-order polymorphism must understand (and choose) these roles. Instead,
  \citet{breitner2016} does not modify the syntax of kinds, safely
  approximating role information with the nominal role when necessary.
\end{itemize}

In practice, the loss of expressiveness due to this simplification has not
been significant and roles have proven to be a popular extension in GHC.
However, we return to this discussion in Section~\ref{sec:modal}, when we
compare our design with the framework provided by modal dependent type theory.

\section{A calculus with dependent types and roles}
\label{sec:key-idea}

\begin{figure}[t]
\centering
Grammar
\[
\begin{array}{llcl}
\mathit{term/type\ variables}        & \ottmv{x} \\
\mathit{constants}                   & \ottmv{F}, \ottmv{T} \\
\mathit{roles}                       & R         & ::=& \ottkw{Nom}  \alt \ottkw{Rep} \\
\mathit{application\ flags}          & \nu    & ::=& \ottnt{R} \alt \ottsym{+} \\
\\
\mathit{terms,\ types} & a, b, A,B   & ::=&  \star  \alt \ottmv{x} \alt \ottmv{F} \alt
                                            \lambda  \ottmv{x} . \ottnt{b}  \alt  \ottnt{a} \  \ottnt{b} ^{ \nu }  \alt  \Pi  \ottmv{x} \!:\! \ottnt{A}  .  \ottnt{B}  \\
\\
\mathit{contexts}       &\Gamma     & ::=& \varnothing \alt  \Gamma ,  \ottmv{x} \!:\! \ottnt{A}  \\

\mathit{signatures}      &\Sigma      & ::=&  \varnothing  \alt  \Sigma  \cup \{ \ottmv{T}  :   \ottnt{A} \  \ottsym{@} \  \overline{R}  \} 
                                             \alt   \Sigma  \cup \{ \ottmv{F}  :   \ottnt{A} \  \ottsym{@} \  \overline{R} \ \mathsf{where}\  \ottnt{p}  \sim_{ \ottnt{R} }  \ottnt{a}  \}  \\
\mathit{patterns}       &\ottnt{p}& ::= &\ottmv{F} \alt  \ottnt{p} \  \ottmv{x} ^{ \ottnt{R} }  \\
\end{array}
\]
\caption{Syntax of core language}
\label{fig:min-syntax}
\end{figure}

\begin{figure}
\[
\begin{array}{c}
\\
\fbox{$ \Gamma  \vDash  \ottnt{a}  :  \ottnt{A} $} \hfill \vspace{6pt}\\
\drule{AE-Star}
\drule{AE-Var}
\drule{AE-Pi}
\drule{AE-Abs} \\ \\
\drule{AE-App}
\drule[width=4in]{AE-TApp} \\ \\
\drule{AE-Conv}
\drule{AE-Const}
\drule[width=3in]{AE-Fam}
\\ \\
\fbox{$ \Gamma  \vDash  \ottnt{a}   \equiv _{ \ottnt{R} }  \ottnt{b}  :  \ottnt{A} $} \hfill \vspace{6pt}\\
\drule{AE-Sub}
\drule{AE-Beta}  \\ \\
\drule[width=3in]{AE-PiCong}
\drule{AE-AbsCong}  \\ \\
\drule{AE-AppCong}
\drule[width=4in]{AE-TAppCong}  \\ \\
\drule{AE-Refl}
\drule{AE-Sym}
\drule[width=3in]{AE-Trans} \\ \\
\fbox{$ \mathsf{Roles}\; ( \ottnt{a} )  =  \overline{R} $} \hfill \vspace{6pt}\\
\drule{RolePath-AbsConst}
\drule{RolePath-Const}
\drule{RolePath-TApp}
\drule{RolePath-App}
\\ \\
\end{array}
\]
\caption{Typing and definitional equality for core language}
\label{fig:min-rules}
\end{figure}

We now introduce System~DR, a dependently-typed calculus with role-indexed
equality. To make our work more approachable, we present this calculus
incrementally, starting with the core ideas. In this section, we
start with a minimal calculus that contains only dependent functions,
constants and axioms.  In Section~\ref{sec:extensions}, we extend the
discussion to full System~DR, including case analysis, irrelevant arguments,
coercion abstraction, and decomposition rules.

System~DR is intended to model an extension of FC~\cite{systemfc}, the
explicitly-typed intermediate language of GHC.  As an intermediate language, it does
not need to specify Haskell's type inference algorithm or include features,
like type classes, that exist only in Haskell's source language.

Furthermore, the goal of our design of System~DR is to describe what terms
type check and how they evaluate. Like System D from prior
work~\cite{weirich:systemd}, this calculus need not have decidable type
checking for this purpose. Instead, once we have determined the language that
we want, we can then figure out how to annotate it in the implementation with
enough information to make type checking simple and syntax directed.  The
connection between System~D and System~DC in prior work provides a roadmap for
this (fairly mechanical) process. Furthermore, this process is also
constrained by implementation details of GHC that are beyond the scope of this
paper, so we do not include an annotated system here.

Therefore, the core calculus that we start with is a Curry-style
dependently-typed lambda-calculus with a single sort $ \star $. The syntax is
shown in Figure~\ref{fig:min-syntax}.  As in pure type
systems~\cite{barendregt-lambda-cube}, we have a shared syntax for terms and
types, but, as we don't require decidable type checking, there are no typing
annotations on function binders. This syntax has been decorated
with role information in two places---applications are marked by flags
($\nu$) and declarations of data type and newtype constants in the signature
($\Sigma$) include role annotations. Application flags are not needed for
source Haskell---they are easily added via elaboration and their presence here
is a mere technical device to make role information easily accessible.

Roles $\ottnt{R}$ are drawn from a lattice, with bottom element $\ottkw{Nom}$ for
Haskell's nominal role and top element $\ottkw{Rep}$ for the representational
role. We use $ \ottnt{R_{{\mathrm{1}}}}  \leq  \ottnt{R_{{\mathrm{2}}}} $ to talk about the ordering within the lattice.
\av{denote the ordering?}
The minimum operation ($ \ottnt{R_{{\mathrm{1}}}} \wedge \ottnt{R_{{\mathrm{2}}}} $) calculates the greatest lower bound
of the two roles.
\av{I'm not sure this sentence makes sense..? To me it sounds like "the GLB calculates
the GLB". My guess is that the intended meaning isn't this, of course --- is
there a way to rephrase it more clearly?}
For concreteness, this paper fixes that lattice to the two
element lattice, which is all that is needed for GHC. However, treating this
lattice generally\av{abstractly?} allows us to define the type system more uniformly.

The rules for typing and definitional equality for this fragment are shown in
Figure~\ref{fig:min-rules}.\footnote{For the purposes of presentation, these
  rules presented in this section are simplified versions of the rules that we
  use in our proofs. The complete listing of rules is available in
  \auxiliarymaterial.}  These rules are implicitly parameterized over a global
signature $\Sigma_{{\mathrm{0}}}$ of type constant declarations.

\paragraph{Typing relation.}

Most rules of the typing relation are standard for dependently-typed
languages.  Because Haskell includes nontermination, we do not need include a
universe hierarchy, instead using the $ \star : \star $
axiom~\cite{cardelli86}.  The novel rules are the application rules
(\rref{AE-App,AE-TApp}), the conversion rule (\rref{AE-Conv}) and the rules
for constants and axioms (\rref{AE-Const,AE-Fam}), all discussed below.

\paragraph{Role-indexed type equality.}


\av{Nitpick, but the following sentence doesn't work if we're talking about
an abstract role hierarchy (not necessarily just Nom and Rep).} %
In System~DR, the equality relation is indexed by a role that determines
whether the equality is \emph{nominal} or \emph{representational}. The
judgment $ \Gamma  \vDash  \ottnt{a}   \equiv _{ \ottnt{R} }  \ottnt{b}  :  \ottnt{A} $ defines when the terms $\ottnt{a}$ and $\ottnt{b}$,
of type $\ottnt{A}$, are equal \emph{at role $\ottnt{R}$}.  The rules for this
judgment appear in the middle of Figure~\ref{fig:min-rules}. This relation is
defined as essentially the language's small-step operational semantics, closed
over reflexivity, symmetry, transitivity, and congruence.
\av{Could shorten by just saying "This relation is the congruence closure of the
language's small-step operational semantics"} %
(Note that because
System~DR allows nontermination, the equality judgment is not a decidable
relation.)

The role sensitivity of the equality relation derives from the fact that
System~DR's small-step operational semantics, written $ \ottnt{a}  \rightarrow_{ \ottnt{R} }^{\beta}  \ottnt{b} $, is
also indexed by a role.
\av{also role sensitive? It sound weird to just say "indexed" --- it sounds
like we just made that choice arbitrarily, while this is of course a fundamental
requirement.} %
Specifically, the role in the small-step relation
determines whether top-level definitions can unfold to their right-hand sides
or are kept abstract (see \rref{ABeta-Axiom}, in
Figure~\ref{fig:min-axioms-roles}).

For example, we have that $ \ottkw{HTML}  \rightarrow_{ \ottkw{Rep} }^{\beta}  \ottkw{String} $, but at role
$\ottkw{Nom}$, the expression $\ottkw{HTML}$ is treated as a value. This step is
reflected into the equality relation via \rref{AE-Beta}.

\paragraph{Conversion.}

Dependently-typed languages use definitional equality for conversion: allowing
the types of terms to be implicitly replaced with equal types.  In source
Haskell, conversion is available for all types that are nominally equal. The
$\ottkw{coerce}$ primitive is required to convert between types that are
representationally equal. This primitive ensures that newtype distinctions are
maintained by default but are erasable when desired.

However, System~DR is intended to define GHC's intermediate language, so we
can assume that the source language type checker has already made sure that
users do not confuse \cd{HTML} and \cd{String}. Instead, the optimizer is free
to conflate these types, for great benefit.

Therefore System~DR does not include a $\ottkw{coerce}$ primitive.  Instead, the
conversion rule, \rref{AE-Conv}, allows conversion using the coarsest
relation, \emph{representational} equality. This choice simplifies the
design because all uses of coercion are implicit; there are no special rules
in the equality relation or operational semantics.
%
%
The downside of this design is that System~DR is not a definition of source
Haskell, an issue that we return to in Section~\ref{sec:src-language}.

\subsection{Role annotations and application congruence}
\label{sec:min-role-annotations}


Haskell allows data type and newtype constants to be
optionally ascribed with \emph{role annotations} for their
parameters. (Definitions without role ascriptions get their roles
inferred~\cite[Section~4.6]{breitner2016}.) These role annotations control
what equalities can be derived about these constants.  For example, \cd{Maybe}
has a representational parameter so \cd{Maybe HTML} is
representationally equal to \cd{Maybe String}. However, the type
\cd{Set HTML} can be prevented from being coercible to \cd{Set String} by
annotating its parameter with the nominal role.

In System~DR, a constant, like $\ottkw{Maybe}$, is an opaque declaration in the
signature $ \Sigma_0 $ of the form $ \ottmv{T} :  \ottnt{A} \  \ottsym{@} \  \overline{R}  $. This declaration specifies
the type $\ottnt{A}$ of the constant $\ottmv{T}$, as well as a list of roles $\overline{R}$
for its parameters. (We assume that role inference has already happened, so
all constants include role annotations.) For example, the signature might
declare constants $\ottkw{Set}:  \star   \to   \star  \  \ottsym{@} \   \ottkw{Nom}  $ and
$\ottkw{Maybe}:  \star   \to   \star  \  \ottsym{@} \   \ottkw{Rep}  $ with their usual types and roles.

The key idea from~\citet{breitner2016} is that the equality rule for
applications headed by constants uses these declared role annotations to determine how
to compare their arguments. We adapt this idea in this context using
\emph{application flags}, $\nu$, marking arguments in applications.  Consistent
usage of flags is checked by the typing judgment using \rref{AE-TApp}.  If
the head of the application is a constant, then this rule ensures that the
flag must be the one calculated by the (partial) function
$ \mathsf{Roles}\; ( \ottnt{a} )  =  \overline{R} $, shown at the bottom of Figure~\ref{fig:min-rules}.  For
example, $ \ottkw{Set} \  \ottkw{String} ^{ \ottkw{Nom} } $ and $ \ottkw{Maybe} \  \ottkw{Int} ^{ \ottkw{Rep} } $ are valid terms.
However, role annotations are optional and can be replaced by the
application flag $\ottsym{+}$, in which case \rref{AE-App} is used to check the
term.

\paragraph{Application congruence.}

The equality \rref{AE-TAppCong} defines when two role-annotated applications
$ \ottnt{a_{{\mathrm{1}}}} \  \ottnt{a_{{\mathrm{2}}}} ^{ \ottnt{R} } $ and $ \ottnt{b_{{\mathrm{1}}}} \  \ottnt{b_{{\mathrm{2}}}} ^{ \ottnt{R} } $ are equal at some (other) role $\ottnt{R'}$.  This
rule is most interesting when $\ottnt{R'}$ is $\ottkw{Rep}$---it explains why
\cd{Maybe HTML} and \cd{Maybe String} are representationally equal but \cd{Set
  String} and \cd{Set HTML} are not. Here, applications such as \cd{Maybe
  String} use role annotations on their arguments to enable this rule. In the
case of \cd{Maybe}, the $R$ above should also be $\ottkw{Rep}$ because \cd{Maybe}
is declared with a representational argument.

The first two premises of the rule specify that the corresponding components
of the application must be equal. Importantly, the role used for equality
between the arguments of the application ($\ottnt{a_{{\mathrm{2}}}}$ and $\ottnt{b_{{\mathrm{2}}}}$) is the
minimum of the current role $\ottnt{R'}$ and the declared role for the argument
$\ottnt{R}$. For example, for \cd{Set} the declared role is $\ottkw{Nom}$, so the
arguments must be nominally equal, as $ \ottkw{Nom} \wedge \ottkw{Rep} $ is $\ottkw{Nom}$, but
for \cd{Maybe} they may be representationally equal.

The use of the minimum role in this rule forces the nominal equality judgment
(i.e. when $\ottnt{R'}$ is $\ottkw{Nom}$), to compare all subterms using nominal
equality while allowing representational equality when both the context
\emph{and} the argument are representational.


The last premise of the rule is a subtle aspect of the combination of roles
and dependent types---it ensures that definitional equality is homogeneous.
In the conclusion of the rule, we want to ensure that both terms have the same
type, even when the type may be dependent. In this case, we know that
$\ottnt{a_{{\mathrm{2}}}}  \equiv  \ottnt{b_{{\mathrm{2}}}}$ at role $ \ottnt{R} \wedge \ottnt{R'} $, but this does not necessarily
imply that the types $\ottnt{B}  \lbrace  \ottnt{a_{{\mathrm{2}}}}  \ottsym{/}  \ottmv{x}  \rbrace$ and $\ottnt{B}  \lbrace  \ottnt{b_{{\mathrm{2}}}}  \ottsym{/}  \ottmv{x}  \rbrace$ are
representationally equal.
For example, given a new type $T$ with dependent type
$ \Pi  \ottmv{x} \!:\!  \star   .   \ottmv{F} \  \ottmv{x} ^{ \ottkw{Nom} }    \to   \star $, then we can show
\[
 \vdash   \ottmv{T} \  \ottkw{String} ^{ \ottkw{Rep} }  : \ottsym{(}   \ottmv{F} \  \ottkw{String} ^{ \ottkw{Nom} }   \ottsym{)}  \to   \star 
\] and
\[
 \vdash   \ottmv{T} \  \ottkw{HTML} ^{ \ottkw{Rep} }  : \ottsym{(}   \ottmv{F} \  \ottkw{HTML} ^{ \ottkw{Nom} }   \ottsym{)}  \to   \star 
\]
In this case, the terms are equal at role $\ottkw{Rep}$ but the types are
not.\footnote{In a system that annotates parameter roles on function types, we
  could check that the parameter is used consistently with the role annotation
  on its type. This would allow us to drop this premise from the application
  rule.}
 \av{This is because B
  could be a roled application, right? I think it might be worth laying out
  more clearly why it's the case. After all, this makes equality a
  non-congruence in general, which is both unusual and the point of roles, as
  we discussed.}  Therefore, the rule ensures that both sides have the same
type by fiat.

In comparison, the (non-roled) application congruence \rref{AE-AppCong} always
uses nominal equality for arguments, following \citet{breitner2016}.  Lacking
any other source of role information about the parameters (such as roles
annotating the function types), this rule defaults to requiring that they be
equal using the finest equality (i.e. $\ottkw{Nom}$).

Whether definitional equality uses \rref{AE-TAppCong} or
\rref{AE-AppCong} depends on the \emph{application flag} annotating the syntax
of the term.  The typing rules (\rref{AE-App} and \rref{AE-TApp}) ensure that the application
flag is appropriate. Some arguments have a choice of application
flag: they can either use the one specified by the roles in the signature, or
they can use \emph{+} (which defaults to $\ottkw{Nom}$ for congruence). Mostly
however, application flags are a technical device for our proofs as they
duplicate information that is already available in the abstract syntax tree.

\subsection{Type families and newtypes via axioms}
\label{sec:axioms}

\begin{figure}
\[
\begin{array}{c}
\fbox{$ \vDash   \ottnt{a}  \rightarrow^{\beta}_{ \ottnt{R} }  \ottnt{b} $} \hfill \vspace{6pt}\\
\drule{ABeta-AppAbs}
\drule[width=5in]{ABeta-Axiom} \\
\\
\fbox{$ \mathsf{MatchSubst}\; (  \ottnt{a}  ,  \ottnt{p}  ,  \ottnt{b_{{\mathrm{1}}}}  ) =  \ottnt{b_{{\mathrm{2}}}} $} \hfill \vspace{6pt}\\
\drule{MatchSubst-Const}
\drule{MatchSubst-AppRelR} \\ \\
\\
\fbox{$\vdash  \Sigma$} \hfill \vspace{6pt}\\
\drule{Sig-Empty}
\drule[width=3in]{Sig-ConsConst} \\ \\
\drule[width=6in]{SSig-ConsAx} \\
\\
\fbox{$ \mathsf{PatCtx}\;( \ottnt{p} ,  \ottmv{F} \!:\! \ottnt{A} ) =   \Gamma  ;  \ottnt{B}  ;  \Omega $} \hfill \vspace{6pt}\\
\drule{SPatCtx-Const}
\drule{SPatCtx-PiRel} \\
\\
\fbox{$ \Omega  \vDash  \ottnt{a}  :  \ottnt{R} $} \hfill \vspace{6pt}\\
\drule{Srole-a-Star}
\drule[width=3in]{Srole-a-Var}
\drule{Srole-a-Const} \\ \\
\drule[width=2in]{Srole-a-Fam}
\drule{Srole-a-Abs}
\drule{Srole-a-App} \\  \\
\drule[width=3in]{Srole-a-TApp}
\drule[width=3in]{Srole-a-Pi} \\
\end{array}
\]
\caption{Axioms and Role-checking}
\label{fig:min-axioms-roles}
\end{figure}

This calculus uses \emph{axioms} to model type families and newtypes in GHC.
An axiom declaration appears in the top-level signature and has the following
form.
\[ \ottmv{F} :  \ottnt{A} \  \ottsym{@} \  \overline{R} \ \mathsf{where}\  \ottnt{p}  \sim_{ \ottnt{R} }  \ottnt{a}  \]
The axiom introduces a contructor $\ottmv{F}$ of type $\ottnt{A}$ with parameter roles
$\overline{R}$. It also declares that the pattern $\ottnt{p}$ (which must be headed by
$\ottmv{F}$) can be equated at role $\ottnt{R}$ to the right-hand side term $\ottnt{a}$.

Patterns come from the following subgrammar of terms and are composed of
a sequence of applications.
\[ \ottnt{p} ::= \ottmv{F} \alt  \ottnt{p} \  \ottmv{x} ^{ \ottnt{R} }  \]
Each variable in the pattern is bound in the right hand side of the axiom. The
variables in the pattern are annotated with their roles, which are repeated in
the list $\overline{R}$ for convenience.

For example, compare the axiom for the type family definition \cd{F} on the
left side below with the one for the newtype declaration \cd{T} on the right.
In each case, the pattern is headed by the corresponding constant and binds
the variable $\ottmv{x}$ at role $\ottkw{Nom}$.

\begin{tabular}[c]{p{7cm}|p{7cm}}
\begin{lstlisting}
type family F a where
   F a = Maybe a
\end{lstlisting}
&
\begin{lstlisting}
newtype T a =
   MkT (F a)
\end{lstlisting} \\
$\ottmv{F} :   \star   \to   \star  \  \ottsym{@} \   \ottkw{Nom}  \ \ \mathsf{where}$
&
$\ottmv{T} :   \star   \to   \star  \  \ottsym{@} \   \ottkw{Nom}  \ \ \mathsf{where}$
\\
\qquad $  \ottmv{F} \  \ottmv{x} ^{ \ottkw{Nom} }    \sim _{ \ottkw{Nom} }  \ottsym{(}   \ottkw{Maybe} \  \ottmv{x} ^{ \ottkw{Rep} }   \ottsym{)} $
&
\qquad $  \ottmv{T} \  \ottmv{x} ^{ \ottkw{Nom} }    \sim _{ \ottkw{Rep} }  \ottsym{(}   \ottmv{F} \  \ottmv{x} ^{ \ottkw{Nom} }   \ottsym{)} $
\\
\\
\end{tabular}
The important distinction is the role marking the $\sim$ in the axiom
declarations: it is $\ottkw{Nom}$ for type families and $\ottkw{Rep}$ for newtypes.
This role determines whether a definition should be treated opaquely or
transparently by definitional equality.  For example, at the nominal role, we
have $ \ottmv{F} \  \ottkw{Int} ^{ \ottkw{Nom} } $ equal to $ \ottkw{Maybe} \  \ottkw{Int} ^{ \ottkw{Rep} } $ and $ \ottmv{T} \  \ottkw{Int} ^{ \ottkw{Nom} } $ distinct
from $ \ottmv{F} \  \ottkw{Int} ^{ \ottkw{Nom} } $. The representational role equates all of these
types.

These equalities are derived via the operational semantics. The types
$ \ottmv{F} \  \ottkw{Int} ^{ \ottkw{Nom} } $ and $ \ottkw{Maybe} \  \ottkw{Int} ^{ \ottkw{Rep} } $ are equal because the former reduces
to the latter. In other words, the operational semantics matches a term of the
form $ \ottmv{F} \  \ottkw{Int} ^{ \ottkw{Nom} } $ against a pattern in an axiom declaration $ \ottmv{F} \  \ottmv{x} ^{ \ottkw{Nom} } $,
producing a substitution $\{ \ottkw{Int} / \ottmv{x} \}$ that is applied to the
right-hand side of the axiom $\ottsym{(}   \ottkw{Maybe} \  \ottmv{x} ^{ \ottkw{Rep} }   \ottsym{)}  \lbrace  \ottkw{Int}  \ottsym{/}  \ottmv{x}  \rbrace$.

More generally, this behavior is specified using \rref{ABeta-Axiom}
This rule
uses an auxiliary relation $ \mathsf{MatchSubst}\; (  \ottnt{a}  ,  \ottnt{p}  ,  \ottnt{b}  ) =  \ottnt{b'} $ to determine whether
the scrutinee $a$ matches the pattern $p$, and if so, substitutes for the
pattern variables in the right-hand side. This rule is also only applicable as long
as the role of the axiom $\ottnt{R_{{\mathrm{1}}}}$ is less-than or equal to the role $\ottnt{R}$ that is
used for evaluation. For example, if $\ottnt{R}$ is $\ottkw{Nom}$ and $\ottnt{R_{{\mathrm{1}}}}$
is $\ottkw{Rep}$, then this rule does not apply (and $\ottnt{a}$ is an opaque
value). Alternatively, such as when $\ottnt{R_{{\mathrm{1}}}}$ is $\ottkw{Nom}$ as in a type family,
then this rule will replace the application headed by a constant with its definition.

Axiom constructors must be \emph{saturated} in order to reduce to their right
hand side. In other words, a constructor $\ottmv{F}$ must be applied to as many
arguments as specified by the pattern in the axiom.
Non-saturated constructor applications are treated as values by
the semantics, no matter their role.

Axioms extend the notion of \emph{definitions} from \citet{weirich:systemd},
which were always transparent and consequently had no need of role
annotations. As before, signatures are unordered and definitions may be
recursive---each right hand side may refer to any name in the entire
signature. As a result, axioms may be used to define a fixed point operator
or other functions and types that use general recursion.

\subsection{Role-checking}
\label{sec:role-checking}

The role checking judgment $ \Omega  \vDash  \ottnt{a}  :  \ottnt{R} $ is used by \rref{SSig-ConsAx},
which checks the well-formedness of axioms. This
rule uses the auxiliary function $ \mathsf{PatCtx}\;( \ottnt{p} ,  \ottmv{F} \!:\! \ottnt{A} ) =   \Gamma  ;  \ottnt{B}  ;  \Omega $ to determine
the context $\Gamma$ and type $\ottnt{B}$ to use to type check the right-hand side
of the axiom. This function also determines $\Omega$, the context to use when
role-checking the right-hand side. The notation $ \mathsf{rng} \Omega $ converts the
role-checking
context into a list of roles that is used when checking application flags.

Unlike opaque constants (which are inert) role annotations for the
parameters to an axiom must be checked by the system. In other words, if
a newtype axiom declares that it has a representational parameter, then there
are restrictions on how that parameter may be used. We check role annotations
using the role-checking judgment $ \Omega  \vDash  \ottnt{a}  :  \ottnt{R} $, shown at the bottom of
Figure~\ref{fig:min-axioms-roles}.

The role-checking context $\Omega$ assigns roles to variables. When we check
the well-formedness of axioms, the role-checking context is derived from the
annotations in pattern $\ottnt{p}$. Note, the roles declared in the pattern need
not be the \emph{most permissive} roles for $\ottnt{a}$. Even if the term would
check at role $\ottkw{Rep}$, the pattern may specify role $\ottkw{Nom}$ instead.

The rules of the role checking judgment appear at the bottom of
Figure~\ref{fig:min-axioms-roles}.  The \rref{Srole-a-Var} specifies that the
role of a variable must be greater-than or equal to its role in the context.
%
In \rref{Srole-a-TApp}, the role marking an annotated, relevant argument
determines how it will be checked. If the role annotation is not present, then
arguments must be checked at role $\ottkw{Nom}$, as in
\rref{Srole-a-App}. Analogously, when role-checking an abstraction, the bound
variable enters the context at role $\ottkw{Nom}$, as this is the most
conservative choice.

\section{Full System~DR}
\label{sec:full}
\label{sec:extensions}

\begin{figure}[t]
\centering
\[
\begin{array}{llcl}
\mathit{relevance}                   & \rho   & ::=& \ottsym{+} \alt \ottsym{-} \\
\mathit{application\ flags\ (terms)}  & \nu    & ::=& \ottnt{R} \alt \rho\\
\mathit{application\ flag\ (any)}  & \upsilon     & ::=&  \nu  \alt  \bullet  \\
\mathit{equality\ constraints}       & \phi   & ::=&  \ottnt{a}   \sim _{ \ottnt{R} }  \ottnt{b}  :  \ottnt{A}  \\
\\
\mathit{terms,\ types} & a, b, A,B   & ::=&  \star  \alt \ottmv{x} \alt \ottmv{F} \alt
                                            \mathrm{\lambda}^{ \rho } \ottmv{x} . \ottnt{b}  \alt  \ottnt{a} \  \ottnt{b} ^{ \nu } 
\alt  \mathrm{\Pi}^ \rho \ottmv{x} \!:\! \ottnt{A} \to \ottnt{B}  \\
& & \alt &  \Box  \alt  \mathrm{\Lambda} \ottmv{c} . \ottnt{b}  \alt  \ottnt{a} \; \bullet  \alt  \forall \ottmv{c} \!:\! \phi . \ottnt{B}  \\
& & \alt &  \mathsf{case} \hspace{3pt}  \ottnt{a}  \hspace{3pt} \mathsf{of} \hspace{3pt}  \ottmv{F} \  \overline{\upsilon}  \rightarrow  \ottnt{b_{{\mathrm{1}}}}  \| \_ \rightarrow  \ottnt{b_{{\mathrm{2}}}}  \\
\\
\mathit{contexts}       &\Gamma     & ::=& \varnothing \alt  \Gamma ,  \ottmv{x} \!:\! \ottnt{A}  \alt
                                               \Gamma ,  \ottmv{c} \!:\! \phi  \\
\mathit{patterns}       &\ottnt{p}& ::= &\ottmv{F} \alt  \ottnt{p} \  \ottmv{x} ^{ \ottnt{R} }  \alt   \ottnt{p} \  \ottmv{x} ^{ \ottsym{+} }  \alt  \ottnt{p} \  \Box ^{ \ottsym{-} }  \alt  \ottnt{p} \; \bullet  \\
\end{array}
\]
\caption{Syntax of full language}
\label{fig:full-syntax}
\end{figure}

\begin{figure}[t]
\begin{center}
\begin{tabular}{lll}
\emph{Typing}                         & $\Gamma  \vDash  \ottnt{a}  \ottsym{:}  \ottnt{A}$  \\
\emph{Definitional equality (terms)}\qquad\qquad & $ \Gamma  ;  \Delta   \vDash   \ottnt{a}   \equiv _{ \ottnt{R} }  \ottnt{b}  :  \ottnt{A} $  \\
\emph{Proposition well-formedness}    & $\Gamma  \vDash  \phi \, \ \mathsf{ok}$  \\
\emph{Definitional equality (propositions)}  & $\Gamma  \ottsym{;}  \Delta  \vDash  \phi_{{\mathrm{1}}}  \equiv  \phi_{{\mathrm{2}}}$
\\
\emph{Context well-formedness}        & $\vDash  \Gamma$ \\
\emph{Signature well-formedness}      & $\vDash  \Sigma$ \\
\\
\emph{Primitive ($\beta$-)reduction}  & $ \vDash   \ottnt{a}  \rightarrow^{\beta}_{ \ottnt{R} }  \ottnt{b} $ \\
\emph{One-step reduction}             & $ \vDash   \ottnt{a}   \leadsto _{ \ottnt{R} }  \ottnt{b} $ \\
\end{tabular}
\end{center}
\caption{Major judgment forms for System~DR.}
\label{fig:judgments}
\end{figure}

The previous section presented the complete details of a minimal calculus to
provide a solid basis for understanding about the interaction between roles
and dependent types in System~DR.
In this section, we zoom out and complete the story at a higher level,
providing an overview of the remaining features of the language. The syntax of
the full language appears in Figure~\ref{fig:full-syntax} and the major
judgment forms are summarized in Figure~\ref{fig:judgments}.
For
reference, the full specification of System~DR is available in \auxiliarymaterial.

\subsection{Coercion abstraction}
\label{sec:coercion-abstraction}

An essential feature of internal languages capable of compiling Haskell is
coercion abstraction, which is used to generalize over equality
propositions~\cite{systemfc}.  Coercion abstraction is the basis for the
implementation of \emph{generalized algebraic
  datatypes}~\cite{xi-gadt,gadt-type-inference} in GHC. For example, a
datatype definition, such as
\begin{lstlisting}
data T :: Type -> Type where MkT :: T Int
\end{lstlisting}
can be encoded by supplying \cd{MkT} with a constraint about
its parameter.
\begin{lstlisting}
data T :: Type -> Type where MkT :: forall a. (a ~ Int) => T a
\end{lstlisting}
Pattern matching an argument of type \cd{T b} brings the equality constraint
 \cd{b ~ Int} into scope.
\begin{lstlisting}
f :: T a -> a
f MkT = 42   -- we have an assumption (a ~ Int) in the context
\end{lstlisting}

In System~DR, definitional equality is indexed by a role $\ottnt{R}$, so we also
allow equality propositions, written $ \ottnt{a}   \sim _{ \ottnt{R} }  \ottnt{b}  :  \ottnt{A} $, to include roles.
When $\ottnt{R}$ is $\ottkw{Nom}$, this proposition corresponds to Haskell's equality
constraint, such as $\ottnt{a}  \sim  \ottkw{Int}$ above.  When the role is $\ottkw{Rep}$, it
corresponds to the \cd{Coercible} (\S\ref{sec:coercible}) constraint.

Coercion abstraction brings equality constraints into the context and
coercion application discharges those assumptions when the equality can be
satisfied.
As in extensional type theory~\cite{martin-lof71}, equality propositions can be used
implicitly by definitional equality. If an equality assumption between two types is
available in the context, then those two types are defined to be
equal.

\[ \drule[width=3in]{E-Assn} \]

For technical reasons, discussed in \citet{weirich:systemd}, the full judgment
for definitional equality in System DR is written
$ \Gamma  ;  \Delta   \vDash   \ottnt{a}   \equiv _{ \ottnt{R} }  \ottnt{b}  :  \ottnt{A} $. The additional component is a set $\Delta$ that
tracks which coercions are actually available, used in the rule above. This
set need not concern us here; feel free simply to assume that $\Delta$ equals
the domain of $\Gamma$, written $ \widetilde { \Gamma } $.

The extension of the System~D rules with roled equality constraints is
straightforward, though care must be taken to ensure that the roles are used
consistently.  Note that when role-checking, all variables that appear in a
nominal equality constraint must have role $\ottkw{Nom}$. This corresponds to the
requirement in GHC that the constrained parameters to GADTs have nominal
arguments.

\subsection{Irrelevant arguments}
\label{sec:irrelevant-arguments}

A dependently-typed intermediate language for GHC must include support for
\emph{irrelevant} arguments as well as relevant arguments~\cite{Miquel:ICC} in
order to implement the type-erasure aspect of parametric polymorphism. In
Haskell, polymorphic functions cannot dispatch on types, so these may be
erased prior to runtime. In (Curry-style) System~DR, irrelevant arguments are
therefore elided from the abstract syntax. We extend the calculus by adding a
new application flag ``$\ottsym{-}$'' to indicate that an argument is irrelevant.
Furthermore, we add a flag $\rho$ to function types to
indicate whether the argument to the
function is relevant or irrelevant.

The typing rules for the introduction of an irrelevant abstraction requires
that the bound variable not actually appear in the body of the term.
When an irrelevant
function is used in an application, the argument must be the trivial term,
$\Box$. Note that the argument is only elided from the term however---it is
still substituted in the result type.\footnote{In System DC, the annotated
  version of the language with syntax-directed type checking, the argument
  does appear in term but is eliminated via an erasure operation,
  following~\citet{barras08}.}

\[ \drule[width=3in]{AE-IAbs} \qquad \drule[width=3in]{E-IApp} \]

Role annotations may \emph{only} apply to relevant arguments, even though
constants and newtypes may have both relevant and irrelevant
parameters. Irrelevant arguments have their own congruence rule for
applications.  Because irrelevant arguments never appear in the syntax of
terms, an equality between two irrelevant applications only need compare the
function components---the arguments are always trivially equal.

\[ \drule[width=5in]{E-IAppCong} \]

Overall, there is little interaction between irrelevant arguments and
roles. However, there is one important benefit of having both capabilities in
the same system. We can use irrelevant quantification to model the
\emph{phantom} role from prior work; details are in Section~\ref{sec:phantom}.

\subsection{Case expressions}
\label{sec:case}

\begin{figure}
\[
\begin{array}{c}
\drule[width=3in]{Beta-PatternTrue} \qquad
\drule[width=3in]{Beta-PatternFalse}
\end{array}
\]
\caption{Case analysis}
\label{fig:case}
\end{figure}

The soundness issue described in Section~\ref{sec:discern} arises through
the use of the \cd{Discern} type family, which returns different results based
on whether its argument is \cd{String} or \cd{HTML}.
To ensure that System~DR is not susceptible to a similar issue, we
include a pattern matching term of the following
form.\footnote{Unlike source Haskell, patterns in System~DR axiom definitions
  may only include variables and thus may not dispatch on their arguments.}
\[  \mathsf{case} \hspace{3pt}  \ottnt{a}  \hspace{3pt} \mathsf{of} \hspace{3pt}  \ottmv{F} \  \overline{\upsilon}  \rightarrow  \ottnt{b_{{\mathrm{1}}}}  \| \_ \rightarrow  \ottnt{b_{{\mathrm{2}}}}  \]
Operationally, the pattern matching term reduces the scrutinee $\ottnt{a}$ to a
value and then compares it against the pattern specified by
$\ottmv{F}\;\overline{\upsilon}$. If there is a match, the expression steps to $\ottnt{b_{{\mathrm{1}}}}$. In
all other cases, the expression steps to $\ottnt{b_{{\mathrm{2}}}}$. Pattern matching is not
nested---only the head constructor can be observed. In Haskell, type families
do both axiom unfolding and discrimination. We separate these features in
System~DR for orthogonality and eventual unification of pattern matching with
Haskell's existing \cd{case} expression. (The semantics of this expression is
not exactly the same as that of Haskell's \cd{case}; more details are in
Section~\ref{sec:extended-pattern-matching}.)

In this syntax, the scrutinee must match the pattern of arguments specified by
$\ottmv{F}\;\overline{\upsilon}$, where $\overline{\upsilon}$ is a list of application flags. Note that
in the full language, these application flags can include roles, +, -, or
indicate a coercion argument.\footnote{This nameless form of
  pattern-matching helps with our formalization.
  The typing rule for case requires the branch
  $\ottnt{b_{{\mathrm{1}}}}$ to start with a sequence of abstractions that matches the form
  specified by the list of flags $\overline{\upsilon}$.}
We specify the behavior of $\mathsf{case}$ with the rules
shown in Figure~\ref{fig:case}.
In the first
rule, the $ \ottnt{a}  \leftrightarrow_{ \ottnt{R} }  \ottmv{F} \; \overline{\upsilon} $ judgment holds when the scrutinee matches the
pattern; i.e. when the scrutinee is an application headed by $\ottmv{F}$ with
arguments specified by $\overline{\upsilon}$. The constructor $\ottmv{F}$ must be a constant
at role $\ottkw{Nom}$; it cannot be a type family axiom.
If this judgment holds, the second premise
passes those arguments to the branch $\ottnt{b_{{\mathrm{1}}}}$. In the conclusion
of the rule, $\ottnt{b'_{{\mathrm{1}}}}$ is further applied to an elided coercion $ \bullet $;
this coercion witnesses the equality between the head of $\ottnt{a}$ and the pattern,
implementing dependent pattern matching.

The second rule, \rref{Beta-PatternFalse}
triggers when the scrutinee is a value, yet the comparison
$ \ottnt{a}  \leftrightarrow_{ \ottnt{R} }  \ottmv{F} \; \overline{\upsilon} $ does not hold.
It steps directly to $\ottnt{b_{{\mathrm{2}}}}$.



Dependent case analysis mean that when the scrutinee matches the head
constructor, not only does the expression step to the first branch, but the
branch is type checked under the assumption of an equality proposition between
the scrutinee and the pattern.

Simple examples of this behavior are possible in source Haskell today, using
the \cd{TypeInType} extension.
\begin{lstlisting}
type family F k (a :: k) :: Bool where
   F Bool  x = x      -- here we have (k ~ Bool)
   F _     x = False
\end{lstlisting}

Because System~DR is dependently-typed, and full-spectrum, the
pattern matching term described in this section can also be used for run-time
type analysis, as well as dispatch during type
checking. We view this as a key benefit of our complete design: we retain the ability
to erase (most) types during compilation by abstracting them via irrelevant quantification,
but can support run-time dispatch on types when desired.
\av{I realize that I don't think we've explained "type erasure semantics", and
as such I'm not fully confident that what I'm thinking of is the intended meaning.}
\av{Also, as of now this is a little dry; either a tiny example or just a bit
more explanations would really help. I agree this is a really cool feature of
the system, and I think it deserves a bigger share of our precious lines.}
\scw{we should write a different paper.}

\paragraph{Case analysis is nominal}

One part of our design that we found surprising is the fact that case analysis
must use the nominal role to evaluate the scrutinee, as we see in the following
rule:\footnote{This rule belongs to a different judgment than the rules in Figure~\ref{fig:case}.
We separate our primitive $\beta$-reduction rules from the congruence rules
for stepping in our semantics.
Only the $\beta$-reduction rules are used in our equality relation,
relying on the equality
relation's congruence rules to correspond to the stepping relation's
congruence rules.}

\[
\drule{E-Pattern}
\]

Indeed, our original draft of the system also allowed a form of
``representational'' case analysis, which first evaluated the scrutinee to a
representational value before pattern matching.  This case analysis could
``see through'' newtype definitions and would match on the underlying
definition.

For example, with representational case analysis, the term
\begin{lstlisting}
case-rep HTML of
   String -> True
   _      -> False
\end{lstlisting}
would evaluate to \cd{True}.

Unfortunately, we found that representational case analysis is unsound in our
system.
%
%
Consider the following term, which uses a representational analysis to
first match the outer structure of its argument, and then uses an inner,
nominal analysis for the parameter. System~DR always assigns nominal roles to
variables bound in a case-match, so this axiom would role-check with a
representational argument.
\begin{lstlisting}
F x = case-rep x of
       [y] -> case-nom y of
                HTML -> True
                _    -> False
        _  -> False
\end{lstlisting}
With this definition, we would be able to show \cd{F [HTML]}
representationally equal to \cd{F [String]} because \cd{F}'s parameter is
representation. However, these two expressions evaluate to different
results. Disaster!

Extending the system to include a safe version of representational case
analysis requires a way to rule out the nominal case analysis of
\cd{y} above. This means that the type system must record \cd{y}'s role
as representational (as it is the argument to the list constructor) and
furthermore use the role-checking judgment to ensure that \cd{y} does not
appear in a nominal context (such as in the scrutinee of a nominal case
analysis). We want to keep role checking completely separate from type
checking (cf. Section~\ref{sec:design-constraints}), so we have not
pursued this extension.
\scw{Mention how the min rule keeps nominal case analysis sound??}

\subsection{Constructor Injectivity}
\label{sec:decomposition}

System~DR is a syntactic type theory. As a result, it supports equality rules
for injectivity.  If two types are equal, then corresponding subterms of those types
should also be equal. In prior work~\cite{weirich:systemd}, the injectivity of
function types was witnessed by rules that allowed an equality between two
function types to be decomposed.

This work augments those rules with the correct role components. For example,
in \rref{E-PiSnd}, shown below, when we pull out an equality between the co-domain types of
a function type, we must provide an equality between the arguments at role
$\ottkw{Nom}$. This is because we have no knowledge about how the parameter
$\ottmv{x}$ is used inside the types $\ottnt{B_{{\mathrm{1}}}}$ and $\ottnt{B_{{\mathrm{2}}}}$ and so we must be
conservative.\footnote{If the function type were annotated with a role, we
  would not be limited to $\ottkw{Nom}$ in this rule.}

\begin{center}
\drule[width=5in]{E-PiSnd}
\end{center}

 \scw{Do we need a special purpose rule for PiSnd for
  nondependent functions?  No need to show that the domain types are inhabited
  in this case.} \rae{I think this is a good point to consider in the
  implementation (I think the answer may be ``yes''), but I don't think we
  really need to belabor it here.}\scw{Leaving this comment for us to think
  about. Won't write anything in the paper.}

This work also extends the reasoning about injectivity to \emph{abstract
  types}. An equality of the form
$F\ \ottnt{a_{{\mathrm{1}}}}^{\ottnt{R_{{\mathrm{1}}}}} \ldots \ottnt{a}_n^{Rn} = F\ \ottnt{b_{{\mathrm{1}}}}^{\ottnt{R_{{\mathrm{1}}}}}\ldots \ottnt{b}_n^{Rn}$ can be decomposed to
equalities between the corresponding arguments at the roles specified for
$\ottmv{F}$. For example, the \rref{E-Right}, shown below shows that we can extract an
equality between $\ottnt{b}$ and $\ottnt{b'}$ when we have an equality between
$ \ottnt{a} \  \ottnt{b} ^{ \ottnt{R} } $ and $ \ottnt{a'} \  \ottnt{b'} ^{ \ottnt{R} } $.

\begin{center}
\drule[width=8in]{E-Right}
\end{center}

The first two premises of this rule require that the equation is between two
applications, headed by the same constructor $\ottmv{F}$, which cannot be matched
to an axiom at role $\ottnt{R_{{\mathrm{2}}}}$.  The next four premises describe the types of
the components of the applications.  These premises ensure that the equality
relation is \emph{homogeneous}, i.e. that only terms of equal types are
related.

The key part of this rule is that the equality role of the conclusion is
determined by both the original role of the equality $\ottnt{R_{{\mathrm{2}}}}$ and the
annotated role of the application $\ottnt{R_{{\mathrm{1}}}}$. This is the dual of
\rref{AE-TAppCong}, the congruence rule for applications. In that rule, we can
use the fact that $\ottnt{b}$ is representationally equivalent to $\ottnt{b'}$ to show
that $ \ottnt{a} \  \ottnt{b} ^{ \ottkw{Rep} } $ is representationally equal to $ \ottnt{a'} \  \ottnt{b'} ^{ \ottkw{Rep} } $. Here, we
can invert that reasoning.


\section{Properties of System~DR}
\label{sec:metatheory}

The main result of our Coq development is the proof of type soundness for full
System~DR. Given the size of the language, the delicate interactions between
its features, and number of iterations we have gone through in its
development, we could not have done it without mechanical assistance.

This type soundness proof follows from the usual preservation and progress
lemmas. Both of these lemmas are useful properties for an intermediate
language. The preservation property holds even for open terms. Therefore, it
implies that simple, reduction-based optimizations, such as inlining, do not
produce ill-typed terms. Our proof of the progress lemma relies on showing
that a particular reduction relation is confluent, which itself provides a
simplification process for terms and (semi-decidable) algorithm for showing
them equivalent.

From the original design of FC~\cite{systemfc}, we inherit a separation between the
proofs of the preservation and progress lemmas that is unusual for
dependently-typed calculi. In this system it is possible to prove preservation
without relying on the consistency of the system. This means that preservation
holds in any context, including ones with contradictory assumptions (such as
\cd{Int ~ Bool}). As a result, GHC can apply, e.g., inlining regardless
of context.

In this section, we provide an overview of the main results of our Coq
development. However, we omit many details. Excluding automatically generated
proofs, our scripts include over 700 lemmas and 250 auxiliary definitions.

\subsection{Values and reduction}

\begin{figure}
\drules[Value]{$ \mathsf{Value}_{ \ottnt{R} }\  \ottnt{a} $}{Values}
{Star,Pi,CPi,UAbsRel,UAbsIrrel,UCAbs,Path}
\caption{Values}
\label{fig:values}
\end{figure}

%
We define the values of this language using the role-indexed relation
$ \mathsf{Value}_{ \ottnt{R} }\  \ottnt{a} $, shown in Figure~\ref{fig:values}.  Whether a term is a value
depends on the role: a newtype \cd{HTML} is a value at role $\ottkw{Nom}$ but
reduces at role $\ottkw{Rep}$. This relation depends on the auxiliary judgment
$ \mathsf{CasePath}_{ \ottnt{R} }\;  \ottnt{a}  =  \ottmv{F} $ (not shown, available in \auxiliarymaterial) which holds
when $\ottnt{a}$ is a path headed by the constant $\ottmv{F}$ that cannot reduce at
role $\ottnt{R}$. (This may be because $\ottmv{F}$ is a constant, or if the role on
$\ottmv{F}$'s definition is greater than than $\ottnt{R}$, or if $\ottmv{F}$ has not been
applied to enough arguments.)

Note that the value relation is contravariant with respect to roles.
If a term is a value at some role, it is a value at all smaller roles.
\begin{lemma}[SubRole-Value\ifsource\footnote{\url{ext_red_one.v: nsub\_Value}}\fi]
If $ \ottnt{R_{{\mathrm{2}}}}  \leq  \ottnt{R_{{\mathrm{1}}}} $ and $ \mathsf{Value}_{ \ottnt{R_{{\mathrm{1}}}} }\  \ottnt{a} $ then $ \mathsf{Value}_{ \ottnt{R_{{\mathrm{2}}}} }\  \ottnt{a} $.
\end{lemma}

Alternatively, if a term steps at some evaluation role, and we make some of
the definitions more transparent, then it will continue to step, but it could
step to a different term. This discrepancy is due to the fact that
$\beta$-reduction only applies when functions are values. Irrelevant functions
are values only when their bodies are values---so changing to a larger role
could allow an abstraction to step further.

\begin{lemma}[SubRole-Step\ifsource\footnote{\url{ext_red_one.v:sub_red_one}}\fi]
If $ \ottnt{R_{{\mathrm{1}}}}  \leq  \ottnt{R_{{\mathrm{2}}}} $ and $ \vDash   \ottnt{a}   \leadsto _{ \ottnt{R_{{\mathrm{1}}}} }  \ottnt{a'} $ then $\exists a'',  \vDash   \ottnt{a}   \leadsto _{ \ottnt{R_{{\mathrm{2}}}} }  \ottnt{a''} $.
\end{lemma}

That said, the operational semantics is deterministic at a fixed role.

\begin{lemma}[Deterministic\ifsource\footnote{\url{ext_red_one.v:reduction_in_one_deterministic}}\fi]
If $ \vDash   \ottnt{a}   \leadsto _{ \ottnt{R} }  \ottnt{a'} $ and $ \vDash   \ottnt{a}   \leadsto _{ \ottnt{R} }  \ottnt{a''} $ then $\ottnt{a'} = \ottnt{a''}$.
\end{lemma}

\subsection{Role-Checking}

The role-checking judgment $ \Omega  \vDash  \ottnt{a}  :  \ottnt{R} $ satisfies a number of important
properties. For example, we can always role-check at a larger role.

\begin{lemma}[SubRole-ing\ifsource\footnote{\url{ett_roleing.v:roleing_sub}}\fi]
If $ \Omega  \vDash  \ottnt{a}  :  \ottnt{R_{{\mathrm{1}}}} $ and $ \ottnt{R_{{\mathrm{1}}}}  \leq  \ottnt{R_{{\mathrm{2}}}} $ then $ \Omega  \vDash  \ottnt{a}  :  \ottnt{R_{{\mathrm{2}}}} $.
\end{lemma}

Furthermore, the following property says that users may always downgrade the
roles of the parameters to their abstract types.

\begin{lemma}[Role assignment narrowing\ifsource\footnote{\url{ett_roleing.v:roleing_ctx_weakening}}\fi]
If $   \Omega_{{\mathrm{1}}} , \ottmv{x} \!:\! \ottnt{R_{{\mathrm{1}}}}  , \Omega_{{\mathrm{2}}}   \vDash  \ottnt{a}  :  \ottnt{R} $ and $ \ottnt{R_{{\mathrm{2}}}}  \leq  \ottnt{R_{{\mathrm{1}}}} $ then
   $   \Omega_{{\mathrm{1}}} , \ottmv{x} \!:\! \ottnt{R_{{\mathrm{2}}}}  , \Omega_{{\mathrm{2}}}   \vDash  \ottnt{a}  :  \ottnt{R} $.
\end{lemma}

Finally, well-typed terms are always well-roled at $\ottkw{Nom}$, when all free
variables have role $\ottkw{Nom}$.

\begin{lemma}[Typing/Roleing\ifsource\footnote{\url{ett_roleing.v:Typing_roleing}}\fi]
If $\Gamma  \vDash  \ottnt{a}  \ottsym{:}  \ottnt{A}$ then $  \Gamma _{\mathsf{Nom} }   \vDash  \ottnt{a}  :  \ottkw{Nom} $, where
$ \Gamma _{\mathsf{Nom} } $ is the role-checking context that assigns $\ottkw{Nom}$ to every
term variable in the domain of $\Gamma$.
\end{lemma}

\subsection{Structural properties}

\begin{lemma}[Typing Regularity\ifsource\footnote{\url{ext_invert.v:Typing_regularity}}\fi]
If $\Gamma  \vDash  \ottnt{a}  \ottsym{:}  \ottnt{A}$ then $\Gamma  \vDash  \ottnt{A}  \ottsym{:}   \star $.
\end{lemma}

\begin{definition}[Context equality]
Define $ \vDash  \Gamma_{{\mathrm{1}}}  \equiv  \Gamma_{{\mathrm{2}}}$ with the following inductive relation:
\[
\begin{array}{c}
\drule{CE-Empty} \qquad
\drule[width=3in]{CE-ConsTm} \qquad
\drule{CE-ConsCo}
\end{array}
\]
\end{definition}

\begin{lemma}[Context Conversion\ifsource\footnote{\url{ext_invert.v:context_DefEq_typing}}\fi]
If $\Gamma_{{\mathrm{1}}}  \vDash  \ottnt{a}  \ottsym{:}  \ottnt{A}$ and $\vDash  \Gamma_{{\mathrm{1}}}  \equiv  \Gamma_{{\mathrm{2}}}$ then $\Gamma_{{\mathrm{2}}}  \vDash  \ottnt{a}  \ottsym{:}  \ottnt{A}$.
\end{lemma}

\begin{lemma}[DefEq Regularity\ifsource\footnote{\url{ext_invert.v:DefEq_regularity}}\fi]
If $ \Gamma  ;  \Delta   \vDash   \ottnt{a}   \equiv _{ \ottnt{R} }  \ottnt{b}  :  \ottnt{A} $ then $\Gamma  \vDash  \ottnt{a}  \ottsym{:}  \ottnt{A}$ and $\Gamma  \vDash  \ottnt{b}  \ottsym{:}  \ottnt{A}$.
\end{lemma}

\subsection{Preservation}

We prove the preservation lemma simultaneously with the property that one-step
reduction is contained within definitional equality. (This property is not
trivial because definitional equality only includes the primitive reductions
directly, and relies on congruence rules for the rest.) The reason that we
need to show these results simultaneously is due to our typing rule for
dependent pattern matching.

\begin{lemma}[Preservation\ifsource\footnote{\url{ext_red.v: reduction_preservation}}\fi]
If $ \vDash   \ottnt{a}   \leadsto _{ \ottnt{R} }  \ottnt{a'} $ and $\Gamma  \vDash  \ottnt{a}  \ottsym{:}  \ottnt{A}$ then
$\Gamma  \vDash  \ottnt{a'}  \ottsym{:}  \ottnt{A}$ and $ \Gamma  ;  \Delta   \vDash   \ottnt{a}   \equiv _{ \ottnt{R} }  \ottnt{a'}  :  \ottnt{A} $.
\end{lemma}


\subsection{Progress}

We prove progress by extending the proof in prior work~\cite{weirich:systemd}
with new rules for axiom reduction and case analysis. This proof, based on a
technique of Tait and Martin-L\"of, proceeds first by
developing a confluent, role-indexed, parallel reduction relation
$\Rightarrow_{\ottnt{R}}$ for terms and then
showing that equal terms must be joinable under parallel reduction~\cite{Barendregt:lambda-calculus}.
Furthermore, this relation
also tracks the roles of free variables using a role-checking context $\Omega$.

We need this role checking context because of the following substitution lemma,
necessary to show the confluence lemma below.
\begin{lemma}[Parallel reduction substitution\ifsource\footnote{\url{ett_par.v:subst1}}\fi]
If $ \Omega'  \vDash  \ottnt{a}   \Rightarrow _{ \ottnt{R_{{\mathrm{1}}}} }  \ottnt{a'} $  and $  \Omega' , \ottmv{x} \!:\! \ottnt{R_{{\mathrm{1}}}}   \vDash  \ottnt{b}  :  \ottnt{R_{{\mathrm{2}}}} $
then $ \Omega'  \vDash  \ottnt{b}  \lbrace  \ottnt{a}  \ottsym{/}  \ottmv{x}  \rbrace   \Rightarrow _{ \ottnt{R_{{\mathrm{2}}}} }  \ottnt{b}  \lbrace  \ottnt{a'}  \ottsym{/}  \ottmv{x}  \rbrace $
\end{lemma}
We know that some term $\ottnt{a}$ reduces and we want to show that we
can reconstruct that reduction after that term has been substituted into some
other term $\ottnt{b}$. However, the variable $\ottmv{x}$ could appear anywhere in
$\ottnt{b}$, perhaps as the argument to a function. As a result, the
role that we use to reduce $\ottnt{a}$ may not be the same role as the one
that we use for $\ottnt{b}$.\scw{Not sure what Pritam was confused about here, or
  how to fix it.}

The parallel reduction relation must be consistent with the role-checking
relation. Although our definition of parallel reduction is not typed (it is
independent of the type system) it maintains a strong connection to the
role-checking judgment.

\begin{lemma}[Parallel Reduction Role-Checking\ifsource\footnote{\url{ett_par.v:Par_roleing_tm_fst}\url{ett_par.v:Par_roleing_tm_snd}}\fi]
If $ \Omega  \vDash  \ottnt{a}   \Rightarrow _{ \ottnt{R} }  \ottnt{a'} $ then $ \Omega  \vDash  \ottnt{a}  :  \ottnt{R} $ and $ \Omega  \vDash  \ottnt{a'}  :  \ottnt{R} $.
\end{lemma}

This property explains why \rref{SSig-ConsAx}, which checks axiom
declarations, uses the role on the declaration to role check the right-hand side of
the axiom. In other words, type families must role check at $\ottkw{Nom}$, whereas
newtypes must role check at $\ottkw{Rep}$. We could imagine trying to weaken this
requirement and role-check all axioms at the most permissive role
$\ottkw{Rep}$. However, then the above property would not hold. We need to know
that when an axiom unfolds, the term remains well-formed at that role.

\begin{lemma}[Confluence\ifsource\footnote{\url{ext_consist.v:confluence}}\fi]
If $ \Omega  \vDash  \ottnt{a}   \Rightarrow _{ \ottnt{R} }  \ottnt{a_{{\mathrm{1}}}} $ and $ \Omega  \vDash  \ottnt{a}   \Rightarrow _{ \ottnt{R} }  \ottnt{a_{{\mathrm{2}}}} $ then there exists
some $\ottnt{b}$ such that $ \Omega  \vDash  \ottnt{a_{{\mathrm{1}}}}   \Rightarrow _{ \ottnt{R} }  \ottnt{b} $ and $ \Omega  \vDash  \ottnt{a_{{\mathrm{2}}}}   \Rightarrow _{ \ottnt{R} }  \ottnt{b} $.
\end{lemma}

The confluence proof allows us to show the usual canonical forms lemmas, which
are the key to showing the progress lemma.

\begin{lemma}[Progress\ifsource\footnote{\url{ext_consist.v:progress}}\fi]
If $\varnothing  \vDash  \ottnt{a}  \ottsym{:}  \ottnt{A}$ then either $ \mathsf{Value}_{ \ottnt{R} }\  \ottnt{a} $ or there exists some $\ottnt{a'}$ such
that $ \vDash   \ottnt{a}   \leadsto _{ \ottnt{R} }  \ottnt{a'} $.
\end{lemma}

\section{Practicalities}

\subsection{Constraint vs. Type}
\label{sec:constraint-vs-type}


Haskell differentiates the kind \cd{Constraint} from the kind \cd{Type}; the former
classifies constraints that appear to the left of an \cd{=>} in Haskell (thus,
we have \cd{Eq a :: Constraint}) while the latter classifies ordinary types, like
\cd{Int}. This separation between \cd{Constraint} and \cd{Type} is necessary for at least
two reasons: we want to disallow users from confusing these two, rejecting types such
as \cd{Int => Int} and \cd{Bool -> Eq Char}; and types in kind \cd{Constraint} have
special rules in Haskell (allowing definition only via classes and instances) to keep
them coherent. \rae{Do we need to define coherence?}

However, we do not want to
distinguish \cd{Constraint} from \cd{Type} in the core language.
Inhabitants of types of both kinds can be passed to and from functions freely, and we
also want to allow (homogeneous)
equalities between elements of \cd{Constraint} and elements of \cd{Type}.
These equalities
come up when the user defines a class with exactly one member, such as
\begin{lstlisting}
class HasDefault a where
  deflt :: a
\end{lstlisting}
Given that the evidence for a \cd{HasDefault a} instance consists only of the \cd{deflt :: a}
member, GHC compiles this class declaration into a newtype definition, producing an
axiom equating \cd{HasDefault a} with \cd{a}, at the representational role. Some packages\footnote{Notably, the \textsf{reflection} package by Edward Kmett.} rely on this encoding, and it would be disruptive to the Haskell ecosystem to
alter this arrangement.

We are thus left with a conundrum: how can we keep \cd{Constraint} distinct from \cd{Type}
in Haskell but identify them in the internal language? This situation clearly has parallels
with the need for newtypes: a newtype is distinct from its representation in Haskell
but is convertible with its representation in the internal language. We find that
we can connect \cd{Constraint} with \cd{Type} by following the same approach, but in kinds
instead of in types.

This would mean defining \cd{Constraint} along with an axiom stating that \cd{Constraint}
is representationally equal to \cd{Type}. That solves the problem: the Haskell type-checker
already knows to keep representationally equal types distinct, and all of the internal
language functionality over \cd{Type}s would now work over \cd{Constraint}s, too. Because
the internal language---System~DR---allows conversion using representational equality,
an axiom relating, say, \cd{HasDefault a :: Constraint} to \cd{a :: Type} would be
homogeneous, as required. The implementors of GHC are eager for System~DR in part
because it solves this thorny problem.\footnote{See \url{https://ghc.haskell.org/trac/ghc/ticket/11715\#comment:64} and \url{https://github.com/ghc-proposals/ghc-proposals/pull/32\#issuecomment-315082072}.}

\subsection{The Phantom Role}
\label{sec:phantom}

Prior work~\cite{breitner2016} includes a third role, the phantom role.
Consider the following newtype definition, which does not make use of its
argument.
\begin{lstlisting}
newtype F a = MkF Int
\end{lstlisting}
All values of type \cd{F a} are representationally equal, for any \cd{a}.
By giving this newtype the phantom role for its parameter,
Haskell programmers can show that \cd{F Int} is representationally equal to
\cd{F Bool} even when the \cd{MkF} constructor is not available.

It is attractive to think about the phantom role when thinking of roles as
indexing a set of equivalence relations, but that doesn't work out for
System~DR. In that interpretation, the phantom role is the coarsest relation
that identifies all terms of the same type, so it should be placed at the top
of the role lattice (above $\ottkw{Rep}$).  However, with this addition, we do not
get the desired semantics for the phantom role.\scw{still not happy with this
  part}

First, we would need a special definition for evaluating at the phantom
role. The difference between nominal and representational evaluation is
determined solely by whether axioms are transparent or opaque. However, we
cannot use this mechanism to define what it means to evaluate at the phantom
role. We would need something else entirely.

Second, arguments with phantom roles require special treatment in the
congruence rule. Logically, phantom would be above representational
in the role hierarchy, as the corresponding equivalence is coarser. However,
\rref{AE-TAppCong} uses $ \ottnt{R_{{\mathrm{1}}}} \wedge \ottnt{R} $ to equate arguments. But the minimum
of $\ottkw{Rep}$ and phantom is $\ottkw{Rep}$, not phantom, so we would need a different
congruence rule for this case.

However, the most compelling reason why we do not include phantom as a role is
because it is already \emph{derivable} using irrelevant arguments. In the
example above, we can implement the desired behavior via two levels of newtype
definition.  First, we define a constant, say $\ottmv{F'}$, with an irrelevant
argument; this is the representation of the newtype above.
\[
  \ottmv{F'} :   \star  \  \ottsym{@} \     \ \mathsf{where}\   \ottmv{F'} \  \Box ^{ \ottsym{-} }   \sim_{ \ottkw{Rep} }  \ottkw{Int} 
\]
(Note that $\ottmv{F'}$ lacks a role annotation. Only relevant arguments are
annotated with roles.) We make this newtype abstract by not exporting this axiom.

Then, we define the phantom type by wrapping the irrelevant argument with a
relevant one, which is ignored.
\[
  \ottmv{F} :   \star  \  \ottsym{@} \   \ottkw{Rep}  \ \mathsf{where}\   \ottmv{F} \  \ottmv{x} ^{ \ottkw{Rep} }   \sim_{ \ottkw{Rep} }   \ottmv{F'} \  \Box ^{ \ottsym{-} }  
\]
When we use $\ottmv{F}$ in a nominal role, we will not be able to show that
$ \ottmv{F} \  \ottkw{Int} ^{ \ottkw{Rep} } $ is equal to $ \ottmv{F} \  \ottkw{Bool} ^{ \ottkw{Rep} } $, as is the case in Haskell.
However, at the representational role, we can unfold the definition of $\ottmv{F}$
in both sides to $ \ottmv{F'} \  \Box ^{ \ottsym{-} } $, equating the two types. Furthermore, the actual
definition of the type can stay hidden, just as in the example above.

\subsection{An explicit \textsf{coerce} term}
\label{sec:no-popl11}

One simplifying idea we use in System~DR, taken from
\citet{breitner2016}, is the separation between the role-checking and
type-checking judgments. This design, overall, leads to a simpler system
because it limits the interactions between the type system and
roles. Furthermore, it is also compatible with the current implementation of
role checking in GHC.

However, one might hope for a more expressive system by combining the
role-checking and type-checking judgments together, as was done in the system
of \citet{weirich:newtypes}. In fact, this was our first approach to this work,
primarily because we wanted to explore a design that factored conversion
into implicit and explicit parts.
\[
\drule{AR-Conv}\qquad\ottdruleARXXCast{}
\]
In the conversion rule on the left, the role on the \emph{typing} judgment
(indexing the typing judgment by a role is new here) determines the
equality that can be used. If this role is $\ottkw{Nom}$, then only nominal
equality is permitted and coercing between representationally equal types
requires an explicit use of coercion, via the rule on the
right. Alternatively, if the role is $\ottkw{Rep}$ then all type conversions are
allowed (and using the $\ottkw{coerce}$ primitive is unnecessary).

This system is attractive because it resembles the design of source
Haskell. In contrast, in the current System~DR, if an expression has type
\cd{HTML}, then it also has type \cd{String}---precisely the situation
newtypes were meant to avoid. We return to the question of what \cd{coerce}
means for source Haskell in the next subsection.

However, after struggling with various designs of the system for some time, we
ultimately abandoned this approach. In particular, we were unhappy with a
number of aspects of the design.
\begin{itemize}
\item How should $ \ottkw{coerce} \  \ottnt{a} $ reduce at role $\ottkw{Nom}$? It cannot
  reduce to $\ottnt{a}$: that would violate type preservation.
  The solution to
  this problem is ``push'' rules, as in System FC~\cite{systemfc}.  These
  complicate the semantics by moving uses of $\ottkw{coerce}$ when they block
  normal reduction. For example, if we have \cd{(coerce (\\x -> x)) 5}, we
  cannot use our normal rule for $\beta$-reduction, as the $\ottkw{coerce}$
  intervenes between the $\lambda$-expression and its argument. Instead, a push
  rule is required to reduce the term to \cd{(\\x -> coerce x) (coerce 5)},
  allowing $\beta$-reduction to proceed.  However, these push rules are
  complex and the complexity increases with each feature added to the language;
  see \citet[Section~5]{weirich:dwk} for a telling example of
  how bad they can be.

\item Push rules prevent $\ottkw{coerce}$ from creating stuck terms, but they are
  not the only evaluation rules for $\ottkw{coerce}$ that we could want.  In
  particular, we would like the operational semantics to eliminate
  \emph{degenerate} coercions, which step $ \ottkw{coerce} \  \ottnt{a} $ to $\ottnt{a}$ in the
  case when the coercion does not actually change the type of $\ottnt{a}$.
  However, this sort of reduction rule would be type-directed: it would apply
  only when the two types involved are definitionally equal. Such an
  operational behavior is at odds with our Curry-style approach and would
  complicate our treatment of irrelevance.

\item Because we are in a dependent setting, we must also consider the impact
  of $\ottkw{coerce}$ on the equality relation. For example, what is the
  relationship between $ \ottkw{coerce} \  \ottnt{a} $ and $\ottnt{a}$? Are they nominally equal?
  Are they representationally equal? In our explorations of the possibilities,
  none worked out well. 
  (See also \citet[Section 5]{overabundance-of-equalities}.)

\end{itemize}

We also hoped that working with the combined role-/type-checking system would
lead to greater expressiveness in other parts of the language. In particular,
the current treatment of roles in GHC was believed to be incompatible with putting
the function \cd{join :: Monad m => m (m a) -> m a} in the \cd{Monad} type-class.\footnote{See \url{https://ghc.haskell.org/trac/ghc/ticket/9123}, which was originally titled ``Need for higher kinded roles''.} 
However, the combined role-/type-checking does not help with this problem.
Fortunately,
the new \cd{QuantifiedConstraints} extension~\cite{quantified-class-constraints},
available in GHC 8.6, provides a new
solution,\footnote{\url{https://ryanglscott.github.io/2018/03/04/how-quantifiedconstraints-can-let-us-put-join-back-in-monad/}} resolving the problem in a much less invasive way.

\subsection{What is Source Haskell?}
\label{sec:src-language}
\label{sec:src-haskell}

As described above, System~DR fails to give a direct semantics for the
$\ottkw{coerce}$ primitive in Haskell. (This is not an issue specific to
System~DR; no prior work does this~\cite{breitner2016,weirich:newtypes}.)
However, all is not lost.  We propose instead that it is better to understand
the $\ottkw{coerce}$ term in the Haskell source language through an
\emph{elaboration semantics}.

More concretely, we can imagine a specification for source Haskell where
source terms can automatically convert
types with nominal equalities and $\ottkw{coerce}$ is needed for representational
equalities.

\[ \drule[width=5in]{S-Conv} \qquad \drule[width=5in]{S-Coerce} \]

However, this language would not have its own small-step semantics (which
would require ``push rules'' as described above). Instead, we would understand its
semantics directly through translation to System~DR. In other words, we would
specify a relation $\Gamma  \vDash_{\mathsf{src} }  \ottnt{a}  \leadsto  \ottnt{a'}  \ottsym{:}  \ottnt{A}$ that translates source Haskell
terms $\ottnt{a}$ to System~DR terms $\ottnt{a'}$. The key step of this translation is
that it compiles away all uses of the $\ottkw{coerce}$ term.
\[ \drule[width=5in]{ST-Coerce} \]
Because we know that System~DR is type-sound, we would automatically get a
type soundness property for well-typed source terms. 

\scw{state more positively}
The drawback of this approach is that without an operational semantics, the
source language type system would fall back to System~DR for conversion,
as in the rules above. Any terms that appear in types would be checked
according to System~DR instead of with the source language rules. However,
this discrepancy would affect only the most advanced Haskell programmers:
those that use \cd{coerce} and dependent types together. In the absence of the use
of either feature, the systems would coincide.


\section{Future work}
\label{sec:future}
\label{sec:extended-pattern-matching}

Our work in this paper lays out a consistent point in the design space of
interactions between roles and dependent types.
However, we plan to continue refining our definitions and extending our system
to enhance its expressiveness. Our efforts will be along the following lines:

\paragraph{Annotated language}
We adopted a Curry-style language design in this paper, where the syntax of
terms includes only computationally relevant terms. This style of language
makes a lot of sense for reasoning about representational
equivalence~\cite{stump2018}. It also simplifies our design process as we
need not worry about propagating annotations at the same time as developing
the semantics.

However, this language, taken at face value, cannot be implemented:
type-checking is undecidable and non-deterministic. In order to extend GHC's
core language along the lines of System~DR, we will need to design a system of
annotations that will make type checking simple and syntax-directed, much like
System~DC~\cite{weirich:systemd}.  Given prior work (including all the work on
System~FC~\cite{systemfc,breitner2016}) as exemplars, this task should be
straightforward; the most challenging parts are choosing a system of
annotations that is easy for GHC to manipulate, and showing that the annotated
language satisfies properties that are useful for the implementation, such as
substitution and preservation.



\paragraph{Other roles}
Our focus in this work has been on the roles $\ottkw{Rep}$ and $\ottkw{Nom}$.
However, the rules are generic enough to support
arbitrary roles in between these two extremes.  Furthermore, perhaps $\ottkw{Nom}$
is not the right role at the bottom of the lattice---it still allows type
families to unfold, for example. Indeed, GHC internally implements an even
finer equality---termed \cd{pickyEqType}---that does not even unfold type
synonyms. Today, this equality is used only to control error message printing, but perhaps
giving users more control over unfolding would open new doors.


\paragraph{Closed types}
The rules of System~DR do not check for the exhaustiveness of case analysis.
Instead, every pattern match must have a ``wildcard'' case to provide an
option when the scrutinee does not match (see \rref{Beta-PatternFalse}).  This
design is appropriate when all types are open and extensible through the use
of declarations in the signature. It ensures that if a term type checks it
will also type check in any extended signature (a necessary property for
separate compilation).  However, we would like to investigate an extension of
the system with \emph{closed types}, those that do not allow the definition of
new constants or axioms in the signature. The types could then be the basis
for a form of case analysis that does not require the wildcard case and can be
shown to be exhaustive.

\paragraph{Extended pattern matching}
Although System~DR allows the definition of \emph{parameterized} datatypes
through the use of constants in the signature, the case analysis term is not
expressive enough to allow pattern matching for these datatypes. For example,
we cannot use this term with the type $ \ottkw{Maybe} \  \ottkw{Int} $.  We omit this
capability only to constrain the scope of this work---our rules for pattern
matching are complex enough already and are targeted to scrutinees of type
$ \star $. However, we would like to extend this system with this ability in
future work and do not foresee any difficulties.


\section{Related work}
\label{sec:related}

\paragraph{Roles}

We have reviewed the connection between this work and its two main
antecedents~\cite{breitner2016,weirich:systemd} throughout. Furthermore,
Sections~\ref{sec:design-constraints} and~\ref{sec:no-popl11} compare with
the approach taken in \citet{weirich:newtypes}.

\citet{overabundance-of-equalities} discusses a way to integrate roles with a
system having some features of dependent types, though that work does not
present a dependently typed system. Instead, the type system under study is an
extension of the non-dependent System~FC used in other prior work on Haskell
(and introduced originally by \citet{systemfc}), but with merged types and
kinds~\cite{weirich:dwk}. The main challenge that work faces is with the
fact that the system described is \emph{heterogeneous}, in that equalities can
relate types of different kinds. By contrast, our work here studies only
\emph{homogeneous} equality, removing much of the complication Eisenberg
discusses. In addition, our work here studies an implicit calculus, allowing
us to simplify the presentation even further over the explicit calculus (with
decidable type-checking) of Eisenberg.

A recent client for roles and \cd{coerce} is \citet{winant2018}.
This work employs roles in a critical way to build a mechanism for explicit
instantiation and manipulation of type-class dictionaries without imperiling
class coherence.

\paragraph{Modal Dependent Type Theory}
\label{sec:modal}

Like this work, \citet{Pfenning01} presents a dependent type theory that
simultaneously supports two different forms of equality in conjunction with
irrelevance. However, Pfenning's equality relations differ from ours (his
system includes $\alpha$-equivalence and $\alpha\beta\eta$-equivalence), and
the type system is a conservative extension of LF~\cite{Harper:1993}.
Furthermore, Pfenning uses a modal typing discipline to internalize these
different concepts.

More recently, modal type theories have refined our understanding of
parametric, erasable and irrelevant arguments in dependent type
theories~\cite{Mishra-LingerS08,Abel12,Abel17,Nuyts17}.\footnote{Note that the
  distinctions between parametric, erasable and irrelevant quantification are
  not present in System DR. Our work does not cover parametricity, so cannot
  determine whether functions take related inputs to related
  outputs. Furthermore, because our equality judgment is not type directed,
  our system does not need to distinguish between irrelevant~\cite{Abel12} and
  shape-irrelevant~\cite{Abel17} quantification.}  In particular,
\citet{Nuyts18} presents a unified framework that includes these and other
modalities.  It is possible that roles could also be understood as a new
modality in this sense, and, once this is accomplished, enjoy a unified
treatment with irrelevant quantification.

A starting point for the first task would be a language that includes roles
(as a modality) annotated on $\Pi$-types and on variables in the context. This
design could draw on the type system of \citet{weirich:newtypes}, extended
with dependent types but without the explicit $\ottkw{coerce}$ term described in
Section~\ref{sec:no-popl11}. This framework would then avoid some of the
approximations made by System DR caused by the lack of role information on
$\Pi$-types (noted in the descriptions of
\rref{AE-AppCong,Srole-a-App,E-PiSnd}) and the separation of type- and role-
checking (noted in the description of the case expression).

The uniform system of \citet{Nuyts18} could form the basis of the second task.
More specifically, this system includes structures that can resolve the
differences in the treatment of co-domain of $\Pi$ types between irrelevant
and roled arguments. Although irrelevant arguments cannot appear in lambda
terms, they can freely appear in the result type of the lambda term.  In
contrast, roled arguments must appear according to their indicated role
everywhere. Furthermore, \citet{Nuyts18} includes a subsumption relationship
between modalities that could generalize sub-roleing.

However, even with the addition of role-annotated $\Pi$-types, there would be
differences in the treatment of roles in System DR and in the general
treatment of modalities in \citet{Nuyts18} that would need to be resolved.
First, the general system would need to be extended by a role-indexed
operational semantics; roles control the unfolding of axioms and must be
considered in any reduction relation.  Second, the application congruence rule
(\rref{AE-TAppCong}) uses the $ \ottnt{R} \wedge \ottnt{R'} $ operator when comparing the
arguments; instead, a modal system would only use the role of argument $\ottnt{R}$
and would be independent of $\ottnt{R'}$, the role of the context. System DR uses
this operator to make nominal equality line up with Haskell's type
equality. It is not clear how the general framework can accommodate this
behavior. 

Overall, it does not make sense in GHC to combine the treatment of irrelevant
quantification and roles. Irrelevant quantification is the basis for the
implementation of parametric polymorphism in System DR. Parametric
polymorphism is used everywhere, even by novice Haskellers, so it cannot be
restricted---so the system must mark relevance on $\Pi$ types. On the other
hand, roles are used to implement the safe-coercions, an advanced feature of
GHC, and are subject to the design considerations described in
Section~\ref{sec:design-constraints}.

\av{These two paragraphs feel to me like they lack punch, but I'm having
trouble coming up with suggestions. I'll be taking some time to think about
them more.}
\av{So, more thoughts:
\begin{itemize}
  \item I don't think we really make the point that their proof works only for
    logically sound languages, and that this would require another proof for
    this hypothetical language
  \item Also, they have no operational semantics \emph{at all} in this paper
    (which I agree is implied by the current sentence about opsem, but it make
    it sound like they already have one, just without role annotation)
  \item The sentence about AE-TAppCong seems to me like it leaves things open --
    yes, we are currently doing things this way, but would it be a problem to it
    \`a la modal TT instead ($\ottnt{R}$ instead of $ \ottnt{R} \wedge \ottnt{R'} $)?
  \item The last sentence ("at the nominal role...") seems to be a big objection
    to using a modal system, actually. I believe this is very much \emph{not} in
    line with modal theories, where the arguments are compared at their declared
    role, not "forced" at the top one. Is this correct? Because, if it is, then
    it implies that Haskell's (type) equality can not be properly represented by
    a modal system. I'm not sure about this, but I'd love to hear other opinions,
    as this seems to me like a potential big deal.
  \item The second paragraph makes sense, but I'm not sure it totally rules out
    the possibility of using (say) Nominal as a default role for everything, and
    only inferring different roles whenever necessary (use of coerce, explicit roling
    by the programmer, etc). It would certainly require a bigger implementation
    effort. However, I wonder if it is possible to avoid the other
    potential downsides, by hiding the default roles in types and kinds as well as in the
    parts of the language specification that don't (really) involve roles. My
    understanding is that this is essentially what the system does already ---
    we default everything to Nom, unless specified otherwise. This would
    mean having a \emph{full} type system where roles are specified everywhere (and
    which would be the one used by the compiler), and a
    \emph{surface} one that would look a lot like the current system (roles specified
    only as needed; this would be the system presented to users, most of the time).
    Internally, the compiler could have a \texttt{default} role
    that would be equal to \texttt{nom}, except that it would also record that the role
    was given by default --- and thus the compiler/ide would avoid printing it
    in error messages, etc. I'm sure this has been debated already, is that a
    real possibility, or are there problems with that approach?
  \end{itemize}}

\scw{You are wright about the ``at the nominal role'' sentence. I've tried to
  rephrase that part so that the issue comes out more forcefully.}
\scw{WRT the last paragraph. What you say makes sense. However, it is useful
  for GHC to not have too much separation between the source language type
  system and that of the intermediate language. (In particular, they share the
  same AST for types!)  Not sure how to say this though. }

\paragraph{Conversions}

Recent work by \citet{stump2018} studies the possibility of zero-cost
conversions in a Curry-style, dependently typed language. However, that work
allows these conversions only when two types differ in their \emph{irrelevant}
parameters only. In effect, their work supports nominal equality and phantom
equality, but not representational equality. Their work does not consider
roles. The main application of their work is code re-use, a challenging and
painful problem in the domain of dependent types, but one we do not explore
here.

The work of \citet{BernardyM13} is similar, and considers color-directed
erasure (where arguments of types can be colored, making them optionally
irrelevant). In turn, the use of colors allows the generation of coercions for
the corresponding arguments. This work provides a mechanism for distinguishing
between different sorts of phantom equalities, thanks to the ability to
combine several colors and perform selective erasure of arguments.

\scw{We need to revise this paragraph. reviewers were not convinced by the
  comparison. (Or asking for Observational Type Theory instead.)}
Our work is tangentially connected to the recent study of \emph{cubical type
  theory}~\cite{cohen2016}. In cubical type theory, equality or isomorphism of
types is described as a path from one to the other.
These equalities can be lifted through type definitions, similar
to how roles permit the lifting of equalities through (some)
datatypes. However, we are proceeding in opposite directions: roles are about
describing \emph{zero-cost} conversions, whereas cubical type theory revolves
around computationally relevant conversions. 

Like this work, 2-level type
theory~\cite{AltenkirchCK16,capriotti:thesis,voevodsky2013HTS} is a variant of
Martin-L\"of type theory with two different equality types. In this case, the
``outer'' equality contains a strict equality type former, with unique
identity proofs, while the ``inner'' one is some version of Homotopy Type
Theory (HoTT). As here, this work reconciles two conflicting definitions of
equality in the same system. However, the specific notions of equality are
radically different from the ones considered here.

\av{I'm tempted to explicitly "unfold" by saying that we make equality more
  syntactic/fine-grained while HoTT enforces the equality (and congruence) of
  more things (needs to be said better, of course). Thoughts on whether this
  is useful or redundant?}\scw{You could also argue that we're making equality
  more coarse by going to representational equality.}

\av{Also, we mention CTT and not HoTT, shouldn't mention HoTT too? The
  argument to me is really about an \emph{implementation} of HoTT, not CTT
  specifically. I'm tempted to say something like "HoTT and its concrete
  models like CTT", thoughts?}  \scw{I'm ok with only comparing with the
  computational versions of HoTT. We are talking about programming after all.}

\av{Note: we should mention the 2-level type theory/HTS here. They have 2
  equalities, the definitional one and the internal, homotopical one. They
  don't directly compare in terms of the meaning of the equalities (though I
  guess definition is our nominal, ish?), but the stratification approach is
  similar}\scw{I added a paragraph. check it out.}

\paragraph{Typecase}

Our $\mathsf{case}$ expression implements a form of \cd{typecase} or
intensional analysis of types~\cite{morrisett:lmli}. In this setting, type
discrimination is available both in types (as in Haskell today) and at
run-time. We do not attempt to summarize the immense literature related to
run-time type analysis here.
\citet{crary:lx} consider a similar problem as we do here: the
sound preservation of typecase through translation.

\section{Conclusion}

In this work, we provide a solid foundation for two popular extensions of
Haskell, dependent types and safe coercions. We have worked out how these
features can combine soundly in the same system and have used a proof
assistant to verify that we have not missed any subtle
interactions. Translating roles into a new framework (dependent type theory,
with irrelevance and coercion abstraction) has given us new insight into their
power and limitations. Our focus has been Haskell, but we believe that this
work is important for more compilers than GHC.

\begin{acks}
  This material is based upon work supported by the
  \grantsponsor{GS100000001}{National Science
    Foundation}{http://dx.doi.org/10.13039/100000001} under Grant
  No.~\grantnum{GS100000001}{1319880} and Grant
  No.~\grantnum{GS100000001}{1521539}.  Any opinions, findings, and
  conclusions or recommendations expressed in this material are those
  of the author and do not necessarily reflect the views of the
  National Science Foundation.
\end{acks}

\newpage
\bibliography{weirich,rae,proposal}


\begin{thebibliography}{46}


\ifx \showCODEN    \undefined \def \showCODEN     #1{\unskip}     \fi
\ifx \showDOI      \undefined \def \showDOI       #1{#1}\fi
\ifx \showISBNx    \undefined \def \showISBNx     #1{\unskip}     \fi
\ifx \showISBNxiii \undefined \def \showISBNxiii  #1{\unskip}     \fi
\ifx \showISSN     \undefined \def \showISSN      #1{\unskip}     \fi
\ifx \showLCCN     \undefined \def \showLCCN      #1{\unskip}     \fi
\ifx \shownote     \undefined \def \shownote      #1{#1}          \fi
\ifx \showarticletitle \undefined \def \showarticletitle #1{#1}   \fi
\ifx \showURL      \undefined \def \showURL       {\relax}        \fi
\providecommand\bibfield[2]{#2}
\providecommand\bibinfo[2]{#2}
\providecommand\natexlab[1]{#1}
\providecommand\showeprint[2][]{arXiv:#2}

\bibitem[\protect\citeauthoryear{Abel and Scherer}{Abel and Scherer}{2012}]%
        {Abel12}
\bibfield{author}{\bibinfo{person}{Andreas Abel} {and} \bibinfo{person}{Gabriel
  Scherer}.} \bibinfo{year}{2012}\natexlab{}.
\newblock \showarticletitle{On Irrelevance and Algorithmic Equality in
  Predicative Type Theory}.
\newblock \bibinfo{journal}{\emph{Logical Methods in Computer Science}}
  \bibinfo{volume}{8}, \bibinfo{number}{1} (\bibinfo{year}{2012}).
\newblock
\urldef\tempurl%
\url{https://doi.org/10.2168/LMCS-8(1:29)2012}
\showDOI{\tempurl}


\bibitem[\protect\citeauthoryear{Abel, Vezzosi, and Winterhalter}{Abel
  et~al\mbox{.}}{2017}]%
        {Abel17}
\bibfield{author}{\bibinfo{person}{Andreas Abel}, \bibinfo{person}{Andrea
  Vezzosi}, {and} \bibinfo{person}{Th{\'{e}}o Winterhalter}.}
  \bibinfo{year}{2017}\natexlab{}.
\newblock \showarticletitle{Normalization by evaluation for sized dependent
  types}.
\newblock \bibinfo{journal}{\emph{{PACMPL}}} \bibinfo{volume}{1},
  \bibinfo{number}{{ICFP}} (\bibinfo{year}{2017}),
  \bibinfo{pages}{33:1--33:30}.
\newblock
\urldef\tempurl%
\url{https://doi.org/10.1145/3110277}
\showDOI{\tempurl}


\bibitem[\protect\citeauthoryear{Altenkirch, Capriotti, and Kraus}{Altenkirch
  et~al\mbox{.}}{2016}]%
        {AltenkirchCK16}
\bibfield{author}{\bibinfo{person}{Thorsten Altenkirch}, \bibinfo{person}{Paolo
  Capriotti}, {and} \bibinfo{person}{Nicolai Kraus}.}
  \bibinfo{year}{2016}\natexlab{}.
\newblock \showarticletitle{Extending Homotopy Type Theory with Strict
  Equality}. In \bibinfo{booktitle}{\emph{25th {EACSL} Annual Conference on
  Computer Science Logic, {CSL} 2016, August 29 - September 1, 2016, Marseille,
  France}} \emph{(\bibinfo{series}{LIPIcs})},
  \bibfield{editor}{\bibinfo{person}{Jean{-}Marc Talbot} {and}
  \bibinfo{person}{Laurent Regnier}} (Eds.), Vol.~\bibinfo{volume}{62}.
  \bibinfo{publisher}{Schloss Dagstuhl - Leibniz-Zentrum fuer Informatik},
  \bibinfo{pages}{21:1--21:17}.
\newblock
\urldef\tempurl%
\url{https://doi.org/10.4230/LIPIcs.CSL.2016.21}
\showDOI{\tempurl}


\bibitem[\protect\citeauthoryear{Aydemir, Chargu\'{e}raud, Pierce, Pollack, and
  Weirich}{Aydemir et~al\mbox{.}}{2008}]%
        {aydemir:popl-binders}
\bibfield{author}{\bibinfo{person}{Brian Aydemir}, \bibinfo{person}{Arthur
  Chargu\'{e}raud}, \bibinfo{person}{Benjamin~C. Pierce},
  \bibinfo{person}{Randy Pollack}, {and} \bibinfo{person}{Stephanie Weirich}.}
  \bibinfo{year}{2008}\natexlab{}.
\newblock \showarticletitle{Engineering Formal Metatheory}. In
  \bibinfo{booktitle}{\emph{{ACM} {SIGPLAN}-{SIGACT} Symposium on Principles of
  Programming Languages (POPL)}}. \bibinfo{pages}{3--15}.
\newblock


\bibitem[\protect\citeauthoryear{Aydemir and Weirich}{Aydemir and
  Weirich}{2010}]%
        {aydemir:lngen}
\bibfield{author}{\bibinfo{person}{Brian Aydemir} {and}
  \bibinfo{person}{Stephanie Weirich}.} \bibinfo{year}{2010}\natexlab{}.
\newblock \bibinfo{booktitle}{\emph{LNgen: Tool Support for Locally Nameless
  Representations}}.
\newblock \bibinfo{type}{{T}echnical {R}eport} MS-CIS-10-24.
  \bibinfo{institution}{Computer and Information Science, University of
  Pennsylvania}.
\newblock


\bibitem[\protect\citeauthoryear{Barendregt}{Barendregt}{1991}]%
        {barendregt-lambda-cube}
\bibfield{author}{\bibinfo{person}{Henk Barendregt}.}
  \bibinfo{year}{1991}\natexlab{}.
\newblock \showarticletitle{Introduction to generalized type systems}.
\newblock \bibinfo{journal}{\emph{J. Funct. Program.}} \bibinfo{volume}{1},
  \bibinfo{number}{2} (\bibinfo{year}{1991}), \bibinfo{pages}{125--154}.
\newblock


\bibitem[\protect\citeauthoryear{Barendregt}{Barendregt}{1984}]%
        {Barendregt:lambda-calculus}
\bibfield{author}{\bibinfo{person}{H.~P. Barendregt}.}
  \bibinfo{year}{1984}\natexlab{}.
\newblock \bibinfo{booktitle}{\emph{The Lambda Calculus: Its Syntax and
  Semantics}}.
\newblock \bibinfo{publisher}{Elsevier}.
\newblock


\bibitem[\protect\citeauthoryear{Barras and Bernardo}{Barras and
  Bernardo}{2008}]%
        {barras08}
\bibfield{author}{\bibinfo{person}{Bruno Barras} {and} \bibinfo{person}{Bruno
  Bernardo}.} \bibinfo{year}{2008}\natexlab{}.
\newblock \showarticletitle{The Implicit Calculus of Constructions as a
  Programming Language with Dependent Types}. In
  \bibinfo{booktitle}{\emph{Foundations of Software Science and Computational
  Structures}}, \bibfield{editor}{\bibinfo{person}{Roberto Amadio}} (Ed.).
  \bibinfo{publisher}{Springer Berlin Heidelberg}, \bibinfo{address}{Berlin,
  Heidelberg}, \bibinfo{pages}{365--379}.
\newblock
\showISBNx{978-3-540-78499-9}


\bibitem[\protect\citeauthoryear{Bernardy and Moulin}{Bernardy and
  Moulin}{2013}]%
        {BernardyM13}
\bibfield{author}{\bibinfo{person}{Jean{-}Philippe Bernardy} {and}
  \bibinfo{person}{Guilhem Moulin}.} \bibinfo{year}{2013}\natexlab{}.
\newblock \showarticletitle{Type-theory in color}. In
  \bibinfo{booktitle}{\emph{{ACM} {SIGPLAN} International Conference on
  Functional Programming, ICFP'13, Boston, MA, {USA} - September 25 - 27,
  2013}}, \bibfield{editor}{\bibinfo{person}{Greg Morrisett} {and}
  \bibinfo{person}{Tarmo Uustalu}} (Eds.). \bibinfo{publisher}{{ACM}},
  \bibinfo{pages}{61--72}.
\newblock
\urldef\tempurl%
\url{https://doi.org/10.1145/2500365.2500577}
\showDOI{\tempurl}


\bibitem[\protect\citeauthoryear{Bl{\"{o}}ndal, L{\"{o}}h, and
  Scott}{Bl{\"{o}}ndal et~al\mbox{.}}{2018}]%
        {deriving-via}
\bibfield{author}{\bibinfo{person}{Baldur Bl{\"{o}}ndal},
  \bibinfo{person}{Andres L{\"{o}}h}, {and} \bibinfo{person}{Ryan Scott}.}
  \bibinfo{year}{2018}\natexlab{}.
\newblock \showarticletitle{Deriving via: or, how to turn hand-written
  instances into an anti-pattern}. In \bibinfo{booktitle}{\emph{Proceedings of
  the 11th {ACM} {SIGPLAN} International Symposium on Haskell, Haskell@ICFP
  2018, St. Louis, MO, USA, September 27-17, 2018}},
  \bibfield{editor}{\bibinfo{person}{Nicolas Wu}} (Ed.).
  \bibinfo{publisher}{{ACM}}, \bibinfo{pages}{55--67}.
\newblock
\urldef\tempurl%
\url{https://doi.org/10.1145/3242744.3242746}
\showDOI{\tempurl}


\bibitem[\protect\citeauthoryear{Bottu, Karachalias, Schrijvers, Oliveira, and
  Wadler}{Bottu et~al\mbox{.}}{2017}]%
        {quantified-class-constraints}
\bibfield{author}{\bibinfo{person}{Gert-Jan Bottu}, \bibinfo{person}{Georgios
  Karachalias}, \bibinfo{person}{Tom Schrijvers}, \bibinfo{person}{Bruno C.
  d.~S. Oliveira}, {and} \bibinfo{person}{Philip Wadler}.}
  \bibinfo{year}{2017}\natexlab{}.
\newblock \showarticletitle{Quantified Class Constraints}. In
  \bibinfo{booktitle}{\emph{Proceedings of the 10th ACM SIGPLAN International
  Symposium on Haskell}} \emph{(\bibinfo{series}{Haskell 2017})}.
  \bibinfo{publisher}{ACM}, \bibinfo{address}{New York, NY, USA},
  \bibinfo{pages}{148--161}.
\newblock
\showISBNx{978-1-4503-5182-9}
\urldef\tempurl%
\url{https://doi.org/10.1145/3122955.3122967}
\showDOI{\tempurl}


\bibitem[\protect\citeauthoryear{Brady}{Brady}{2013}]%
        {idris}
\bibfield{author}{\bibinfo{person}{Edwin Brady}.}
  \bibinfo{year}{2013}\natexlab{}.
\newblock \showarticletitle{Idris, a general-purpose dependently typed
  programming language: {D}esign and implementation}.
\newblock \bibinfo{journal}{\emph{J.~Funct.~Prog.}}  \bibinfo{volume}{23}
  (\bibinfo{year}{2013}).
\newblock


\bibitem[\protect\citeauthoryear{Breitner, Eisenberg, Jones, and
  Weirich}{Breitner et~al\mbox{.}}{2016}]%
        {breitner2016}
\bibfield{author}{\bibinfo{person}{Joachim Breitner},
  \bibinfo{person}{Richard~A. Eisenberg}, \bibinfo{person}{Simon~Peyton Jones},
  {and} \bibinfo{person}{Stephanie Weirich}.} \bibinfo{year}{2016}\natexlab{}.
\newblock \showarticletitle{Safe zero-cost coercions for {Haskell}}.
\newblock \bibinfo{journal}{\emph{Journal of Functional Programming}}
  \bibinfo{volume}{26} (\bibinfo{year}{2016}).
\newblock
\urldef\tempurl%
\url{https://doi.org/10.1017/S0956796816000150}
\showDOI{\tempurl}


\bibitem[\protect\citeauthoryear{Capriotti}{Capriotti}{2016}]%
        {capriotti:thesis}
\bibfield{author}{\bibinfo{person}{Paolo Capriotti}.}
  \bibinfo{year}{2016}\natexlab{}.
\newblock \emph{\bibinfo{title}{Models of type theory with strict equality}}.
\newblock \bibinfo{thesistype}{Ph.D. Dissertation}. \bibinfo{school}{School of
  Computer Science, University of Nottingham}, \bibinfo{address}{Nottingham,
  UK}.
\newblock
\urldef\tempurl%
\url{http://arxiv.org/abs/1702.04912}
\showURL{%
\tempurl}


\bibitem[\protect\citeauthoryear{Cardelli}{Cardelli}{1986}]%
        {cardelli86}
\bibfield{author}{\bibinfo{person}{Luca Cardelli}.}
  \bibinfo{year}{1986}\natexlab{}.
\newblock \bibinfo{booktitle}{\emph{A Polymorphic Lambda Calculus with
  Type:Type}}.
\newblock \bibinfo{type}{{T}echnical {R}eport}~10.
  \bibinfo{institution}{Digital Equipment Corporation, SRC}.
\newblock


\bibitem[\protect\citeauthoryear{Chakravarty, Keller, and {Peyton
  Jones}}{Chakravarty et~al\mbox{.}}{2005}]%
        {DBLP:conf/icfp/ChakravartyKJ05}
\bibfield{author}{\bibinfo{person}{Manuel M.~T. Chakravarty},
  \bibinfo{person}{Gabriele Keller}, {and} \bibinfo{person}{Simon~L. {Peyton
  Jones}}.} \bibinfo{year}{2005}\natexlab{}.
\newblock \showarticletitle{Associated type synonyms}. In
  \bibinfo{booktitle}{\emph{Proceedings of the 10th {ACM} {SIGPLAN}
  International Conference on Functional Programming, {ICFP} 2005, Tallinn,
  Estonia, September 26-28, 2005}}, \bibfield{editor}{\bibinfo{person}{Olivier
  Danvy} {and} \bibinfo{person}{Benjamin~C. Pierce}} (Eds.).
  \bibinfo{publisher}{{ACM}}, \bibinfo{pages}{241--253}.
\newblock
\urldef\tempurl%
\url{https://doi.org/10.1145/1086365.1086397}
\showDOI{\tempurl}


\bibitem[\protect\citeauthoryear{Cohen, Coquand, Huber, and M{\"o}rtberg}{Cohen
  et~al\mbox{.}}{2018}]%
        {cohen2016}
\bibfield{author}{\bibinfo{person}{Cyril Cohen}, \bibinfo{person}{Thierry
  Coquand}, \bibinfo{person}{Simon Huber}, {and} \bibinfo{person}{Anders
  M{\"o}rtberg}.} \bibinfo{year}{2018}\natexlab{}.
\newblock \showarticletitle{{Cubical Type Theory: A Constructive Interpretation
  of the Univalence Axiom}}. In \bibinfo{booktitle}{\emph{21st International
  Conference on Types for Proofs and Programs (TYPES 2015)}}
  \emph{(\bibinfo{series}{Leibniz International Proceedings in Informatics
  (LIPIcs)})}, \bibfield{editor}{\bibinfo{person}{Tarmo Uustalu}} (Ed.),
  Vol.~\bibinfo{volume}{69}. \bibinfo{publisher}{Schloss
  Dagstuhl--Leibniz-Zentrum fuer Informatik}, \bibinfo{address}{Dagstuhl,
  Germany}, \bibinfo{pages}{5:1--5:34}.
\newblock
\showISBNx{978-3-95977-030-9}
\showISSN{1868-8969}
\urldef\tempurl%
\url{https://doi.org/10.4230/LIPIcs.TYPES.2015.5}
\showDOI{\tempurl}


\bibitem[\protect\citeauthoryear{Crary and Weirich}{Crary and Weirich}{1999}]%
        {crary:lx}
\bibfield{author}{\bibinfo{person}{Karl Crary} {and} \bibinfo{person}{Stephanie
  Weirich}.} \bibinfo{year}{1999}\natexlab{}.
\newblock \showarticletitle{Flexible Type Analysis}. In
  \bibinfo{booktitle}{\emph{Proceedings of the fourth {ACM SIGPLAN}
  International Conference on Functional Programming (ICFP)}}.
  \bibinfo{address}{Paris, France}, \bibinfo{pages}{233--248}.
\newblock


\bibitem[\protect\citeauthoryear{Devriese and Piessens}{Devriese and
  Piessens}{2011}]%
        {agda-classes}
\bibfield{author}{\bibinfo{person}{Dominique Devriese} {and}
  \bibinfo{person}{Frank Piessens}.} \bibinfo{year}{2011}\natexlab{}.
\newblock \showarticletitle{On the bright side of type classes: instance
  arguments in Agda}. In \bibinfo{booktitle}{\emph{Proceeding of the 16th {ACM}
  {SIGPLAN} international conference on Functional Programming, {ICFP} 2011,
  Tokyo, Japan, September 19-21, 2011}},
  \bibfield{editor}{\bibinfo{person}{Manuel M.~T. Chakravarty},
  \bibinfo{person}{Zhenjiang Hu}, {and} \bibinfo{person}{Olivier Danvy}}
  (Eds.). \bibinfo{publisher}{{ACM}}, \bibinfo{pages}{143--155}.
\newblock
\urldef\tempurl%
\url{https://doi.org/10.1145/2034773.2034796}
\showDOI{\tempurl}


\bibitem[\protect\citeauthoryear{Diehl, Firsov, and Stump}{Diehl
  et~al\mbox{.}}{2018}]%
        {stump2018}
\bibfield{author}{\bibinfo{person}{Larry Diehl}, \bibinfo{person}{Denis
  Firsov}, {and} \bibinfo{person}{Aaron Stump}.}
  \bibinfo{year}{2018}\natexlab{}.
\newblock \showarticletitle{Generic zero-cost reuse for dependent types}.
\newblock \bibinfo{journal}{\emph{{PACMPL}}} \bibinfo{volume}{2},
  \bibinfo{number}{{ICFP}} (\bibinfo{year}{2018}),
  \bibinfo{pages}{104:1--104:30}.
\newblock
\urldef\tempurl%
\url{https://doi.org/10.1145/3236799}
\showDOI{\tempurl}


\bibitem[\protect\citeauthoryear{Eisenberg}{Eisenberg}{2015}]%
        {overabundance-of-equalities}
\bibfield{author}{\bibinfo{person}{Richard~A. Eisenberg}.}
  \bibinfo{year}{2015}\natexlab{}.
\newblock \bibinfo{booktitle}{\emph{An Overabundance of Equality: Implementing
  Kind Equalities into Haskell}}.
\newblock \bibinfo{type}{{T}echnical {R}eport} MS-CIS-15-10.
  \bibinfo{institution}{University of Pennsylvania}.
\newblock
\urldef\tempurl%
\url{http://www.cis.upenn.edu/~eir/papers/2015/equalities/equalities.pdf}
\showURL{%
\tempurl}


\bibitem[\protect\citeauthoryear{Eisenberg}{Eisenberg}{2016}]%
        {eisenberg:phd}
\bibfield{author}{\bibinfo{person}{Richard~A. Eisenberg}.}
  \bibinfo{year}{2016}\natexlab{}.
\newblock \emph{\bibinfo{title}{Dependent Types in Haskell: Theory and
  Practice}}.
\newblock \bibinfo{thesistype}{Ph.D. Dissertation}. \bibinfo{school}{University
  of Pennsylvania}.
\newblock


\bibitem[\protect\citeauthoryear{Eisenberg, Vytiniotis, Peyton~Jones, and
  Weirich}{Eisenberg et~al\mbox{.}}{2014}]%
        {closed-type-families}
\bibfield{author}{\bibinfo{person}{Richard~A. Eisenberg},
  \bibinfo{person}{Dimitrios Vytiniotis}, \bibinfo{person}{Simon Peyton~Jones},
  {and} \bibinfo{person}{Stephanie Weirich}.} \bibinfo{year}{2014}\natexlab{}.
\newblock \showarticletitle{Closed Type Families with Overlapping Equations}.
  In \bibinfo{booktitle}{\emph{Principles of Programming Languages}}
  \emph{(\bibinfo{series}{POPL '14})}. \bibinfo{publisher}{ACM}.
\newblock


\bibitem[\protect\citeauthoryear{Gundry}{Gundry}{2013}]%
        {gundry:phd}
\bibfield{author}{\bibinfo{person}{Adam Gundry}.}
  \bibinfo{year}{2013}\natexlab{}.
\newblock \emph{\bibinfo{title}{Type Inference, Haskell and Dependent Types}}.
\newblock \bibinfo{thesistype}{Ph.D. Dissertation}. \bibinfo{school}{University
  of Strathclyde}.
\newblock


\bibitem[\protect\citeauthoryear{Harper, Honsell, and Plotkin}{Harper
  et~al\mbox{.}}{1993}]%
        {Harper:1993}
\bibfield{author}{\bibinfo{person}{Robert Harper}, \bibinfo{person}{Furio
  Honsell}, {and} \bibinfo{person}{Gordon Plotkin}.}
  \bibinfo{year}{1993}\natexlab{}.
\newblock \showarticletitle{A Framework for Defining Logics}.
\newblock \bibinfo{journal}{\emph{J. ACM}} \bibinfo{volume}{40},
  \bibinfo{number}{1} (\bibinfo{date}{Jan.} \bibinfo{year}{1993}),
  \bibinfo{pages}{143--184}.
\newblock
\showISSN{0004-5411}
\urldef\tempurl%
\url{https://doi.org/10.1145/138027.138060}
\showDOI{\tempurl}


\bibitem[\protect\citeauthoryear{Harper and Morrisett}{Harper and
  Morrisett}{1995}]%
        {morrisett:lmli}
\bibfield{author}{\bibinfo{person}{Robert Harper} {and}
  \bibinfo{person}{J.~Gregory Morrisett}.} \bibinfo{year}{1995}\natexlab{}.
\newblock \showarticletitle{Compiling Polymorphism Using Intensional Type
  Analysis}. In \bibinfo{booktitle}{\emph{Conference Record of POPL'95: 22nd
  {ACM} {SIGPLAN-SIGACT} Symposium on Principles of Programming Languages, San
  Francisco, California, USA, January 23-25, 1995}},
  \bibfield{editor}{\bibinfo{person}{Ron~K. Cytron} {and}
  \bibinfo{person}{Peter Lee}} (Eds.). \bibinfo{publisher}{{ACM} Press},
  \bibinfo{pages}{130--141}.
\newblock
\urldef\tempurl%
\url{https://doi.org/10.1145/199448.199475}
\showDOI{\tempurl}


\bibitem[\protect\citeauthoryear{Martin-L\"of}{Martin-L\"of}{1971}]%
        {martin-lof71}
\bibfield{author}{\bibinfo{person}{Per Martin-L\"of}.}
  \bibinfo{year}{1971}\natexlab{}.
\newblock \bibinfo{title}{A Theory of Types}.  (\bibinfo{year}{1971}).
\newblock
\newblock
\shownote{Unpublished manuscript.}


\bibitem[\protect\citeauthoryear{Miquel}{Miquel}{2001}]%
        {Miquel:ICC}
\bibfield{author}{\bibinfo{person}{Alexandre Miquel}.}
  \bibinfo{year}{2001}\natexlab{}.
\newblock \bibinfo{booktitle}{\emph{The Implicit Calculus of Constructions
  Extending Pure Type Systems with an Intersection Type Binder and Subtyping}}.
\newblock \bibinfo{publisher}{Springer Berlin Heidelberg},
  \bibinfo{address}{Berlin, Heidelberg}, \bibinfo{pages}{344--359}.
\newblock
\showISBNx{978-3-540-45413-7}
\urldef\tempurl%
\url{https://doi.org/10.1007/3-540-45413-6_27}
\showDOI{\tempurl}


\bibitem[\protect\citeauthoryear{Mishra{-}Linger and Sheard}{Mishra{-}Linger
  and Sheard}{2008}]%
        {Mishra-LingerS08}
\bibfield{author}{\bibinfo{person}{Nathan Mishra{-}Linger} {and}
  \bibinfo{person}{Tim Sheard}.} \bibinfo{year}{2008}\natexlab{}.
\newblock \showarticletitle{Erasure and Polymorphism in Pure Type Systems}. In
  \bibinfo{booktitle}{\emph{Foundations of Software Science and Computational
  Structures, 11th International Conference, {FOSSACS} 2008, Held as Part of
  the Joint European Conferences on Theory and Practice of Software, {ETAPS}
  2008, Budapest, Hungary, March 29 - April 6, 2008. Proceedings}}
  \emph{(\bibinfo{series}{Lecture Notes in Computer Science})},
  \bibfield{editor}{\bibinfo{person}{Roberto~M. Amadio}} (Ed.),
  Vol.~\bibinfo{volume}{4962}. \bibinfo{publisher}{Springer},
  \bibinfo{pages}{350--364}.
\newblock
\showISBNx{978-3-540-78497-5}
\urldef\tempurl%
\url{https://doi.org/10.1007/978-3-540-78499-9\_25}
\showDOI{\tempurl}


\bibitem[\protect\citeauthoryear{Nuyts and Devriese}{Nuyts and
  Devriese}{2018}]%
        {Nuyts18}
\bibfield{author}{\bibinfo{person}{Andreas Nuyts} {and}
  \bibinfo{person}{Dominique Devriese}.} \bibinfo{year}{2018}\natexlab{}.
\newblock \showarticletitle{Degrees of Relatedness: {A} Unified Framework for
  Parametricity, Irrelevance, Ad Hoc Polymorphism, Intersections, Unions and
  Algebra in Dependent Type Theory}. In \bibinfo{booktitle}{\emph{Proceedings
  of the 33rd Annual {ACM/IEEE} Symposium on Logic in Computer Science, {LICS}
  2018, Oxford, UK, July 09-12, 2018}}, \bibfield{editor}{\bibinfo{person}{Anuj
  Dawar} {and} \bibinfo{person}{Erich Gr{\"{a}}del}} (Eds.).
  \bibinfo{publisher}{{ACM}}, \bibinfo{pages}{779--788}.
\newblock
\urldef\tempurl%
\url{https://doi.org/10.1145/3209108.3209119}
\showDOI{\tempurl}


\bibitem[\protect\citeauthoryear{Nuyts, Vezzosi, and Devriese}{Nuyts
  et~al\mbox{.}}{2017}]%
        {Nuyts17}
\bibfield{author}{\bibinfo{person}{Andreas Nuyts}, \bibinfo{person}{Andrea
  Vezzosi}, {and} \bibinfo{person}{Dominique Devriese}.}
  \bibinfo{year}{2017}\natexlab{}.
\newblock \showarticletitle{Parametric quantifiers for dependent type theory}.
\newblock \bibinfo{journal}{\emph{{PACMPL}}} \bibinfo{volume}{1},
  \bibinfo{number}{{ICFP}} (\bibinfo{year}{2017}),
  \bibinfo{pages}{32:1--32:29}.
\newblock
\urldef\tempurl%
\url{https://doi.org/10.1145/3110276}
\showDOI{\tempurl}


\bibitem[\protect\citeauthoryear{Oury and Swierstra}{Oury and
  Swierstra}{2008}]%
        {swierstra:power-of-pi}
\bibfield{author}{\bibinfo{person}{Nicolas Oury} {and} \bibinfo{person}{Wouter
  Swierstra}.} \bibinfo{year}{2008}\natexlab{}.
\newblock \showarticletitle{The power of Pi}. In
  \bibinfo{booktitle}{\emph{Proceeding of the 13th {ACM} {SIGPLAN}
  international conference on Functional programming, {ICFP} 2008, Victoria,
  BC, Canada, September 20-28, 2008}}, \bibfield{editor}{\bibinfo{person}{James
  Hook} {and} \bibinfo{person}{Peter Thiemann}} (Eds.).
  \bibinfo{publisher}{{ACM}}, \bibinfo{pages}{39--50}.
\newblock
\urldef\tempurl%
\url{https://doi.org/10.1145/1411204.1411213}
\showDOI{\tempurl}


\bibitem[\protect\citeauthoryear{{Peyton Jones}, Vytiniotis, Weirich, and
  Washburn}{{Peyton Jones} et~al\mbox{.}}{2006}]%
        {gadt-type-inference}
\bibfield{author}{\bibinfo{person}{Simon {Peyton Jones}},
  \bibinfo{person}{Dimitrios Vytiniotis}, \bibinfo{person}{Stephanie Weirich},
  {and} \bibinfo{person}{Geoffrey Washburn}.} \bibinfo{year}{2006}\natexlab{}.
\newblock \showarticletitle{Simple unification-based type inference for
  {GADTs}}. In \bibinfo{booktitle}{\emph{International Conference on Functional
  Programming}} \emph{(\bibinfo{series}{ICFP '06})}. \bibinfo{publisher}{ACM}.
\newblock


\bibitem[\protect\citeauthoryear{Pfenning}{Pfenning}{2001}]%
        {Pfenning01}
\bibfield{author}{\bibinfo{person}{Frank Pfenning}.}
  \bibinfo{year}{2001}\natexlab{}.
\newblock \showarticletitle{Intensionality, Extensionality, and Proof
  Irrelevance in Modal Type Theory}. In \bibinfo{booktitle}{\emph{Proceedings
  of the 16th Annual Symposium on Logic in Computer Science (LICS'01)}},
  \bibfield{editor}{\bibinfo{person}{J.~Halpern}} (Ed.).
  \bibinfo{publisher}{IEEE Computer Society Press}, \bibinfo{address}{Boston,
  Massachusetts}, \bibinfo{pages}{221--230}.
\newblock


\bibitem[\protect\citeauthoryear{Sewell, {Zappa Nardelli}, Owens, Peskine,
  Ridge, Sarkar, and Strni\v{s}a}{Sewell et~al\mbox{.}}{2010}]%
        {ott}
\bibfield{author}{\bibinfo{person}{Peter Sewell}, \bibinfo{person}{Francesco
  {Zappa Nardelli}}, \bibinfo{person}{Scott Owens}, \bibinfo{person}{Gilles
  Peskine}, \bibinfo{person}{Thomas Ridge}, \bibinfo{person}{Susmit Sarkar},
  {and} \bibinfo{person}{Rok Strni\v{s}a}.} \bibinfo{year}{2010}\natexlab{}.
\newblock \showarticletitle{Ott: Effective tool support for the working
  semanticist}.
\newblock \bibinfo{journal}{\emph{Journal of Functional Programming}}
  \bibinfo{volume}{20}, \bibinfo{number}{1} (\bibinfo{date}{Jan.}
  \bibinfo{year}{2010}).
\newblock


\bibitem[\protect\citeauthoryear{Sozeau and Oury}{Sozeau and Oury}{2008}]%
        {coq-classes}
\bibfield{author}{\bibinfo{person}{Matthieu Sozeau} {and}
  \bibinfo{person}{Nicolas Oury}.} \bibinfo{year}{2008}\natexlab{}.
\newblock \showarticletitle{First-Class Type Classes}. In
  \bibinfo{booktitle}{\emph{Theorem Proving in Higher Order Logics, 21st
  International Conference, TPHOLs 2008, Montreal, Canada, August 18-21, 2008.
  Proceedings}} \emph{(\bibinfo{series}{Lecture Notes in Computer Science})},
  \bibfield{editor}{\bibinfo{person}{Otmane~A{\"{\i}}t Mohamed},
  \bibinfo{person}{C{\'{e}}sar~A. Mu{\~{n}}oz}, {and}
  \bibinfo{person}{Sofi{\`{e}}ne Tahar}} (Eds.), Vol.~\bibinfo{volume}{5170}.
  \bibinfo{publisher}{Springer}, \bibinfo{pages}{278--293}.
\newblock
\urldef\tempurl%
\url{https://doi.org/10.1007/978-3-540-71067-7\_23}
\showDOI{\tempurl}


\bibitem[\protect\citeauthoryear{Sulzmann, Chakravarty, {Peyton Jones}, and
  Donnelly}{Sulzmann et~al\mbox{.}}{2007}]%
        {systemfc}
\bibfield{author}{\bibinfo{person}{Martin Sulzmann}, \bibinfo{person}{Manuel
  M.~T. Chakravarty}, \bibinfo{person}{Simon {Peyton Jones}}, {and}
  \bibinfo{person}{Kevin Donnelly}.} \bibinfo{year}{2007}\natexlab{}.
\newblock \showarticletitle{System {F} with type equality coercions}. In
  \bibinfo{booktitle}{\emph{Types in languages design and implementation}}
  \emph{(\bibinfo{series}{TLDI '07})}. \bibinfo{publisher}{ACM}.
\newblock


\bibitem[\protect\citeauthoryear{Voevodsky}{Voevodsky}{2013}]%
        {voevodsky2013HTS}
\bibfield{author}{\bibinfo{person}{Vladimir Voevodsky}.}
  \bibinfo{year}{2013}\natexlab{}.
\newblock \showarticletitle{A simple type system with two identity types}.
\newblock \bibinfo{journal}{\emph{Unpublished note}} (\bibinfo{year}{2013}),
  \bibinfo{pages}{8 pages}.
\newblock
\urldef\tempurl%
\url{https://www.math.ias.edu/vladimir/sites/math.ias.edu.vladimir/files/HTS.pdf}
\showURL{%
\tempurl}


\bibitem[\protect\citeauthoryear{Weirich}{Weirich}{2014}]%
        {weirich:icfp14}
\bibfield{author}{\bibinfo{person}{Stephanie Weirich}.}
  \bibinfo{year}{2014}\natexlab{}.
\newblock \bibinfo{title}{Depending on Types}.
\newblock
\newblock
\newblock
\shownote{Invited keynote given at ICFP 2014.}


\bibitem[\protect\citeauthoryear{Weirich}{Weirich}{2017}]%
        {weirich:popl17}
\bibfield{author}{\bibinfo{person}{Stephanie Weirich}.}
  \bibinfo{year}{2017}\natexlab{}.
\newblock \bibinfo{title}{The Influence of Dependent Types}.
\newblock
\newblock
\newblock
\shownote{Invited keynote given at POPL 2017.}


\bibitem[\protect\citeauthoryear{Weirich, Hsu, and Eisenberg}{Weirich
  et~al\mbox{.}}{2013}]%
        {weirich:dwk}
\bibfield{author}{\bibinfo{person}{Stephanie Weirich}, \bibinfo{person}{Justin
  Hsu}, {and} \bibinfo{person}{Richard~A. Eisenberg}.}
  \bibinfo{year}{2013}\natexlab{}.
\newblock \showarticletitle{System {FC} with explicit kind equality}. In
  \bibinfo{booktitle}{\emph{Proceedings of The 18th ACM SIGPLAN International
  Conference on Functional Programming}} \emph{(\bibinfo{series}{ICFP '13})}.
  \bibinfo{address}{Boston, MA}, \bibinfo{pages}{275--286}.
\newblock


\bibitem[\protect\citeauthoryear{Weirich, Voizard, de~Amorim, and
  Eisenberg}{Weirich et~al\mbox{.}}{2017}]%
        {weirich:systemd}
\bibfield{author}{\bibinfo{person}{Stephanie Weirich}, \bibinfo{person}{Antoine
  Voizard}, \bibinfo{person}{Pedro Henrique~Avezedo de Amorim}, {and}
  \bibinfo{person}{Richard~A. Eisenberg}.} \bibinfo{year}{2017}\natexlab{}.
\newblock \showarticletitle{A Specification for Dependent Types in {Haskell}}.
\newblock \bibinfo{journal}{\emph{Proc. ACM Program. Lang.}}
  \bibinfo{volume}{1}, \bibinfo{number}{ICFP}, Article \bibinfo{articleno}{31}
  (\bibinfo{date}{Aug.} \bibinfo{year}{2017}), \bibinfo{numpages}{29}~pages.
\newblock
\showISSN{2475-1421}
\urldef\tempurl%
\url{https://doi.org/10.1145/3110275}
\showDOI{\tempurl}


\bibitem[\protect\citeauthoryear{Weirich, Vytiniotis, {Peyton Jones}, and
  Zdancewic}{Weirich et~al\mbox{.}}{2011}]%
        {weirich:newtypes}
\bibfield{author}{\bibinfo{person}{Stephanie Weirich},
  \bibinfo{person}{Dimitrios Vytiniotis}, \bibinfo{person}{Simon {Peyton
  Jones}}, {and} \bibinfo{person}{Steve Zdancewic}.}
  \bibinfo{year}{2011}\natexlab{}.
\newblock \showarticletitle{Generative Type Abstraction and Type-level
  Computation}. In \bibinfo{booktitle}{\emph{POPL 11: 38th {ACM SIGACT-SIGPLAN}
  Symposium on Principles of Programming Languages, January 26--28, 2011.
  Austin, TX, USA.}} \bibinfo{pages}{227--240}.
\newblock


\bibitem[\protect\citeauthoryear{Winant and Devriese}{Winant and
  Devriese}{2018}]%
        {winant2018}
\bibfield{author}{\bibinfo{person}{Thomas Winant} {and}
  \bibinfo{person}{Dominique Devriese}.} \bibinfo{year}{2018}\natexlab{}.
\newblock \showarticletitle{Coherent explicit dictionary application for
  Haskell}. In \bibinfo{booktitle}{\emph{Proceedings of the 11th {ACM}
  {SIGPLAN} International Symposium on Haskell, Haskell@ICFP 2018, St. Louis,
  MO, USA, September 27-17, 2018}}, \bibfield{editor}{\bibinfo{person}{Nicolas
  Wu}} (Ed.). \bibinfo{publisher}{{ACM}}, \bibinfo{pages}{81--93}.
\newblock
\urldef\tempurl%
\url{https://doi.org/10.1145/3242744.3242752}
\showDOI{\tempurl}


\bibitem[\protect\citeauthoryear{Xi, Chen, and Chen}{Xi et~al\mbox{.}}{2003}]%
        {xi-gadt}
\bibfield{author}{\bibinfo{person}{Hongwei Xi}, \bibinfo{person}{Chiyan Chen},
  {and} \bibinfo{person}{Gang Chen}.} \bibinfo{year}{2003}\natexlab{}.
\newblock \showarticletitle{Guarded recursive datatype constructors}. In
  \bibinfo{booktitle}{\emph{Principles of Programming Languages}}
  \emph{(\bibinfo{series}{POPL '03})}. \bibinfo{publisher}{ACM}.
\newblock


\bibitem[\protect\citeauthoryear{Xie and Eisenberg}{Xie and Eisenberg}{2018}]%
        {xie-coercion-quantification}
\bibfield{author}{\bibinfo{person}{Ningning Xie} {and}
  \bibinfo{person}{Richard~A. Eisenberg}.} \bibinfo{year}{2018}\natexlab{}.
\newblock \showarticletitle{Coercion Quantification}. In
  \bibinfo{booktitle}{\emph{Haskell Implementors' Workshop}}.
\newblock


\end{thebibliography}

\newpage

\ifextended
\appendix

\admissibletrue

\section{Complete specification of System~DR}
\label{app:ext}

The complete specification of System~DR appears in this appendix, and includes
the actual rules that we used to generate our Coq definitions, via Ott. Some
rules in the paper have been modified for presentation purposes; they appear
in their complete form here.

On modification that we made was to remove redundant hypotheses from rules in
the main body of the paper when they were implied via regularity.  We have
proven (in Coq) that these additional premises are admissible, so their
removal does not change the type
system.\ifsource\footnote{\url{ext_invert.v:E_Pi2,E_Abs2,E_CPi2,E_CAbs2,E_Fam2}}\fi
\ifsource\footnote{\url{ext_invert.v:E_Wff2,E_PiCong2,E_AbsCong2,E_CPiCong2,E_CAbsCong2}}\fi
\ifsource\footnote{\url{ext_red.v:E_Beta2}}\fi These redundant hypotheses are
marked via square brackets in the complete system below.

We include these redundant hypotheses in our rules for two reasons. First,
sometimes these hypotheses simplify the reasoning and allow us to prove
properties more independently of one another.  For example, in the
\rref{E-Beta} rule, we require $\ottnt{a_{{\mathrm{2}}}}$ to have the same type as
$\ottnt{a_{{\mathrm{1}}}}$. However, this type system supports the preservation lemma so this
typing premise will always be derivable. But, it is convenient to prove the
regularity property early, so we include that hypothesis.

Another source of redundancy comes from our use of the Coq proof assistant.
Some of our proofs require the use of induction on judgments that are not
direct premises, but are derived from other premises via regularity. These
derivations are always the same height or shorter than the original, so this
use of induction is justified.  However, while Coq natively supports proofs by
induction on derivations, it does not natively support induction on the
\emph{heights} of derivations. Therefore, to make these induction hypotheses
available for reasoning, we include them as additional premises.

\section{Grammar}

Metavariables
\[
\begin{array}{ll@{\qquad}ll@{\qquad}ll}
\ottmv{x} & \text{term/type variable} & \ottmv{c} & \text{coercion variable} &
\ottmv{F}, \ottmv{T} & \text{constant}
\end{array}
\]

Grammar
\[
\begin{array}{llcl}
\mathit{role}              & R         & ::=& \ottkw{Nom}  \alt \ottkw{Rep} \\
\mathit{relevance}         & \rho   & ::=& + \alt - \\
\mathit{application\ flag\ (terms)} & \nu    & ::=& \ottnt{R} \alt \rho \\
\mathit{application\ flag\ (any)}  & \upsilon   & ::=&  \nu  \alt  \bullet  \\
\\
\\
\mathit{terms,\ types} & a, b,    & ::=&  \star  \alt \ottmv{x} \alt \ottmv{F} \alt
                                              \mathrm{\lambda}^{ \rho } \ottmv{x} . \ottnt{b}  \alt  \ottnt{a} \  \ottnt{b} ^{ \nu }  \alt \Box \\
                       & A, B             &\alt&  \mathrm{\Pi}^ \rho \ottmv{x} \!:\! \ottnt{A} \to \ottnt{B}  \alt
                                              \mathrm{\Lambda} \ottmv{c} . \ottnt{a} 
                                        \alt  \ottnt{a} \; \bullet  \alt  \forall \ottmv{c} \!:\! \phi . \ottnt{A}  \\
                       &              &\alt&  \mathsf{case} \hspace{3pt}  \ottnt{a}  \hspace{3pt} \mathsf{of} \hspace{3pt}  \ottmv{F} \  \overline{\upsilon}  \rightarrow  \ottnt{b_{{\mathrm{1}}}}  \| \_ \rightarrow  \ottnt{b_{{\mathrm{2}}}}  \\
\mathit{propositions}  &\phi       & ::=&  \ottnt{a}   \sim _{ \ottnt{R} }  \ottnt{b}  :  \ottnt{A}  \\
\\
\\
\mathit{contexts}       &\Gamma        & ::=& \varnothing \alt  \Gamma ,  \ottmv{x} \!:\! \ottnt{A}  \alt
                                              \Gamma ,  \ottmv{c} \!:\! \phi \\
\mathit{variable\ set}  &\Delta        & ::=&  \varnothing  \alt \Delta  \ottsym{,}  \ottmv{c} \alt \Delta  \ottsym{,}  \ottmv{x}\\

\mathit{applicator}     &\mu & ::=&  \ottnt{a} ^{ \nu }  \alt  \bullet  \\

%
\mathit{application\ path} &\ottnt{w}           & ::=& \ottmv{F} \alt  \ottnt{w} \  \ottnt{b} ^{ \ottnt{R} }  \alt  \ottnt{w} \  \ottnt{b} ^{ \ottsym{+} }  \alt
                                                 \ottnt{w} \  \Box ^{ \ottsym{-} }  \alt  \ottnt{w} \; \bullet   \\

\mathit{pattern}        &\ottnt{p}           & ::=& \ottmv{F} \alt  \ottnt{p} \  \ottmv{x} ^{ \ottnt{R} }  \alt  \ottnt{p} \  \ottmv{x} ^{ \ottsym{+} }  \alt
                                                 \ottnt{p} \  \Box ^{ \ottsym{-} }  \alt  \ottnt{p} \; \bullet   \\

\mathit{signature}      &\Sigma           & ::=&  \varnothing  \alt  \Sigma  \cup \{ \ottmv{F}  :   \ottnt{A} \  \ottsym{@} \  \overline{R}  \}  \alt
                                                 \Sigma  \cup \{ \ottmv{F}  :   \ottnt{A} \  \ottsym{@} \  \overline{R} \ \mathsf{where}\  \ottnt{p}  \sim_{ \ottnt{R} }  \ottnt{a}  \}  \\
\end{array}
\]
Notation
\begin{itemize}
\item $\ottnt{A}  \to  \ottnt{B}$ is shorthand for $ \mathrm{\Pi}^ \ottsym{+} \ottmv{x} \!:\! \ottnt{A} \to \ottnt{B} $, where $\ottmv{x}$ is not free in $\ottnt{B}$.
\item $\phi  \Rightarrow  \ottnt{A}$ is shorthand for $ \forall \ottmv{c} \!:\! \phi . \ottnt{A} $, where $\ottmv{c}$ is not free in $\ottnt{A}$.
\item $\ottmv{F}\ \overline{\mu}$ can be written as an abbreviation for a path $\ottnt{w}$.
\end{itemize}

The key features of System~DR language include
\begin{itemize}
\item a single sort ($ \star $) for classifying types (including itself);
\item relevant functions ($ \mathrm{\lambda}^{ \ottsym{+} } \ottmv{x} . \ottnt{a} $) with dependent types
  ($ \mathrm{\Pi}^ \ottsym{+} \ottmv{x} \!:\! \ottnt{A} \to \ottnt{B} $), and their associated application form ($ \ottnt{a} \  \ottnt{b} ^{ \ottsym{+} } $);
\item irrelevant functions ($ \mathrm{\lambda}^{ \ottsym{-} } \ottmv{x} . \ottnt{a} $) with dependent types
  ($ \mathrm{\Pi}^ \ottsym{-} \ottmv{x} \!:\! \ottnt{A} \to \ottnt{B} $), and their associated application form ($ \ottnt{a} \  \Box ^{ \ottsym{-} } $);
\item coercion abstractions ($ \mathrm{\Lambda} \ottmv{c} . \ottnt{a} $), their types
  ($ \forall \ottmv{c} \!:\! \phi . \ottnt{B} $), and instantiation form ($ \ottnt{a} \; \bullet $);
\item constants and top level definitions ($\ottmv{F}$), and an associated
  role-annotated application form ($ \ottnt{a} \  \ottnt{b} ^{ \ottnt{R} } $);
\item and a simple form of case analysis
  ($ \mathsf{case} \hspace{3pt}  \ottnt{a}  \hspace{3pt} \mathsf{of} \hspace{3pt}  \ottmv{F} \  \overline{\upsilon}  \rightarrow  \ottnt{b_{{\mathrm{1}}}}  \| \_ \rightarrow  \ottnt{b_{{\mathrm{2}}}} $)
  that matches against the head constructor of a value.
\end{itemize}

In this grammar, term and type variables, $\ottmv{x}$, are bound in the bodies of
functions and their types. Similarly, coercion variables, $\ottmv{c}$, are bound
in the bodies of coercion abstractions and their types.  (Technically,
irrelevant term/type variables and coercion variables are prevented by the
typing rules from actually appearing in the bodies of their respective
abstractions.)

We use the tools Ott~\cite{ott} and LNgen~\cite{aydemir:lngen} to represent
the binding structure using a \emph{locally nameless
  representation}~\cite{aydemir:popl-binders}.  However, these tools do not
extend to structures that bind more than one variable at a time. This causes
difficulty in newtype axioms and case expressions.  In newtype axioms, which
are expressed in the signature as $ \ottnt{A} \  \ottsym{@} \  \overline{R} \ \mathsf{where}\  \ottnt{p}  \sim_{ \ottnt{R} }  \ottnt{a} $, variables that
occur in the pattern $\ottnt{p}$ should be bound in $\ottnt{a}$. We implement this
binding structure by manually renaming the right-hand side when necessary.
Binding in case expressions is done using a nameless pattern and higher-order
binding.

\section{Toplevel Signature}

The system is defined with respect to a toplevel signature, written
$ \Sigma_0 $, that satisfies the signature checking judgment $\vDash   \Sigma_0 $.

\section{Roles}

\subsection{Roles}

Roles are ordered, with bottom $\ottkw{Nom}$ and top $\ottkw{Rep}$.

\ottdefnsJSubRole{}

The operation $ \ottnt{R_{{\mathrm{1}}}} \wedge \ottnt{R_{{\mathrm{2}}}} $ computes the minimum role and has the
following definition in the system with two roles:

\[
\begin{array}{ll}
 \ottkw{Nom} \wedge \ottnt{R}    &= \ottkw{Nom} \\
 \ottnt{R} \wedge \ottkw{Nom}    &= \ottkw{Nom} \\
 \ottkw{Rep} \wedge \ottkw{Rep}  &= \ottkw{Rep} \\
\end{array}
\]

\section{Values and Reduction relations}

\subsection{Values}
\label{app:values}

Values include type forms, abstractions and certain application paths.
\ottdefnsJValue{}

Application paths are values when they are of the right form (i.e. when
$ \mathsf{Head}( \ottnt{a} )  \mathrel{+\!\!\!\!\!\!\rightarrow}   \ottmv{F} $ is derivable) and when they cannot step at the current
role (i.e. when $ \mathsf{CasePath}_{ \ottnt{R} }\;  \ottnt{a}  =  \ottmv{F} $ is derivable).
Paths could fail to step because they are headed by constnats, by newtype axioms
(cf. \rref{CasePath-Const}) or because their head does not match the pattern
of the axiom (cf. \rref{CasePath-UnMatch}).

\ottdefnsJValuePath{}
\ottdefnsJCasePath{}

A term matches a pattern when there is some prefix of the term that can be
unified with the pattern.

\ottdefnsJTmPatternAgree{}
\ottdefnsJSubTmPatternAgree{}

\subsection{Reduction}

System DR separates primitive reductions, written $ \vDash   \ottnt{a}  \rightarrow^{\beta}_{ \ottnt{R} }  \ottnt{b} $ from the
full small-step reduction relation, written $ \vDash   \ottnt{a}   \leadsto _{ \ottnt{R} }  \ottnt{b} $.  The former
includes call-by-name beta reduction, axiom unfolding and the two primitive
reduction rules for case. The latter relation extends the former with
congruence rules.

\paragraph{Primitive reduction}
\label{app:beta}
Axiom matching requires the following two auxiliary relations, for
renaming the variables bound in the pattern and for matching a path against
a pattern, producing a substitution and then applying that substitution to
the right-hand side of the axiom.
\ottdefnsJRename{}
\ottdefnsJMatchSubst{}
Case analysis requires the following auxiliary relations, for determining
whether the path matches a list of applicators for the pattern, and
for applying the arguments of the scrutinee to the branch. The case expression
itself should be saturated with respect to the relevant, role-annotated
arguments of the type constructor.
\ottdefnAppsPath{}
\ottdefnsJApplyArgs{}
\ottdefnAppRoles{}
\ottdefnSatApp{}

Beta reduction includes the usual application rules for abstractions (note
that irrelevant quantification must reduce the body of the abstraction to a
value before reduction can occur). Axiom matching occurs when the term matches
against the pattern. Pattern matching occurs when the scrutinee matches
against the list of application flags.
\ottdefnBeta{}

\subsection{One-step reduction}
\label{app:reduction_in_one}
\ottdefnreductionXXinXXone{}

\subsection{Typing}
\label{app:typing}
Auxiliary definitions for typing.
\ottdefnBranchTyping{}
\ottdefnRolePath{}
\ottdefnAppsPath{}
\ottdefnsJChk{}

The following four judgments are mutually defined.
\ottdefnTyping{}
\ottdefnPropWff{}
\ottdefnIso{}
\ottdefnDefEq{}
\ottdefnCtx{}

\subsection{Role-Checking judgment}
\label{app:roleing}
\ottdefnsJroleing{}

\subsection{Signatures}
\ottdefnPatternContexts{}
\ottdefnSig{}

\subsection{Parallel reduction}
\label{app:parallel}
\ottdefnPar{}

\fi 

\end{document}